\documentclass[useAMS,usenatbib]{mn2e}

\usepackage{graphicx}
\usepackage{amsmath}
\usepackage{mathrsfs}
\usepackage{natbib}
\usepackage{float}

\def   \araa {{\rm { ARA\&{}A}}}
\def   \apj {{\rm {ApJ}}}
\def   \apjl {{\rm { ApJL}}}
\def   \apjs {{\rm { ApJS}}}
\def   \apss {{\rm { Ap\&{}SS}}}
\def   \aap {{\rm { A\&{}A}}}

\def   \aaps {{\rm { A\&{}AS}}}

\def   \mnras {{\rm { MNRAS}}}

\def   \solphys {{\rm \emph{Sol. Phys.}}}

\title[A spectroscopic survey of Herbig Ae/Be stars I]{
A spectroscopic survey of Herbig Ae/Be stars with X-Shooter I: Stellar parameters and accretion rates \thanks{Based on observations using the ESO Very Large Telescope, at Cerro Paranal
, under the observing program 084.C-0952A} }
\author[J.~R.~Fairlamb et al.]{J.~R.~Fairlamb$^{1}$\thanks{E-mail; pyjrf@leeds.ac.uk} R.~D.~Oudmaijer$^{1}$ I.~Mendigut\'{i}a$^{1}$ J.~D.~Ilee$^{2}$ 
\newauthor
M.~E.~van den Ancker$^{3}$\\
$^{1}$School of Physics and Astronomy, EC Stoner Building, University of Leeds, Leeds, LS2 9JT, UK\\
$^{2}$ SUPA, School of Physics and Astronomy, University of St Andrews, North Haugh, St Andrews, Scotland, KY16 9SS, UK\\
$^{3}$European Southern Observatory (ESO), Karl-Schwarzschild-Str. 2, 85748 Garching, Germany}

\begin{document}

\defcitealias{Calvet1998}{CG98}

\date{Accepted 2015 July 13.  Received 2015 July 13; in original form 2015 March 3}

\pagerange{\pageref{firstpage}--\pageref{lastpage}} \pubyear{20XX}

\maketitle

\label{firstpage}


\begin{abstract}

Herbig Ae/Be stars span a key mass range that links low and high mass stars, and thus provide an ideal window from which to explore their formation. This paper presents VLT/X-Shooter spectra of 91 Herbig Ae/Be stars, HAeBes; the largest spectroscopic study of HAeBe accretion to date.  A homogeneous approach to determining stellar parameters is undertaken for the majority of the sample.  Measurements of the ultra-violet (UV) are modelled within the context of magnetospheric accretion, allowing a direct determination of mass accretion rates.  Multiple correlations are observed across the sample between accretion and stellar properties: the youngest and often most massive stars are the strongest accretors, and there is an almost 1:1 relationship between the accretion luminosity and stellar luminosity.  Despite these overall trends of increased accretion rates in HAeBes when compared to classical T Tauri stars, we also find noticeable differences in correlations when considering the Herbig Ae and Herbig Be subsets. This, combined with the difficulty in applying a magnetospheric accretion model to some of the Herbig Be stars, could suggest that another form of accretion may be occurring within the Herbig Be mass range.

\end{abstract}

\begin{keywords}
stars: early-type --
stars: variables: Herbig Ae/Be -- 
stars: pre-main sequence --
stars: formation --
accretion, accretion discs --
techniques: spectroscopic.
\end{keywords}


\section{Introduction}
\label{sec:intro}

Herbig Ae/Be stars are pre-main sequence, PMS, stars that bridge the key mass range of 2--10~M$_\odot$, between low and high mass stars. Their importance lies in linking the reasonably well understood formation of low mass stars, like the numerous Classical T Tauri stars, CTTs (which have $M_\star<2$~M$_\odot$ and will go on to form stars like our own Sun),  to the rarer, more deeply embedded high mass stars (or massive young stellar objects, MYSOs). The greater number of HAeBes compared to forming high mass stars, along with them being closer and optically visible, makes them a powerful link and important tool in furthering our understanding of star formation upto the high mass star formation regime.

HAeBes were originally identified by \citet{Herbig1960} in an attempt to push the mass boundaries in understanding of CTTs, and had to meet the criteria of: ``spectral type A or earlier with emission lines; lies in an obscured region; the star illuminates fairly bright nebulosity in its immediate vicinity". The latter two criteria have been relaxed since then in order to find more potential targets \citep[see ][]{Finkenzeller1984,The1994,Vieira2003}. In these surveys more attention has been drawn towards the colours of the objects, particularly in the infra-red, IR. This is because the IR often displays an excess of emission, compared to main-sequence, MS, stars of the same spectral type, and is the product of the circumstellar disc around HAeBes. This has been confirmed in numerous studies \citep{vandenAncker2000,Meeus2001}, and by direct observations in the optical \citep{McCaughrean1996,Grady2001}, sub-mm \citep{Mannings1997}, and of scattered, polarised light \citep{Vink2002,Vink2005}. It is these indicators which ultimately point to the stars being young and in the PMS phase. This young nature, suspected by Herbig, was confirmed by \citet{Strom1972} who observed the HAeBes to have lower surface gravities than MS stars. With their PMS nature established, the next obvious questions are: what happens with the disc-star interaction; how do they both evolve; and what physical mechanisms are present in these interactions? And of course, how are these aspects related to each other between the CTTs and massive young stellar objects?

With regards to the disc-star interaction, an ultra-violet, UV, excess in the CTTs \citep{Garrison1978,Gullbring1998} has been shown to match well with the theory of disc-to-star accretion within the Magnetospheric Accretion, MA, regime \citep{Calvet1998, Gullbring2000, Ingleby2013}. Under this paradigm the disc is truncated by the stellar magnetic field lines, and from here the material is funnelled by the field lines, in free-fall, onto the star. The accreted material shocks the photosphere causing X-ray emission, the majority of which is then absorbed by the surroundings, heating them, and is re-emitted at longer wavelengths; giving rise to an observable  UV excess \citep{Calvet1998}. Therefore, measurement of the UV excess can be directly related to accretion from the disc to the star. The accretion rate has large implications on PMS systems; it affects the achievable final mass of not just the star, but also any possible planets forming within the disc and the timescales upon which they can evolve \citep{Lubow2006,Dunhill2015}. Establishing an accretion rate also requires accurate stellar parameters which, until now, have mostly been performed on an ad-hoc basis dependent on the particular study. The results of this work will also help future studies to disentangle the complexities in the environments of these such as: outflows, infalling material, structure, and even possible ongoing planet formation.

However, for MA to be applicable the star must have a magnetic field with sufficient strength to truncate the disc; for CTTs this is of the order of kilo-Gauss \citep{Ghosh1979, Koenigl1991, Shu1994, Johns-Krull2007, Bouvier2007}. For MS stars, magnetic fields driven by convection are not predicted to exist for stars with $T_{\rm eff} > 8300$~K \citep{Simon2002}. However, this may not be the case for PMS stars as there have been a few detections of magnetic fields in HAeBes \citep{Wade2005, Catala2007, Hubrig2009}, but the origin of their fields remains unknown (be they dynamo generated or the result of fossil fields). The largest survey into the magnetic fields of HAeBes was performed recently by \citet{Alecian2013}, and yields clear detections of only 5 stars out of 70. When applying the theory of MA \citep{Koenigl1991,Shu1994} to HAeBes, a weak dipole magnetic field of only a few hundred gauss, or even less, is needed for MA to occur \citep{Wade2007,Cauley2014}. These strengths are below current detection limits, meaning that MA acting in HAeBes is still a possibility.

A key aim of this paper is to provide the largest survey on direct accretion tracers in HAeBes to date. To do this, measurements of the UV excess are made and fitted within the context of MA shock modelling. This method of accretion-shock modelling has been successfully adapted and applied, in the majority of cases, to small sets of HAeBes in recent years \citep{Muzerolle2004, Donehew2011, Mendigutia2011b, Pogodin2012, Mendigutia2013, Mendigutia2014}. However, the number of HBes analysed in previous works is often small, particularly for early-type HBes, and needs to be tested further.
The derivation of accretion rates, and other properties, depend heavily upon basic stellar parameters. Thus far, most works on accretion in HAeBes build upon previous stellar parameter determinations from a wide range of sources and methodologies. This can result in parameters which are not directly comparable within a sample. Therefore, our approach is to provide a homogeneous determination of stellar parameters for as many of the targets as possible. This not only helps in our accretion determination but also  helps in future works which require basic stellar parameters i.e. detailed modelling of the circumstellar disc, or energy balance in the SED.

The overall aim of this paper is to provide a quantitative look into the properties of HAeBes, with a particular emphasis on how the accretion rate varies as a function of stellar parameters; along with an assessment of the applicability of using MA to obtain the accretion rate. To do this we present 91 HAeBe objects observed with the X-Shooter spectrograph, VLT, Chile. The paper is broken down into the following sections: Section \ref{sec:xs} details the sample, observations, and data reduction; Section \ref{sec:st_title} details the methods, and results, of deriving the basic stellar parameters; Section \ref{sec:be} presents the methods of measuring the UV excess; Section \ref{sec:anal} presents the derived accretion rates, including a detailed description of how MA is applied to the HAeBe sample; Section \ref{sec:dis} forms the discussion; and Section \ref{sec:conc} provides the conclusions of this work. Finally, photometric data and literature information on the sample are provided in the appendix.

\section[]{Observations and Data Reduction}
\label{sec:xs}


\begin{table*}
 \centering
 \begin{minipage}{190mm}
  \caption{Column 1 shows the target names, columns 2 and 3 are RA and DEC, column 4 gives the observation date, columns 5-7 give the exposure times for each arm, column 8 is the number of Detector Integration Times for the NIR arm, and finally columns 9-11 give the Signal-to-Noise Ratio in each arm.}
  \label{tab:obs}
  \begin{tabular}{lccc rrrc ccc}
  \hline
   Name   &  RA & DEC & Obs Date & \multicolumn{4}{c}{Exposure Time (s)} & \multicolumn{3}{c}{SNR}\\
          &  \multicolumn{2}{c}{(J2000)}  & (yyyy/mm/dd) & UVB & VIS & NIR & NDIT & UVB & VIS & NIR \\ 
  \hline
 
UX Ori &  05:04:29.9 & -03:47:16.8   &  2009-10-05 & $ 90\times4$ & $ 90\times4$ & $  5\times4$ &  18 &   42 & 257 & 306 \\
PDS 174 &  05:06:55.4 & -03:21:16.0   &  2009-10-05 & $300\times4$ & $300\times4$ & $( 20\times  6)\times4$ &   2 &  127 &  64 & 225 \\
V1012 Ori &  05:11:36.5 & -02:22:51.1   &  2009-10-05 & $250\times4$ & $(125\times  2)\times4$ & $ 20\times4$ &  12 &  128 & 157 & 229 \\
HD 34282 &  05:16:00.4 & -09:48:38.5   &  2009-12-06 & $( 15\times  2)\times4$ & $( 15\times  2)\times4$ & $ 10\times4$ &   3 &  171 & 215 & 170 \\
HD 287823 &  05:24:08.1 &  02:27:44.4   &  2009-12-06 & $( 15\times  2)\times4$ & $( 15\times  2)\times4$ & $ 15\times4$ &   2 &   99 & 171 & 256 \\
HD 287841 &  05:24:42.8 &  01:43:45.4   &  2009-12-06 & $ 90\times4$ & $ 90\times4$ & $ 15\times4$ &   6 &   78 & 200 & 323 \\
HD 290409 &  05:27:05.3 &  00:25:04.9   &  2010-01-02 & $( 15\times  2)\times4$ & $( 15\times  2)\times4$ & $ 15\times4$ &   2 &  195 & 157 & 130 \\
HD 35929 &  05:27:42.6 & -08:19:40.8   &  2009-12-17 & $( 10\times  2)\times4$ & $( 10\times  2)\times4$ & $  5\times4$ &   8 &   36 & 125 & 161 \\
HD 290500 &  05:29:48.0 & -00:23:45.8   &  2009-12-17 & $150\times4$ & $150\times4$ & $ 75\times4$ &   2 &  208 & 341 & 331 \\
HD 244314 &  05:30:18.9 &  11:20:18.2   &  2010-01-02 & $( 15\times  2)\times4$ & $( 15\times  2)\times4$ & $ 15\times4$ &   2 &   64 & 140 & 180 \\
HK Ori &  05:31:28.1 &  12:09:07.6   &  2009-12-17 & $250\times4$ & $250\times4$ & $ 20\times4$ &  12 &   49 & 120 & 192 \\
HD 244604 &  05:31:57.3 &  11:17:38.8   &  2009-12-17 & $( 15\times  2)\times4$ & $( 15\times  2)\times4$ & $  5\times4$ &   8 &   76 & 170 & 213 \\
UY Ori &  05:32:00.4 & -04:55:54.6   &  2009-12-26 & $300\times4$ & $300\times4$ & $( 50\times  6)\times4$ &   1 &  229 & 326 & 143 \\
HD 245185 &  05:35:09.7 &  10:01:49.9   &  2009-12-17 & $( 15\times  2)\times4$ & $( 15\times  2)\times4$ & $ 15\times4$ &   2 &  240 & 171 & 213 \\
T Ori &  05:35:50.6 & -05:28:36.9   &  2009-12-17 & $( 25\times  2)\times4$ & $( 25\times  2)\times4$ & $(  2\times  5)\times4$ &   5 &  156 & 198 & 132 \\
V380 Ori &  05:36:25.5 & -06:42:58.9   &  2009-12-17 & $( 80\times  2)\times4$ & $( 80\times  2)\times4$ & $(  3\times  2)\times4$ &  20 &    9 & 178 & 161 \\
HD 37258 &  05:36:59.1 & -06:09:17.9   &  2010-01-02 & $( 15\times  2)\times4$ & $( 15\times  2)\times4$ & $ 10\times4$ &   3 &   62 & 219 & 168 \\
HD 290770 &  05:37:02.5 & -01:37:21.3   &  2009-12-26 & $( 15\times  2)\times4$ & $( 15\times  2)\times4$ & $  7\times4$ &   4 &  261 & 244 & 262 \\
BF Ori &  05:37:13.2 & -06:35:03.3   &  2010-01-02 & $ 90\times4$ & $ 90\times4$ & $ 10\times4$ &   9 &   25 & 185 & 329 \\
HD 37357 &  05:37:47.2 & -06:42:31.7   &  2010-02-05 & $( 10\times  2)\times4$ & $( 10\times  2)\times4$ & $ 10\times4$ &   4 &  123 & 209 & 161 \\
HD 290764 &  05:38:05.3 & -01:15:22.2   &  2009-12-26 & $( 30\times  2)\times4$ & $( 30\times  2)\times4$ & $  7\times4$ &   8 &   61 & 199 & 233 \\
HD 37411 &  05:38:14.6 & -05:25:14.4   &  2010-02-05 & $( 25\times  2)\times4$ & $( 25\times  2)\times4$ & $ 15\times4$ &   3 &  215 & 183 & 138 \\
V599 Ori &  05:38:58.4 & -07:16:49.2   &  2010-01-06 & $360\times4$ & $(180\times  2)\times4$ & $( 10\times  2)\times4$ &  10 &   61 & 149 & 328 \\
V350 Ori &  05:40:11.9 & -09:42:12.2   &  2010-02-05 & $150\times4$ & $150\times4$ & $ 25\times4$ &   6 &   77 & 260 & 150 \\
HD 250550 &  06:01:59.9 &  16:30:53.4   &  2010-01-02 & $( 15\times  2)\times4$ & $( 15\times  2)\times4$ & $  3\times4$ &  10 &  103 & 218 & 200 \\
V791 Mon &  06:02:15.0 & -10:01:01.4   &  2010-02-24 & $ 90\times4$ & $ 90\times4$ & $ 15\times4$ &   6 &   67 & 343 & 147 \\
PDS 124 &  06:06:58.5 & -05:55:09.2   &  2010-02-10 & $300\times4$ & $300\times4$ & $( 50\times  6)\times4$ &   1 &  188 &  62 & 119 \\
LkHa 339 &  06:10:57.7 & -06:14:41.8   &  2010-01-17 & $300\times4$ & $300\times4$ & $ 60\times4$ &   5 &  122 &  63 & 297 \\
VY Mon &  06:31:06.8 &  10:26:02.9   &  2010-02-08 & $(150\times  2)\times4$ & $(100\times  3)\times4$ & $(  2\times  6)\times4$ &  20 &   11 &  49 & 102 \\
R Mon &  06:39:10.0 &  08:44:08.2   &  2010-02-01 & $300\times4$ & $(150\times  2)\times4$ & $(  2\times  6)\times4$ &  20 &   11 &  49 & 141 \\
V590 Mon &  06:40:44.7 &  09:47:59.7   &  2010-02-01 & $300\times4$ & $300\times4$ & $ 50\times4$ &   6 &  143 & 166 & 288 \\
PDS 24 &  06:48:41.8 & -16:48:06.0   &  2009-12-16 & $300\times4$ & $300\times4$ & $ 90\times4$ &   3 &  145 &  48 & 260 \\
PDS 130 &  06:49:58.7 & -07:38:52.1   &  2009-12-16 & $300\times4$ & $300\times4$ & $ 60\times4$ &   5 &  125 & 122 & 273 \\
PDS 229N &  06:55:40.1 & -03:09:53.1   &  2010-02-10 & $300\times4$ & $300\times4$ & $100\times4$ &   3 &  111 & 105 & 191 \\
GU CMa &  07:01:49.6 & -11:18:03.9   &  2009-12-16 & $(  2\times  3)\times4$ & $(  2\times  3)\times4$ & $(  2\times  2)\times4$ &   6 &  127 & 180 & 142 \\
HT CMa &  07:02:42.7 & -11:26:12.3   &  2010-01-30 & $300\times4$ & $300\times4$ & $ 30\times4$ &  10 &  201 & 245 & 274 \\
Z CMa &  07:03:43.2 & -11:33:06.7   &  2010-02-24 & $( 75\times  2)\times4$ & $( 10\times  3)\times4$ & $(  0.665\times  5)\times4$ &  20 &   28 &  87 & 102 \\
HU CMa &  07:04:06.8 & -11:26:08.0   &  2010-01-17 & $300\times4$ & $300\times4$ & $ 50\times4$ &   6 &  185 & 218 & 217 \\
HD 53367 &  07:04:25.6 & -10:27:15.8   &  2010-02-24 & $(  2\times  3)\times4$ & $(  2\times  3)\times4$ & $(  2\times  2)\times4$ &   6 &   62 & 166 & 220 \\
PDS 241 &  07:08:38.8 & -04:19:07.0   &  2009-12-21 & $300\times4$ & $300\times4$ & $100\times4$ &   3 &  228 & 295 & 250 \\
NX Pup &  07:19:28.4 & -44:35:08.8   &  2010-02-01 & $120\times4$ & $( 60\times  2)\times4$ & $(  2\times  2)\times4$ &  20 &   44 & 191 & 125 \\
PDS 27 &  07:19:36.1 & -17:39:17.9   &  2010-02-24 & $300\times4$ & $300\times4$ & $(  2\times  6)\times4$ &  20 &    9 & 118 & 228 \\
PDS 133 &  07:25:05.1 & -25:45:49.1   &  2010-02-24 & $300\times4$ & $300\times4$ & $( 40\times  4)\times4$ &   2 &    3 &  22 &  50 \\
HD 59319 &  07:28:36.9 & -21:57:48.4   &  2010-02-24 & $( 10\times  2)\times4$ & $( 10\times  2)\times4$ & $ 10\times4$ &   4 &  265 &  14 & 245 \\
PDS 134 &  07:32:26.8 & -21:55:35.3   &  2010-02-24 & $300\times4$ & $300\times4$ & $150\times4$ &   2 &  196 & 210 & 157 \\
HD 68695 &  08:11:44.3 & -44:05:07.5   &  2009-12-21 & $( 20\times  2)\times4$ & $( 20\times  2)\times4$ & $ 15\times4$ &   3 &  159 & 179 & 314 \\
HD 72106 &  08:29:35.0 & -38:36:18.5   &  2009-12-19 & $( 10\times  2)\times4$ & $( 10\times  2)\times4$ & $ 10\times4$ &   4 &   49 & 136 & 154 \\
TYC 8581-2002-1 &  08:44:23.5 & -59:56:55.8   &  2009-12-21 & $150\times4$ & $150\times4$ & $ 50\times4$ &   3 &  181 & 205 & 224 \\
PDS 33 &  08:48:45.4 & -40:48:20.1   &  2009-12-21 & $300\times4$ & $300\times4$ & $150\times4$ &   2 &  278 & 181 & 186 \\
HD 76534 &  08:55:08.8 & -43:27:57.3   &  2010-01-30 & $( 10\times  3)\times4$ & $( 15\times  3)\times4$ & $ 15\times4$ &   3 &  149 & 259 & 150 \\
PDS 281 &  08:55:45.9 & -44:25:11.4   &  2009-12-21 & $  (15 \times 2)\times 4$ & $  (15 \times 2)\times 4$ & $  7.5 \times 4$ &   4 &  193 & 174 & 203 \\
PDS 286 &  09:05:59.9 & -47:18:55.2   &  2009-12-21 & $300\times4$ & $(150\times  2)\times4$ & $(  2\times  6)\times4$ &  20 &   96 & 205 & 157 \\
PDS 297 &  09:42:40.0 & -56:15:32.2   &  2010-01-04 & $300\times4$ & $300\times4$ & $(150\times  2)\times4$ &   2 &  213 & 210 & 188 \\
HD 85567 &  09:50:28.3 & -60:57:59.5   &  2010-03-06 & $( 10\times  3)\times4$ & $( 15\times  3)\times4$ & $  2\times4$ &  20 &  142 & 183 & 283 \\
HD 87403 &  10:02:51.3 & -59:16:52.7   &  2010-03-06 & $( 15\times  2)\times4$ & $( 15\times  2)\times4$ & $ 15\times4$ &   2 &  137 & 137 & 160 \\
PDS 37 &  10:10:00.3 & -57:02:04.4   &  2010-03-31 & $300\times4$ & $300\times4$ & $(  3\times  6)\times4$ &  15 &   18 & 182 & 166 \\
HD 305298 &  10:33:05.0 & -60:19:48.6   &  2010-03-31 & $ 90\times4$ & $ 90\times4$ & $ 45\times4$ &   2 &  155 & 209 & 207 \\
HD 94509 &  10:53:27.2 & -58:25:21.4   &  2010-02-05 & $( 20\times  2)\times4$ & $( 20\times  2)\times4$ & $ 15\times4$ &   3 &   12 & 171 & 208 \\
HD 95881 &  11:01:57.1 & -71:30:46.9   &  2010-01-04 & $( 10\times  3)\times4$ & $( 15\times  3)\times4$ & $  2\times4$ &  20 &   56 & 223 & 296 \\
HD 96042 &  11:03:40.6 & -59:25:55.9   &  2010-02-05 & $( 10\times  2)\times4$ & $( 10\times  2)\times4$ & $ 10\times4$ &   4 &   97 & 107 & 176 \\

  \hline
  \end{tabular}
\end{minipage}
\end{table*}

\begin{table*}
 \centering
 \begin{minipage}{190mm}
  \contcaption{}
  \begin{tabular}{lccc rrrc ccc}
  \hline
   Name   &  RA & DEC & Obs Date & \multicolumn{4}{c}{Exposure Time (s)} & \multicolumn{3}{c}{SNR}\\
          &  \multicolumn{2}{c}{(J2000)}  & (yyyy/mm/dd) & UVB & VIS & NIR & NDIT & UVB & VIS & NIR \\ 
  \hline

HD 97048 &  11:08:03.0 & -77:39:16.0   &  2010-02-05 & $( 15\times  3)\times4$ & $( 15\times  3)\times4$ & $  2\times4$ &  20 &  208 & 254 & 200 \\
HD 98922 &  11:22:31.5 & -53:22:09.0   &  2010-03-30 & $(  3\times  3)\times4$ & $(  3\times  3)\times4$ & $(  0.75 \times  2)\times4$ &  20 &   85 & 196 & 103 \\
HD 100453 &  11:33:05.3 & -54:19:26.1   &  2010-03-29 & $( 10\times  3)\times4$ & $( 15\times  3)\times4$ & $  2\times4$ &  20 &   35 & 110 & 289 \\
HD 100546 &  11:33:25.1 & -70:11:39.6   &  2010-03-30 & $(  3\times  3)\times4$ & $(  3\times  3)\times4$ & $(  1\times  2)\times4$ &  20 &  283 & 206 & 348 \\
HD 101412 &  11:39:44.3 & -60:10:25.1   &  2010-03-30 & $( 15\times  2)\times4$ & $( 15\times  2)\times4$ & $ 10\times4$ &   3 &   89 & 125 & 169 \\
PDS 344 &  11:40:32.8 & -64:32:03.0   &  2010-03-31 & $300\times4$ & $300\times4$ & $150\times4$ &   2 &  236 & 190 & 157 \\
HD 104237 &  12:00:04.8 & -78:11:31.9   &  2010-03-30 & $(  3\times  3)\times4$ & $(  3\times  3)\times4$ & $(  0.75\times  2)\times4$ &  20 &   25 &  88 & 216 \\
V1028 Cen &  13:01:17.6 & -48:53:17.0   &  2010-03-29 & $ 90\times4$ & $ 90\times4$ & $ 10\times4$ &   9 &  101 & 165 & 148 \\
PDS 361S &  13:03:21.6 & -62:13:23.5   &  2010-03-31 & $300\times4$ & $300\times4$ & $150\times4$ &   2 &  118 & 192 & 181 \\
HD 114981 &  13:14:40.4 & -38:39:05.0   &  2010-03-29 & $(  3\times  3)\times4$ & $(  3\times  3)\times4$ & $(  3\times  2)\times4$ &   6 &  250 & 185 &  82 \\
PDS 364 &  13:20:03.5 & -62:23:51.7   &  2010-03-31 & $300\times4$ & $300\times4$ & $ 90\times4$ &   3 &  130 & 116 & 168 \\
PDS 69 &  13:57:44.0 & -39:58:47.0   &  2010-03-29 & $( 15\times  2)\times4$ & $( 15\times  2)\times4$ & $  5\times4$ &   6 &   45 &  66 & 149 \\
DG Cir &  15:03:23.4 & -63:22:57.2   &  2010-03-31 & $360\times4$ & $(180\times  2)\times4$ & $ 10\times4$ &  20 &   14 & 103 & 138 \\
HD 132947 &  15:04:56.2 & -63:07:50.0   &  2010-03-12 & $( 20\times  2)\times4$ & $( 20\times  2)\times4$ & $ 15\times4$ &   3 &  232 & 265 & 236 \\
HD 135344B &  15:15:48.2 & -37:09:16.7   &  2010-03-31 & $(  5\times  2)\times4$ & $(  5\times  2)\times4$ & $  2\times4$ &   6 &   42 & 116 & 307 \\
HD 139614 &  15:40:46.3 & -42:29:51.4   &  2010-03-28 & $( 10\times  2)\times4$ & $( 10\times  2)\times4$ & $  5\times4$ &   8 &   35 & 139 & 127 \\
PDS 144S &  15:49:15.4 & -26:00:52.8   &  2010-03-31 & $300\times4$ & $300\times4$ & $(  5\times  5)\times4$ &  10 &   49 & 127 & 124 \\
HD 141569 &  15:49:57.8 & -03:55:18.6   &  2010-03-28 & $(  2\times  3)\times4$ & $(  2\times  3)\times4$ & $(  2\times  2)\times4$ &   6 &  177 & 151 & 144 \\
HD 141926 &  15:54:21.5 & -55:19:41.3   &  2010-03-12 & $( 15\times  2)\times4$ & $( 15\times  2)\times4$ & $  3\times4$ &  10 &   66 & 198 &  78 \\
HD 142666 &  15:56:40.2 & -22:01:39.5   &  2010-03-28 & $( 15\times  3)\times4$ & $( 15\times  3)\times4$ & $  2\times4$ &  20 &   53 & 126 & 114 \\
HD 142527 &  15:56:41.8 & -42:19:21.0   &  2010-04-01 & $(  5\times  2)\times4$ & $(  5\times  2)\times4$ & $  1\times4$ &  20 &   23 &  88 & 305 \\
HD 144432 &  16:06:57.8 & -27:43:07.4   &  2010-03-12 & $( 10\times  3)\times4$ & $( 15\times  3)\times4$ & $  2\times4$ &  20 &   55 & 126 & 109 \\
HD 144668 &  16:08:34.0 & -39:06:19.4   &  2010-03-30 & $(  3\times  3)\times4$ & $(  3\times  3)\times4$ & $(  0.75 \times  2)\times4$ &  20 &   96 & 191 & 276 \\
HD 145718 &  16:13:11.4 & -22:29:08.3   &  2010-03-29 & $( 15\times  2)\times4$ & $( 15\times  2)\times4$ & $  3\times4$ &  10 &   85 & 165 & 165 \\
PDS 415N &  16:18:37.4 & -24:05:22.0   &  2010-03-31 & $300\times4$ & $300\times4$ & $( 10\times  5)\times4$ &   5 &   13 &  66 & 103 \\
HD 150193 &  16:40:17.7 & -23:53:47.0   &  2010-03-30 & $( 25\times  3)\times4$ & $( 15\times  3)\times4$ & $  2\times4$ &  20 &  100 & 152 & 322 \\
AK Sco &  16:54:45.0 & -36:53:17.1   &  2009-10-05 & $( 10\times  2)\times4$ & $( 10\times  2)\times4$ & $  5\times4$ &   8 &   26 &  91 & 198 \\
PDS 431 &  16:54:58.9 & -43:21:47.7   &  2010-04-01 & $300\times4$ & $300\times4$ & $150\times4$ &   2 &  174 & 171 &  84 \\
KK Oph &  17:10:07.9 & -27:15:18.6   &  2010-03-26 & $200\times4$ & $(100\times  2)\times4$ & $(  2\times  6)\times4$ &  20 &   58 & 274 & 231 \\
HD 163296 &  17:56:21.4 & -21:57:21.7   &  2009-10-05 & $(  3\times  3)\times4$ & $(  3\times  3)\times4$ & $(  1\times  2)\times4$ &  20 &   87 & 241 & 127 \\
MWC 297 &  18:27:39.7 & -03:49:53.1   &  2009-10-06 & $300\times4$ & $ 10\times4$ & $(  0.665 \times 15)\times4$ &  20 &  162 &  96 & 127 \\

  \hline
  \end{tabular}
\end{minipage}
\end{table*}


\begin{figure*}
 \includegraphics[trim=0.5cm 0.5cm 1.4cm 0.5cm, width=\textwidth]{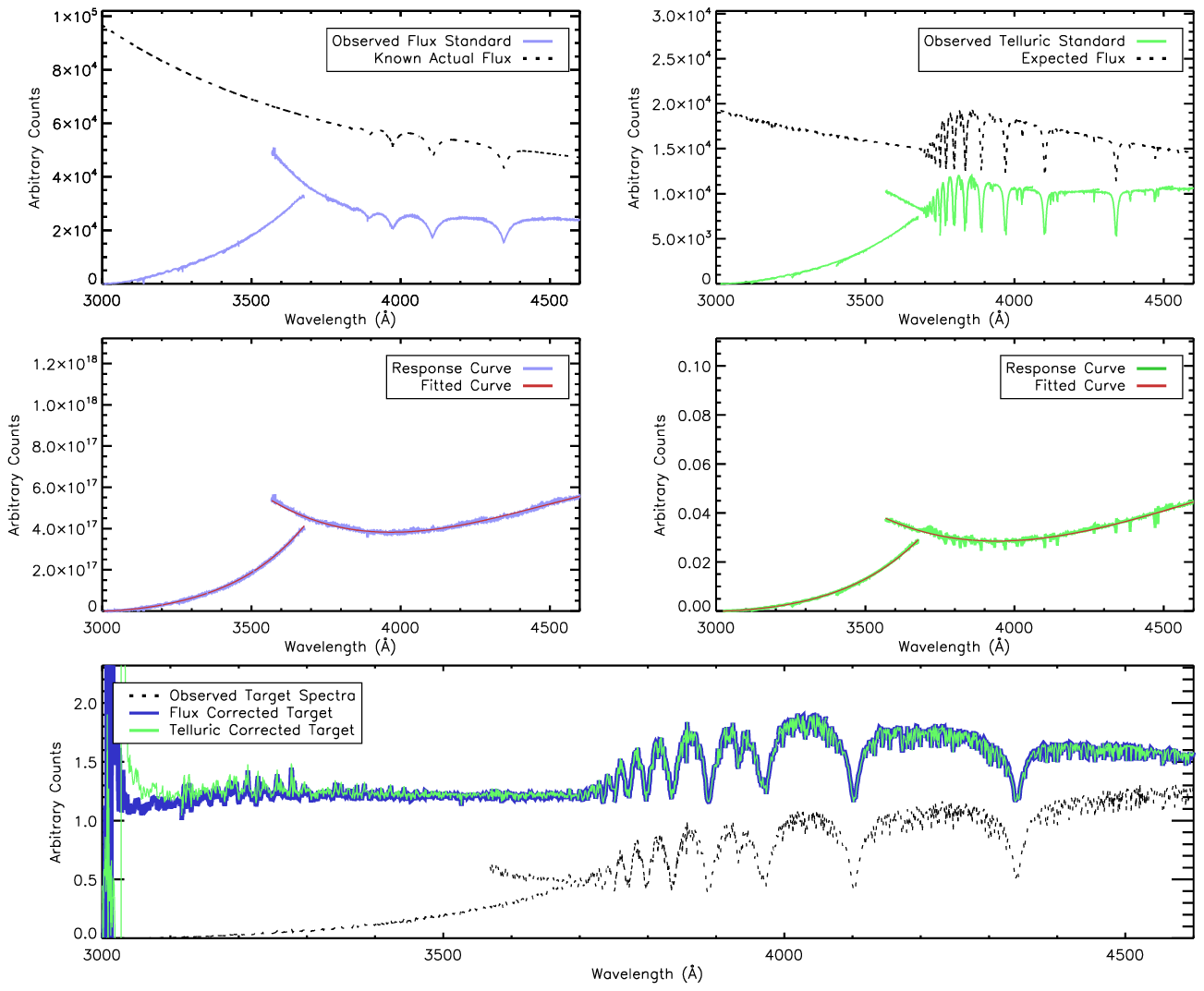}
 \caption{Shown above are the two different cases of correcting target spectra via either: a flux standard - shown on the left; or a telluric standard - shown on the right. The top-left panel shows the observed flux standard (blue) and the true flux of the standard star (dashed-black). Similarly the top-right panel shows the observed telluric standard (green) and its expected spectra (dashed-black). The middle two panels show a division of the observed standard stars by their expected spectra from the top panels (blue and green). A two part curve (red) is also shown as a fit to this division. The bottom panel shows the result of applying the two fits from the middle panel to the target spectra (dashed-black). It can be seen that the two methods of correction are equivalent by lying on top of each other. All spectra have been arbitrarily scaled in order to be visible on each plot.}
\label{fig:response}
\end{figure*}


\subsection{X-Shooter and Target Selection}
\label{sec:xs_targ}

Observations were performed over a period of 6 months between October 2009 and April 2010 using the X-Shooter echelle spectrograph -- mounted at the VLT, Cerro Paranal, Chile \citep{Vernet2011}. X-Shooter provides spectra covering a large wavelength range of 3000--23000~$\rm \AA$, split into three arms and taken simultaneously. The arms are split into the following: the UVB arm, 3000--5600~\AA; the VIS arm, 5500--10200~${\rm \AA}$; and the NIR arm, 10200-24800~${\rm \AA}$. The smallest slit widths available of 0.5'', 0.4'' and 0.4'' were used to provide the highest possible resolutions of R~$\sim$~10000, 18000 and 10500 for the respective UVB, VIS and NIR arms. In total 91 science targets were observed in nodding mode using an ABBA sequence. Table \ref{tab:obs} includes details of each targets RA and DEC, exposure times, and the signal-to-noise in each arm. The SNR is calculated by analysing a 30~$\rm \AA$ region of spectra centred about the wavelengths of 4600, 6750, and 16265~${\rm \AA}$ for the UVB, VIS, and NIR arm respectively. These regions were chosen as they are generally the flattest continuum regions in each star. Although emission lines, and absorption lines of cooler objects, can artificially lower the measured SNR, for a fair treatment we stick with the above regions to provide a rough guide to the quality of each spectrum. 

The targets were selected from the catalogues of \citet{The1994}, and \citet{Vieira2003}. 51 targets were selected from \citet{The1994}, and 40 from the \citet{Vieira2003} catalogue, bringing the total number of targets to 91. The observations cover around 70\% of the southern HAeBes identified by \citet{The1994}, and about 50\% of the targets observed by \citet{Vieira2003}. For many of these targets little information is known about them, particularly in regard to multiplicity \citep[see ][ for a review]{Duchene2015}. It is known that HAeBes have high binary fractions \citep{Baines2006,Wheelwright2010}, and as a consequence, any close separation binaries will contribute towards observed spectra. In this work our focus is on the UV and optical portions of the spectra, where we assume that the primary star, the HAeBe target, provides the largest contribution to the brightness. Contributions from secondary stars will be greater in the observed literature photometry as they use either larger slit widths or an aperture greater than the slit widths used here. Photometry of all the targets are sourced from the literature, and are provided in Table \ref{tab:phot} in Appendix \ref{sec:app_phot}.

Telluric standards were observed either just before or after each science exposure. They were observed in stare mode for short exposures, $\sim$10~s, due to their brightness. Flux standard stars were observed on approximately half of the evenings in offset mode (offset mode allows accurate sky subtraction to be performed).

\subsection{Data Reduction}
\label{sec:xs_data}

 All data were reduced following standard procedures of the X-Shooter pipeline v0.9.7 \citep{Modigliani2010}. Only one aspect is not included in the standard procedures and that is flux calibration. However, we do not require a flux calibration for the work presented in this paper, we focus instead on determining the spectral shape. This spectral shape is needed in the UVB arm; specifically across the Balmer Jump region where the difference between the $U$-band and $B$-band is required in order to measure any excess UV-emission. The analysis of the spectral lines and the derivation of their luminosities is deferred to Paper II.

Ordinarily, to obtain the correct spectral shape a flux calibration is performed using the observed flux standards of each night. However, as mentioned, flux standards were not observed on all evenings. A solution to this is to instead use the telluric standards, for which there is at least one per target, as a means of correction. This will ensure a uniform treatment to all of the targets. Caution should be noted of using non-flux standard stars for calibration; to mitigate any problems that could arise from this, consistency checks are made against the flux standards for the nights where they are available and will be discussed at the end of this section.
For this method of spectral shape calibration accurate knowledge of the spectral type of each telluric is required. To ensure a homogeneous reduction we adopted our own spectral typing of each telluric in this work. This helps to minimise any reduction errors, and will also allow us to place a systematic error on this reduction method. Full details of the spectral typing, along with a discussion of how they compare with literature values, will be provided in Section \ref{sec:st_title}. 
Once the spectral type is determined the observed telluric spectrum is divided through by a model atmosphere of the same spectral type in order to obtain an instrumental \textit{response} curve. The model atmospheres adopted here, and throughout this work, are sets of Kurucz-Castelli models \citep{Kurucz1993,Castelli2004} computed by \citet{Munari2005}, due to their small dispersion of 1~\AA~over the UVB wavelength range (these will be referred to as KC-models hereafter). The resulting response curve from this division is then fit with two curves: one for the echelle orders where $\lambda<3600~{\rm \AA}$ and another for the orders where $\lambda>3600~{\rm \AA}$. This is because the response at $\sim3600~{\rm \AA}$ is not the same between the two over-lapping echelle orders. Figure \ref{fig:response} shows the procedure of the above method, for a target star for which both flux standard and telluric standards were observed, and highlights the two different response curves intersecting around the $U$-band region in the middle panels. The figure also provides a consistency check by comparing the flux standard reduction, on the left, to the telluric standard reduction, on the right. The bottom panel of the figure demonstrates the similarity of both results with a difference of $<3$\% across the spectra. Larger deviations are seen between the two spectra close to 3000~$\rm \AA$, due to low levels of counts. This region is not used in this work and can be disregarded. This same check is performed on other stars for which both a telluric and flux standard are available, and the maximum deviation observed is only 5\% across the spectra. Overall, it can be seen that this method of using the telluric for instrumental response correction provides a satisfactory calibration of the data, and is therefore performed on all targets.


\section[]{Determining the Distance and Stellar Parameters}
\label{sec:st_title}



\begin{table*}
 \centering
 \begin{minipage}{160mm}
  \caption{Details of the derived stellar parameters. Stars for which a distance estimate from the literature is used, or if the star was moved to the ZAMS, are noted in the final column (Column 9, a legend is provided as a footnote).}
  \label{tab:params}
  \begin{tabular}{l ccccc cc cl}
  \hline
   Name  &  $T_{\rm eff}$   & log(g) & log($L_\star $) & $M_\star $ & $R_\star $ & $A_V$ & Age & Distance & Notes \\
         &  (K) &  [cm/s$^2$]& [$\rm L_\odot$] & ($\rm M_\odot$) & ($\rm R_\odot$) & (mag) & (Myr) & (pc) & \\
  \hline

UX Ori &  8500 $\pm$  250 & 3.90$^{+0.25}_{-0.25}$ &  1.54$^{+ 0.38}_{- 0.33}$ &  2.1$^{+ 0.7}_{- 0.3}$ &  2.7$^{+ 0.4}_{- 0.2}$ &  0.48$^{+ 0.07}_{- 0.03}$ &   4.24$^{+  3.24}_{-  2.35}$ &  600$^{+  96}_{-  50}$ &  \\ [3pt]
PDS 174 & 17000 $\pm$ 2000 & 4.10$^{+0.40}_{-0.40}$ &  2.91$^{+ 0.56}_{- 0.56}$ &  5.0$^{+ 2.3}_{- 2.3}$ &  3.3$^{+ 0.7}_{- 0.7}$ &  3.51$^{+ 0.07}_{- 0.07}$ &   0.60$^{+  0.43}_{-  0.43}$ & 1126$^{+ 238}_{- 237}$ &  \\ [3pt]
V1012 Ori &  8500 $\pm$  250 & 4.38$^{+0.15}_{-0.15}$ &  0.94$^{+ 0.28}_{- 0.37}$ &  1.6$^{+ 0.3}_{- 0.3}$ &  1.4$^{+ 0.4}_{- 0.4}$ &  1.32$^{+ 0.02}_{- 0.04}$ &  15.16$^{+  7.58}_{-  7.58}$ &  445$^{+ 137}_{- 130}$ &  $\dagger$  $\ast$  \\ [3pt]
HD 34282 &  9500 $\pm$  250 & 4.40$^{+0.15}_{-0.15}$ &  1.17$^{+ 0.28}_{- 0.36}$ &  1.9$^{+ 0.4}_{- 0.4}$ &  1.4$^{+ 0.4}_{- 0.4}$ &  0.01$^{+ 0.02}_{- 0.00}$ &  10.00$^{+  5.00}_{-  5.00}$ &  366$^{+ 111}_{- 109}$ &  \\ [3pt]
HD 287823 &  8375 $\pm$  125 & 4.23$^{+0.11}_{-0.15}$ &  1.09$^{+ 0.17}_{- 0.20}$ &  1.7$^{+ 0.1}_{- 0.1}$ &  1.7$^{+ 0.3}_{- 0.3}$ &  0.00$^{+ 0.05}_{- 0.00}$ &   9.01$^{+  4.11}_{-  2.28}$ &  340$^{+  68}_{-  68}$ &  $\dagger$ a \\ [3pt]
HD 287841 &  7750 $\pm$  250 & 4.27$^{+0.12}_{-0.12}$ &  0.87$^{+ 0.17}_{- 0.21}$ &  1.5$^{+ 0.1}_{- 0.1}$ &  1.5$^{+ 0.2}_{- 0.2}$ &  0.00$^{+ 0.05}_{- 0.00}$ &  14.07$^{+  4.10}_{-  4.10}$ &  340$^{+  68}_{-  68}$ &  $\dagger$ a \\ [3pt]
HD 290409 &  9750 $\pm$  500 & 4.25$^{+0.25}_{-0.25}$ &  1.42$^{+ 0.29}_{- 0.29}$ &  2.1$^{+ 0.2}_{- 0.2}$ &  1.8$^{+ 0.1}_{- 0.1}$ &  0.00$^{+ 0.05}_{- 0.00}$ &   5.50$^{+  2.03}_{-  2.03}$ &  514$^{+  36}_{-  29}$ &  \\ [3pt]
HD 35929 &  7000 $\pm$  250 & 3.47$^{+0.11}_{-0.11}$ &  1.76$^{+ 0.17}_{- 0.21}$ &  2.9$^{+ 0.4}_{- 0.4}$ &  5.2$^{+ 0.7}_{- 0.8}$ &  0.00$^{+ 0.05}_{- 0.00}$ &   1.65$^{+  0.81}_{-  0.55}$ &  360$^{+  72}_{-  72}$ &  $\dagger$ b \\ [3pt]
HD 290500 &  9500 $\pm$  500 & 3.80$^{+0.40}_{-0.40}$ &  1.94$^{+ 0.61}_{- 0.53}$ &  2.8$^{+ 1.8}_{- 0.7}$ &  3.5$^{+ 1.0}_{- 0.5}$ &  0.00$^{+ 0.05}_{- 0.00}$ &   2.26$^{+  3.39}_{-  1.72}$ & 1522$^{+ 436}_{- 211}$ &  \\ [3pt]
HD 244314 &  8500 $\pm$  250 & 4.15$^{+0.11}_{-0.15}$ &  1.21$^{+ 0.18}_{- 0.23}$ &  1.8$^{+ 0.1}_{- 0.1}$ &  1.9$^{+ 0.3}_{- 0.3}$ &  0.10$^{+ 0.02}_{- 0.05}$ &   7.52$^{+  1.97}_{-  1.96}$ &  440$^{+  88}_{-  88}$ &  $\dagger$ a \\ [3pt]
HK Ori &  8500 $\pm$  500 & 4.22$^{+0.13}_{-0.13}$ &  1.13$^{+ 0.24}_{- 0.27}$ &  1.7$^{+ 0.2}_{- 0.2}$ &  1.7$^{+ 0.3}_{- 0.3}$ &  1.21$^{+ 0.12}_{- 0.14}$ &   8.73$^{+  2.73}_{-  2.73}$ &  440$^{+  88}_{-  88}$ &  $\dagger$ a \\ [3pt]
HD 244604 &  9000 $\pm$  250 & 3.99$^{+0.15}_{-0.13}$ &  1.54$^{+ 0.18}_{- 0.23}$ &  2.1$^{+ 0.3}_{- 0.2}$ &  2.4$^{+ 0.4}_{- 0.5}$ &  0.14$^{+ 0.00}_{- 0.04}$ &   4.56$^{+  1.53}_{-  1.32}$ &  440$^{+  88}_{-  88}$ &  $\dagger$ a \\ [3pt]
UY Ori &  9750 $\pm$  250 & 4.30$^{+0.20}_{-0.20}$ &  1.36$^{+ 0.23}_{- 0.23}$ &  2.0$^{+ 0.1}_{- 0.1}$ &  1.7$^{+ 0.1}_{- 0.1}$ &  1.11$^{+ 0.02}_{- 0.00}$ &   6.35$^{+  1.84}_{-  1.84}$ & 1027$^{+  38}_{-  35}$ &  \\ [3pt]
HD 245185 & 10000 $\pm$  500 & 4.25$^{+0.25}_{-0.25}$ &  1.49$^{+ 0.29}_{- 0.29}$ &  2.2$^{+ 0.2}_{- 0.2}$ &  1.8$^{+ 0.1}_{- 0.1}$ &  0.00$^{+ 0.05}_{- 0.00}$ &   4.91$^{+  1.70}_{-  1.70}$ &  519$^{+  34}_{-  28}$ &  \\ [3pt]
T Ori &  9000 $\pm$  500 & 3.60$^{+0.30}_{-0.30}$ &  2.12$^{+ 0.47}_{- 0.46}$ &  3.3$^{+ 1.5}_{- 1.0}$ &  4.8$^{+ 1.0}_{- 0.8}$ &  1.50$^{+ 0.08}_{- 0.05}$ &   1.35$^{+  2.27}_{-  0.93}$ &  750$^{+ 159}_{- 123}$ &  \\ [3pt]
V380 Ori &  9750 $\pm$  750 & 4.00$^{+0.35}_{-0.35}$ &  1.71$^{+ 0.52}_{- 0.26}$ &  2.3$^{+ 1.1}_{- 0.2}$ &  2.5$^{+ 0.5}_{- 0.1}$ &  2.21$^{+ 0.05}_{- 0.07}$ &   3.73$^{+  2.46}_{-  2.49}$ &  330$^{+  73}_{-  17}$ &  \\ [3pt]
HD 37258 &  9750 $\pm$  500 & 4.25$^{+0.25}_{-0.25}$ &  1.42$^{+ 0.29}_{- 0.29}$ &  2.1$^{+ 0.2}_{- 0.2}$ &  1.8$^{+ 0.1}_{- 0.1}$ &  0.06$^{+ 0.05}_{- 0.04}$ &   5.50$^{+  2.03}_{-  2.03}$ &  424$^{+  25}_{-  22}$ &  \\ [3pt]
HD 290770 & 10500 $\pm$  250 & 4.20$^{+0.30}_{-0.30}$ &  1.64$^{+ 0.38}_{- 0.38}$ &  2.3$^{+ 0.5}_{- 0.5}$ &  2.0$^{+ 0.2}_{- 0.2}$ &  0.00$^{+ 0.05}_{- 0.00}$ &   4.16$^{+  1.89}_{-  1.89}$ &  440$^{+  43}_{-  41}$ &  \\ [3pt]
BF Ori &  9000 $\pm$  250 & 3.97$^{+0.15}_{-0.13}$ &  1.57$^{+ 0.19}_{- 0.22}$ &  2.1$^{+ 0.3}_{- 0.2}$ &  2.5$^{+ 0.4}_{- 0.5}$ &  0.33$^{+ 0.03}_{- 0.02}$ &   4.34$^{+  1.55}_{-  1.32}$ &  510$^{+ 102}_{- 102}$ &  $\dagger$ a \\ [3pt]
HD 37357 &  9500 $\pm$  250 & 4.10$^{+0.10}_{-0.10}$ &  1.52$^{+ 0.13}_{- 0.11}$ &  2.1$^{+ 0.1}_{- 0.1}$ &  2.1$^{+ 0.1}_{- 0.0}$ &  0.00$^{+ 0.05}_{- 0.00}$ &   4.93$^{+  0.87}_{-  0.87}$ &  344$^{+  15}_{-   4}$ &  \\ [3pt]
HD 290764 &  7875 $\pm$  375 & 3.90$^{+0.17}_{-0.15}$ &  1.36$^{+ 0.22}_{- 0.26}$ &  1.9$^{+ 0.4}_{- 0.2}$ &  2.6$^{+ 0.5}_{- 0.5}$ &  0.16$^{+ 0.12}_{- 0.14}$ &   5.25$^{+  2.58}_{-  1.90}$ &  470$^{+  94}_{-  94}$ &  $\dagger$ a \\ [3pt]
HD 37411 &  9750 $\pm$  250 & 4.35$^{+0.15}_{-0.15}$ &  1.28$^{+ 0.28}_{- 0.36}$ &  1.9$^{+ 0.4}_{- 0.4}$ &  1.5$^{+ 0.5}_{- 0.5}$ &  0.21$^{+ 0.01}_{- 0.00}$ &   9.00$^{+  4.50}_{-  4.50}$ &  358$^{+ 109}_{- 107}$ &  \\ [3pt]
V599 Ori &  8000 $\pm$  250 & 3.72$^{+0.13}_{-0.12}$ &  1.68$^{+ 0.19}_{- 0.23}$ &  2.5$^{+ 0.4}_{- 0.4}$ &  3.6$^{+ 0.6}_{- 0.7}$ &  4.65$^{+ 0.06}_{- 0.07}$ &   2.82$^{+  1.40}_{-  1.00}$ &  510$^{+ 102}_{- 102}$ &  $\dagger$ a \\ [3pt]
V350 Ori &  9000 $\pm$  250 & 4.18$^{+0.11}_{-0.16}$ &  1.31$^{+ 0.19}_{- 0.22}$ &  1.9$^{+ 0.1}_{- 0.1}$ &  1.9$^{+ 0.3}_{- 0.3}$ &  0.69$^{+ 0.02}_{- 0.03}$ &   6.41$^{+  1.97}_{-  1.67}$ &  510$^{+ 102}_{- 102}$ &  $\dagger$ a \\ [3pt]
HD 250550 & 11000 $\pm$  500 & 3.80$^{+0.40}_{-0.40}$ &  2.28$^{+ 0.61}_{- 0.53}$ &  3.4$^{+ 2.1}_{- 0.9}$ &  3.8$^{+ 1.0}_{- 0.5}$ &  0.00$^{+ 0.05}_{- 0.00}$ &   1.42$^{+  2.21}_{-  1.09}$ &  973$^{+ 267}_{- 136}$ &  \\ [3pt]
V791 Mon & 15000 $\pm$ 1500 & 4.30$^{+0.16}_{-0.16}$ &  2.35$^{+ 0.40}_{- 0.51}$ &  3.6$^{+ 0.7}_{- 0.7}$ &  2.2$^{+ 0.7}_{- 0.7}$ &  1.17$^{+ 0.06}_{- 0.04}$ &   1.80$^{+  0.90}_{-  0.90}$ &  648$^{+ 204}_{- 184}$ &  \\ [3pt]
PDS 124 & 10250 $\pm$  250 & 4.30$^{+0.20}_{-0.20}$ &  1.47$^{+ 0.23}_{- 0.23}$ &  2.2$^{+ 0.2}_{- 0.2}$ &  1.7$^{+ 0.1}_{- 0.1}$ &  1.23$^{+ 0.03}_{- 0.00}$ &   5.48$^{+  1.64}_{-  1.64}$ &  894$^{+  34}_{-  33}$ &  \\ [3pt]
LkHa 339 & 10500 $\pm$  250 & 4.20$^{+0.20}_{-0.20}$ &  1.64$^{+ 0.24}_{- 0.24}$ &  2.3$^{+ 0.2}_{- 0.2}$ &  2.0$^{+ 0.1}_{- 0.1}$ &  3.54$^{+ 0.01}_{- 0.01}$ &   4.16$^{+  1.18}_{-  1.18}$ &  597$^{+  27}_{-  25}$ &  \\ [3pt]
VY Mon & 12000 $\pm$ 4000 & 3.75$^{+0.50}_{-0.50}$ &  2.56$^{+ 0.77}_{- 0.65}$ &  4.0$^{+ 3.7}_{- 1.3}$ &  4.4$^{+ 1.6}_{- 0.7}$ &  5.68$^{+ 0.17}_{- 0.45}$ &   0.89$^{+  2.61}_{-  0.80}$ &  439$^{+ 181}_{- 113}$ &  \\ [3pt]
R Mon & 12000 $\pm$ 2000 & 4.00$^{+0.11}_{-0.24}$ &  2.19$^{+ 0.34}_{- 0.40}$ &  3.1$^{+ 0.8}_{- 0.6}$ &  2.9$^{+ 0.3}_{- 0.3}$ &  2.42$^{+ 0.08}_{- 0.13}$ &   1.92$^{+  1.23}_{-  0.89}$ &  800$^{+ 160}_{- 160}$ &  $\dagger$ c \\ [3pt]
V590 Mon & 12500 $\pm$ 1000 & 4.20$^{+0.30}_{-0.30}$ &  2.06$^{+ 0.37}_{- 0.37}$ &  3.1$^{+ 0.6}_{- 0.6}$ &  2.3$^{+ 0.2}_{- 0.2}$ &  1.03$^{+ 0.04}_{- 0.05}$ &   2.19$^{+  1.00}_{-  1.00}$ & 1722$^{+ 171}_{- 160}$ &  \\ [3pt]
PDS 24 & 10500 $\pm$  500 & 4.20$^{+0.30}_{-0.30}$ &  1.64$^{+ 0.38}_{- 0.38}$ &  2.3$^{+ 0.5}_{- 0.5}$ &  2.0$^{+ 0.2}_{- 0.2}$ &  1.11$^{+ 0.04}_{- 0.03}$ &   4.16$^{+  1.91}_{-  1.91}$ & 1646$^{+ 162}_{- 157}$ &  \\ [3pt]
PDS 130 & 10500 $\pm$  250 & 3.90$^{+0.20}_{-0.20}$ &  2.02$^{+ 0.30}_{- 0.27}$ &  2.8$^{+ 0.7}_{- 0.4}$ &  3.1$^{+ 0.4}_{- 0.2}$ &  2.07$^{+ 0.01}_{- 0.00}$ &   2.25$^{+  1.32}_{-  1.07}$ & 1748$^{+ 216}_{- 129}$ &  \\ [3pt]
PDS 229N & 12500 $\pm$  250 & 4.20$^{+0.20}_{-0.20}$ &  2.06$^{+ 0.23}_{- 0.23}$ &  3.1$^{+ 0.2}_{- 0.2}$ &  2.3$^{+ 0.1}_{- 0.1}$ &  2.03$^{+ 0.01}_{- 0.01}$ &   2.19$^{+  0.63}_{-  0.63}$ & 1379$^{+  57}_{-  56}$ &  \\ [3pt]
GU CMa & 22500 $\pm$ 1500 & 3.90$^{+0.40}_{-0.40}$ &  3.87$^{+ 0.62}_{- 0.62}$ &  9.5$^{+ 6.8}_{- 6.8}$ &  5.7$^{+ 1.7}_{- 1.7}$ &  0.57$^{+ 0.03}_{- 0.02}$ &   0.11$^{+  0.10}_{-  0.10}$ &  531$^{+ 163}_{- 163}$ &  \\ [3pt]
HT CMa & 10500 $\pm$  500 & 4.00$^{+0.20}_{-0.20}$ &  1.88$^{+ 0.29}_{- 0.24}$ &  2.6$^{+ 0.6}_{- 0.2}$ &  2.6$^{+ 0.3}_{- 0.1}$ &  0.23$^{+ 0.05}_{- 0.02}$ &   2.96$^{+  1.19}_{-  1.32}$ & 1634$^{+ 187}_{-  72}$ &  \\ [3pt]
Z CMa &  8500 $\pm$  500 & 2.53$^{+0.17}_{-0.17}$ &  3.62$^{+ 0.24}_{- 0.28}$ & 11.0$^{+ 1.7}_{- 1.7}$ & 29.8$^{+ 5.2}_{- 5.3}$ &  3.37$^{+ 0.12}_{- 0.16}$ &   0.03$^{+  0.02}_{-  0.02}$ & 1050$^{+ 210}_{- 210}$ &  $\dagger$ d \\ [3pt]
HU CMa & 13000 $\pm$  250 & 4.20$^{+0.20}_{-0.20}$ &  2.16$^{+ 0.23}_{- 0.23}$ &  3.2$^{+ 0.2}_{- 0.2}$ &  2.4$^{+ 0.1}_{- 0.1}$ &  0.80$^{+ 0.02}_{- 0.01}$ &   1.88$^{+  0.52}_{-  0.52}$ & 1240$^{+  47}_{-  46}$ &  \\ [3pt]
HD 53367 & 29500 $\pm$ 1000 & 4.25$^{+0.25}_{-0.25}$ &  4.11$^{+ 0.37}_{- 0.37}$ & 12.3$^{+ 4.2}_{- 4.2}$ &  4.3$^{+ 0.7}_{- 0.7}$ &  1.88$^{+ 0.02}_{- 0.01}$ &   0.08$^{+  0.08}_{-  0.08}$ &  340$^{+  53}_{-  54}$ &  \\ [3pt]
PDS 241 & 26000 $\pm$ 1500 & 4.00$^{+0.30}_{-0.30}$ &  4.11$^{+ 0.46}_{- 0.46}$ & 11.6$^{+ 5.5}_{- 5.5}$ &  5.6$^{+ 1.2}_{- 1.2}$ &  2.60$^{+ 0.04}_{- 0.01}$ &   0.08$^{+  0.07}_{-  0.07}$ & 2907$^{+ 614}_{- 617}$ &  \\ [3pt]
NX Pup &  7000 $\pm$  250 & 3.78$^{+0.13}_{-0.13}$ &  1.28$^{+ 0.20}_{- 0.21}$ &  1.9$^{+ 0.3}_{- 0.3}$ &  3.0$^{+ 0.5}_{- 0.5}$ &  0.00$^{+ 0.07}_{- 0.00}$ &   4.92$^{+  2.37}_{-  1.67}$ &  410$^{+  82}_{-  82}$ &  $\dagger$ a \\ [3pt]
PDS 27 & 17500 $\pm$ 3500 & 3.16$^{+0.27}_{-0.27}$ &  4.39$^{+ 0.40}_{- 0.40}$ & 15.3$^{+ 5.4}_{- 4.4}$ & 17.0$^{+ 4.0}_{- 4.0}$ &  5.03$^{+ 0.13}_{- 0.13}$ &   0.10$^{+  0.10}_{-  0.10}$ & 3170$^{+ 660}_{- 620}$ &  $\dagger$ e \\ [3pt]
PDS 133 & 14000 $\pm$ 2000 & 4.08$^{+0.12}_{-0.11}$ &  2.46$^{+ 0.33}_{- 0.38}$ &  3.7$^{+ 0.9}_{- 0.7}$ &  2.9$^{+ 0.4}_{- 0.4}$ &  1.43$^{+ 0.09}_{- 0.10}$ &   1.27$^{+  0.94}_{-  0.56}$ & 2500$^{+ 500}_{- 500}$ &  $\dagger$ f \\ [3pt]
HD 59319 & 12500 $\pm$  500 & 3.50$^{+0.20}_{-0.20}$ &  3.03$^{+ 0.31}_{- 0.30}$ &  5.7$^{+ 1.6}_{- 1.2}$ &  7.0$^{+ 0.9}_{- 0.8}$ &  0.00$^{+ 0.05}_{- 0.00}$ &   0.32$^{+  0.34}_{-  0.17}$ & 1218$^{+ 162}_{- 137}$ &  \\ [3pt]
PDS 134 & 14000 $\pm$  500 & 3.40$^{+0.30}_{-0.30}$ &  3.45$^{+ 0.46}_{- 0.45}$ &  7.6$^{+ 3.5}_{- 2.3}$ &  9.1$^{+ 1.9}_{- 1.5}$ &  1.22$^{+ 0.03}_{- 0.02}$ &   0.15$^{+  0.28}_{-  0.10}$ & 5687$^{+1178}_{- 931}$ &  \\ [3pt]
HD 68695 &  9250 $\pm$  250 & 4.40$^{+0.15}_{-0.15}$ &  1.11$^{+ 0.28}_{- 0.37}$ &  1.8$^{+ 0.4}_{- 0.4}$ &  1.4$^{+ 0.4}_{- 0.4}$ &  0.00$^{+ 0.05}_{- 0.00}$ &  10.00$^{+  5.00}_{-  5.00}$ &  344$^{+ 106}_{- 103}$ &  \\ [3pt]

  \hline
  \end{tabular}
\end{minipage}
\end{table*}

\begin{table*}
 \centering
 \begin{minipage}{190mm}
  \contcaption{}
  \begin{tabular}{ l ccccc cc cl}
  \hline
   Name  &  $T_{\rm eff}$   & log(g) & log($L_\star $) & $M_\star $ & $R_\star $ & $A_V$ & Age & Distance & Notes \\
         &  (K) &  [cm/s$^2$]& [$\rm L_\odot$] & ($\rm M_\odot$) & ($\rm R_\odot$) & (mag) & (Myr) & (pc)  &\\
  \hline

HD 72106 &  8750 $\pm$  250 & 3.89$^{+0.13}_{-0.12}$ &  1.63$^{+ 0.18}_{- 0.21}$ &  2.3$^{+ 0.3}_{- 0.3}$ &  2.8$^{+ 0.5}_{- 0.5}$ &  0.00$^{+ 0.05}_{- 0.00}$ &   3.76$^{+  1.47}_{-  1.18}$ &  370$^{+  74}_{-  74}$ &  $\dagger$ a \\ [3pt]
TYC 8581-2002-1 &  9750 $\pm$  250 & 4.00$^{+0.10}_{-0.10}$ &  1.71$^{+ 0.14}_{- 0.12}$ &  2.3$^{+ 0.2}_{- 0.1}$ &  2.5$^{+ 0.1}_{- 0.1}$ &  0.94$^{+ 0.04}_{- 0.00}$ &   3.73$^{+  0.78}_{-  0.89}$ &  902$^{+  47}_{-  25}$ &  \\ [3pt]
PDS 33 &  9750 $\pm$  250 & 4.40$^{+0.15}_{-0.15}$ &  1.23$^{+ 0.27}_{- 0.36}$ &  1.9$^{+ 0.4}_{- 0.4}$ &  1.4$^{+ 0.4}_{- 0.4}$ &  0.52$^{+ 0.04}_{- 0.00}$ &   9.00$^{+  4.50}_{-  4.50}$ &  932$^{+ 282}_{- 279}$ &  $\ast$  \\ [3pt]
HD 76534 & 19000 $\pm$  500 & 4.10$^{+0.20}_{-0.20}$ &  3.18$^{+ 0.26}_{- 0.20}$ &  6.0$^{+ 0.9}_{- 0.6}$ &  3.6$^{+ 0.3}_{- 0.2}$ &  0.62$^{+ 0.02}_{- 0.01}$ &   0.37$^{+  0.19}_{-  0.14}$ &  568$^{+  43}_{-  29}$ &  \\ [3pt]
PDS 281 & 16000 $\pm$ 1500 & 3.50$^{+0.30}_{-0.30}$ &  3.62$^{+ 0.47}_{- 0.45}$ &  8.3$^{+ 4.0}_{- 2.5}$ &  8.5$^{+ 1.8}_{- 1.4}$ &  1.89$^{+ 0.07}_{- 0.10}$ &   0.12$^{+  0.23}_{-  0.09}$ &  936$^{+ 207}_{- 168}$ &  \\ [3pt]
PDS 286 & 30000 $\pm$ 3000 & 4.25$^{+0.16}_{-0.16}$ &  4.18$^{+ 0.41}_{- 0.52}$ & 13.5$^{+ 2.7}_{- 2.7}$ &  4.6$^{+ 1.5}_{- 1.5}$ &  6.27$^{+ 0.05}_{- 0.04}$ &   0.10$^{+  0.05}_{-  0.05}$ &  521$^{+ 167}_{- 146}$ &  \\ [3pt]
PDS 297 & 10750 $\pm$  250 & 4.00$^{+0.20}_{-0.20}$ &  1.93$^{+ 0.29}_{- 0.24}$ &  2.6$^{+ 0.6}_{- 0.2}$ &  2.7$^{+ 0.3}_{- 0.1}$ &  0.81$^{+ 0.01}_{- 0.02}$ &   2.77$^{+  1.12}_{-  1.22}$ & 1465$^{+ 166}_{-  59}$ &  \\ [3pt]
HD 85567 & 13000 $\pm$  500 & 3.50$^{+0.30}_{-0.30}$ &  3.13$^{+ 0.46}_{- 0.45}$ &  6.0$^{+ 2.7}_{- 1.8}$ &  7.2$^{+ 1.5}_{- 1.2}$ &  0.89$^{+ 0.03}_{- 0.02}$ &   0.27$^{+  0.52}_{-  0.18}$ &  907$^{+ 183}_{- 146}$ &  \\ [3pt]
HD 87403 & 10000 $\pm$  250 & 3.30$^{+0.10}_{-0.10}$ &  2.83$^{+ 0.15}_{- 0.15}$ &  5.5$^{+ 0.7}_{- 0.6}$ &  8.7$^{+ 0.6}_{- 0.5}$ &  0.00$^{+ 0.05}_{- 0.00}$ &   0.32$^{+  0.15}_{-  0.10}$ & 1801$^{+ 125}_{- 109}$ &  \\ [3pt]
PDS 37 & 17500 $\pm$ 3500 & 2.94$^{+0.35}_{-0.35}$ &  4.75$^{+ 0.39}_{- 0.39}$ & 21.1$^{+ 11.0}_{- 5.3}$ & 25.8$^{+ 5.0}_{- 5.0}$ &  5.81$^{+ 0.13}_{- 0.13}$ &   0.10$^{+  0.10}_{-  0.10}$ & 4310$^{+670}_{-670}$ &  $\dagger$ e \\ [3pt]
HD 305298 & 34000 $\pm$ 1000 & 4.31$^{+0.16}_{-0.16}$ &  4.46$^{+ 0.23}_{- 0.41}$ & 15.7$^{+ 3.1}_{- 3.1}$ &  4.6$^{+ 1.4}_{- 1.4}$ &  1.30$^{+ 0.00}_{- 0.02}$ &   0.02$^{+  0.01}_{-  0.01}$ & 3366$^{+1010}_{- 979}$ &  $\ast$  \\ [3pt]
HD 94509 & 11500 $\pm$ 1000 & 2.90$^{+0.40}_{-0.40}$ &  3.76$^{+ 0.65}_{- 0.62}$ & 10.8$^{+ 9.0}_{- 4.3}$ & 19.2$^{+ 6.8}_{- 4.3}$ &  0.00$^{+ 0.05}_{- 0.00}$ &   0.05$^{+  0.16}_{-  0.05}$ & 4384$^{+1585}_{-1009}$ &  \\ [3pt]
HD 95881 & 10000 $\pm$  250 & 3.20$^{+0.10}_{-0.10}$ &  2.98$^{+ 0.15}_{- 0.15}$ &  6.2$^{+ 0.8}_{- 0.7}$ & 10.3$^{+ 0.7}_{- 0.6}$ &  0.00$^{+ 0.05}_{- 0.00}$ &   0.21$^{+  0.10}_{-  0.07}$ & 1290$^{+  90}_{-  78}$ &  \\ [3pt]
HD 96042 & 25500 $\pm$ 1500 & 3.80$^{+0.20}_{-0.20}$ &  4.36$^{+ 0.33}_{- 0.29}$ & 14.0$^{+ 5.1}_{- 2.8}$ &  7.8$^{+ 1.3}_{- 0.8}$ &  0.78$^{+ 0.03}_{- 0.01}$ &   0.02$^{+  0.05}_{-  0.02}$ & 1792$^{+ 302}_{- 197}$ &  \\ [3pt]
HD 97048 & 10500 $\pm$  500 & 4.30$^{+0.20}_{-0.20}$ &  1.52$^{+ 0.23}_{- 0.23}$ &  2.2$^{+ 0.2}_{- 0.2}$ &  1.7$^{+ 0.1}_{- 0.1}$ &  0.90$^{+ 0.05}_{- 0.02}$ &   5.12$^{+  1.52}_{-  1.52}$ &  171$^{+   7}_{-   7}$ &  \\ [3pt]
HD 98922 & 10500 $\pm$  250 & 3.60$^{+0.10}_{-0.10}$ &  2.48$^{+ 0.15}_{- 0.15}$ &  4.0$^{+ 0.5}_{- 0.5}$ &  5.2$^{+ 0.3}_{- 0.3}$ &  0.09$^{+ 0.01}_{- 0.00}$ &   0.84$^{+  0.35}_{-  0.26}$ &  346$^{+  22}_{-  20}$ &  \\ [3pt]
HD 100453 &  7250 $\pm$  250 & 4.08$^{+0.15}_{-0.13}$ &  0.93$^{+ 0.17}_{- 0.21}$ &  1.5$^{+ 0.2}_{- 0.1}$ &  1.9$^{+ 0.3}_{- 0.3}$ &  0.00$^{+ 0.05}_{- 0.00}$ &   9.97$^{+  6.29}_{-  2.79}$ &  122$^{+  24}_{-  25}$ &  $\dagger$ b \\ [3pt]
HD 100546 &  9750 $\pm$  500 & 4.34$^{+0.06}_{-0.06}$ &  1.29$^{+ 0.14}_{- 0.14}$ &  1.9$^{+ 0.1}_{- 0.1}$ &  1.5$^{+ 0.1}_{- 0.1}$ &  0.00$^{+ 0.05}_{- 0.00}$ &   7.02$^{+  1.49}_{-  1.49}$ &   97$^{+  10}_{-  10}$ &  $\dagger$ b \\ [3pt]
HD 101412 &  9750 $\pm$  250 & 4.30$^{+0.20}_{-0.20}$ &  1.36$^{+ 0.23}_{- 0.23}$ &  2.0$^{+ 0.1}_{- 0.1}$ &  1.7$^{+ 0.1}_{- 0.1}$ &  0.21$^{+ 0.03}_{- 0.00}$ &   6.35$^{+  1.84}_{-  1.84}$ &  301$^{+  11}_{-  10}$ &  \\ [3pt]
PDS 344 & 15250 $\pm$  500 & 4.30$^{+0.20}_{-0.20}$ &  2.39$^{+ 0.25}_{- 0.25}$ &  3.7$^{+ 0.5}_{- 0.5}$ &  2.3$^{+ 0.1}_{- 0.1}$ &  0.86$^{+ 0.01}_{- 0.02}$ &   1.48$^{+  0.52}_{-  0.52}$ & 2756$^{+ 172}_{- 165}$ &  \\ [3pt]
HD 104237 &  8000 $\pm$  250 & 3.89$^{+0.12}_{-0.12}$ &  1.41$^{+ 0.17}_{- 0.21}$ &  2.0$^{+ 0.3}_{- 0.2}$ &  2.6$^{+ 0.4}_{- 0.4}$ &  0.00$^{+ 0.05}_{- 0.00}$ &   4.92$^{+  1.87}_{-  1.46}$ &  115$^{+  23}_{-  23}$ &  $\dagger$ b \\ [3pt]
V1028 Cen & 14000 $\pm$  500 & 3.80$^{+0.30}_{-0.30}$ &  2.85$^{+ 0.45}_{- 0.41}$ &  4.7$^{+ 2.0}_{- 1.0}$ &  4.5$^{+ 0.9}_{- 0.5}$ &  0.57$^{+ 0.01}_{- 0.03}$ &   0.59$^{+  0.69}_{-  0.39}$ & 1843$^{+ 355}_{- 215}$ &  \\ [3pt]
PDS 361S & 18500 $\pm$ 1000 & 3.80$^{+0.30}_{-0.30}$ &  3.53$^{+ 0.46}_{- 0.41}$ &  7.4$^{+ 3.2}_{- 1.7}$ &  5.7$^{+ 1.1}_{- 0.7}$ &  1.90$^{+ 0.04}_{- 0.01}$ &   0.19$^{+  0.23}_{-  0.12}$ & 4385$^{+ 872}_{- 541}$ &  \\ [3pt]
HD 114981 & 16000 $\pm$  500 & 3.60$^{+0.20}_{-0.20}$ &  3.47$^{+ 0.30}_{- 0.30}$ &  7.3$^{+ 2.0}_{- 1.5}$ &  7.1$^{+ 0.9}_{- 0.8}$ &  0.00$^{+ 0.05}_{- 0.00}$ &   0.18$^{+  0.17}_{-  0.09}$ &  908$^{+ 118}_{-  99}$ &  \\ [3pt]
PDS 364 & 12500 $\pm$ 1000 & 4.20$^{+0.20}_{-0.20}$ &  2.06$^{+ 0.23}_{- 0.23}$ &  3.1$^{+ 0.2}_{- 0.2}$ &  2.3$^{+ 0.1}_{- 0.1}$ &  1.87$^{+ 0.05}_{- 0.03}$ &   2.19$^{+  0.58}_{-  0.58}$ & 1715$^{+  97}_{-  91}$ &  \\ [3pt]
PDS 69 & 15000 $\pm$ 2000 & 4.00$^{+0.35}_{-0.35}$ &  2.72$^{+ 0.52}_{- 0.76}$ &  4.3$^{+ 2.0}_{- 1.5}$ &  3.4$^{+ 0.7}_{- 0.6}$ &  1.60$^{+ 0.07}_{- 0.07}$ &   0.84$^{+  2.02}_{-  0.62}$ &  630$^{+ 141}_{- 126}$ &  \\ [3pt]
DG Cir & 11000 $\pm$ 3000 & 4.41$^{+0.18}_{-0.18}$ &  1.49$^{+ 0.68}_{- 0.93}$ &  2.2$^{+ 0.4}_{- 0.4}$ &  1.5$^{+ 0.5}_{- 0.5}$ &  3.94$^{+ 0.13}_{- 0.54}$ &   5.95$^{+  2.97}_{-  2.97}$ &  713$^{+ 250}_{- 184}$ &  $\dagger$  $\ast$  \\ [3pt]
HD 132947 & 10250 $\pm$  250 & 3.90$^{+0.10}_{-0.10}$ &  1.97$^{+ 0.15}_{- 0.14}$ &  2.7$^{+ 0.3}_{- 0.3}$ &  3.1$^{+ 0.2}_{- 0.1}$ &  0.00$^{+ 0.05}_{- 0.00}$ &   2.44$^{+  0.77}_{-  0.65}$ &  565$^{+  35}_{-  26}$ &  \\ [3pt]
HD 135344B &  6375 $\pm$  125 & 3.94$^{+0.12}_{-0.12}$ &  0.85$^{+ 0.18}_{- 0.22}$ &  1.5$^{+ 0.2}_{- 0.2}$ &  2.2$^{+ 0.4}_{- 0.4}$ &  0.23$^{+ 0.05}_{- 0.06}$ &   7.99$^{+  3.24}_{-  2.34}$ &  140$^{+  28}_{-  28}$ &  $\dagger$ a \\ [3pt]
HD 139614 &  7750 $\pm$  250 & 4.31$^{+0.12}_{-0.12}$ &  0.82$^{+ 0.17}_{- 0.21}$ &  1.5$^{+ 0.1}_{- 0.1}$ &  1.4$^{+ 0.2}_{- 0.2}$ &  0.00$^{+ 0.05}_{- 0.00}$ &  15.64$^{+  4.29}_{-  4.29}$ &  140$^{+  28}_{-  28}$ &  $\dagger$ a \\ [3pt]
PDS 144S &  7750 $\pm$  250 & 4.13$^{+0.14}_{-0.16}$ &  1.02$^{+ 0.20}_{- 0.23}$ &  1.6$^{+ 0.2}_{- 0.1}$ &  1.8$^{+ 0.3}_{- 0.3}$ &  0.57$^{+ 0.07}_{- 0.08}$ &   9.45$^{+  4.81}_{-  2.88}$ & 1000$^{+ 200}_{- 200}$ &  $\dagger$ f \\ [3pt]
HD 141569 &  9750 $\pm$  250 & 4.35$^{+0.15}_{-0.15}$ &  1.28$^{+ 0.28}_{- 0.37}$ &  1.9$^{+ 0.4}_{- 0.4}$ &  1.5$^{+ 0.5}_{- 0.5}$ &  0.01$^{+ 0.01}_{- 0.00}$ &   9.00$^{+  4.50}_{-  4.50}$ &  112$^{+  34}_{-  33}$ &  \\ [3pt]
HD 141926 & 28000 $\pm$ 1500 & 3.75$^{+0.25}_{-0.25}$ &  4.70$^{+ 0.26}_{- 0.37}$ & 19.4$^{+ 4.5}_{- 5.0}$ &  9.7$^{+ 1.1}_{- 1.3}$ &  2.40$^{+ 0.03}_{- 0.04}$ &   0.00$^{+  0.03}_{-  0.00}$ & 1254$^{+ 143}_{- 175}$ &  \\ [3pt]
HD 142666 &  7500 $\pm$  250 & 4.13$^{+0.11}_{-0.16}$ &  0.96$^{+ 0.20}_{- 0.24}$ &  1.6$^{+ 0.2}_{- 0.1}$ &  1.8$^{+ 0.3}_{- 0.3}$ &  0.50$^{+ 0.08}_{- 0.09}$ &  10.43$^{+  6.21}_{-  3.34}$ &  145$^{+  29}_{-  29}$ &  $\dagger$ a \\ [3pt]
HD 142527 &  6500 $\pm$  250 & 3.93$^{+0.08}_{-0.08}$ &  0.90$^{+ 0.12}_{- 0.13}$ &  1.6$^{+ 0.1}_{- 0.1}$ &  2.2$^{+ 0.1}_{- 0.2}$ &  0.00$^{+ 0.05}_{- 0.00}$ &   8.08$^{+  1.94}_{-  1.63}$ &  140$^{+  20}_{-  20}$ &  $\dagger$ g \\ [3pt]
HD 144432 &  7500 $\pm$  250 & 4.05$^{+0.17}_{-0.14}$ &  1.04$^{+ 0.19}_{- 0.21}$ &  1.6$^{+ 0.2}_{- 0.1}$ &  2.0$^{+ 0.3}_{- 0.3}$ &  0.00$^{+ 0.06}_{- 0.00}$ &   8.72$^{+  4.81}_{-  2.50}$ &  160$^{+  32}_{-  32}$ &  $\dagger$ b \\ [3pt]
HD 144668 &  8500 $\pm$  250 & 3.75$^{+0.13}_{-0.12}$ &  1.76$^{+ 0.19}_{- 0.22}$ &  2.5$^{+ 0.4}_{- 0.4}$ &  3.5$^{+ 0.6}_{- 0.6}$ &  0.33$^{+ 0.05}_{- 0.04}$ &   2.70$^{+  1.32}_{-  0.93}$ &  160$^{+  32}_{-  32}$ &  $\dagger$ b \\ [3pt]
HD 145718 &  8000 $\pm$  250 & 4.37$^{+0.15}_{-0.15}$ &  0.82$^{+ 0.29}_{- 0.37}$ &  1.5$^{+ 0.3}_{- 0.3}$ &  1.3$^{+ 0.4}_{- 0.4}$ &  0.74$^{+ 0.06}_{- 0.05}$ &  19.54$^{+  9.77}_{-  9.77}$ &  134$^{+  41}_{-  39}$ &  $\dagger$  $\ast$  \\ [3pt]
PDS 415N &  6250 $\pm$  250 & 4.47$^{+0.15}_{-0.15}$ &  0.13$^{+ 0.30}_{- 0.39}$ &  1.1$^{+ 0.2}_{- 0.2}$ &  1.0$^{+ 0.3}_{- 0.3}$ &  1.11$^{+ 0.11}_{- 0.15}$ & 336.02$^{+168.01}_{-168.01}$ &  197$^{+  60}_{-  58}$ &  $\dagger$  $\ast$  \\ [3pt]
HD 150193 &  9000 $\pm$  250 & 4.27$^{+0.17}_{-0.17}$ &  1.21$^{+ 0.19}_{- 0.23}$ &  1.9$^{+ 0.1}_{- 0.1}$ &  1.7$^{+ 0.3}_{- 0.3}$ &  1.55$^{+ 0.02}_{- 0.04}$ &   7.22$^{+  1.89}_{-  1.89}$ &  120$^{+  24}_{-  24}$ &  $\dagger$ h \\ [3pt]
AK Sco &  6250 $\pm$  250 & 4.26$^{+0.10}_{-0.10}$ &  0.38$^{+ 0.18}_{- 0.20}$ &  1.2$^{+ 0.1}_{- 0.1}$ &  1.3$^{+ 0.2}_{- 0.2}$ &  0.00$^{+ 0.05}_{- 0.00}$ &  17.71$^{+  4.71}_{-  3.42}$ &  103$^{+  20}_{-  21}$ &  $\dagger$ b \\ [3pt]
PDS 431 & 10500 $\pm$  500 & 3.70$^{+0.20}_{-0.20}$ &  2.32$^{+ 0.31}_{- 0.30}$ &  3.5$^{+ 1.0}_{- 0.7}$ &  4.4$^{+ 0.6}_{- 0.5}$ &  1.76$^{+ 0.03}_{- 0.03}$ &   1.19$^{+  1.07}_{-  0.61}$ & 2875$^{+ 384}_{- 316}$ &  \\ [3pt]
KK Oph &  8500 $\pm$  500 & 4.38$^{+0.15}_{-0.15}$ &  0.94$^{+ 0.33}_{- 0.43}$ &  1.6$^{+ 0.3}_{- 0.3}$ &  1.4$^{+ 0.4}_{- 0.4}$ &  2.70$^{+ 0.10}_{- 0.15}$ &  15.16$^{+  7.58}_{-  7.58}$ &  279$^{+  86}_{-  81}$ &  $\dagger$  $\ast$  \\ [3pt]
HD 163296 &  9250 $\pm$  250 & 4.30$^{+0.20}_{-0.20}$ &  1.23$^{+ 0.23}_{- 0.23}$ &  1.9$^{+ 0.1}_{- 0.1}$ &  1.6$^{+ 0.0}_{- 0.0}$ &  0.00$^{+ 0.05}_{- 0.00}$ &   7.56$^{+  2.17}_{-  2.17}$ &  101$^{+   4}_{-   3}$ &  \\ [3pt]
MWC 297 & 24500 $\pm$ 1500 & 4.00$^{+0.30}_{-0.30}$ &  3.95$^{+ 0.46}_{- 0.46}$ & 10.2$^{+ 4.6}_{- 4.6}$ &  5.3$^{+ 1.1}_{- 1.1}$ &  8.47$^{+ 0.04}_{- 0.03}$ &   0.10$^{+  0.08}_{-  0.08}$ &  170$^{+  34}_{-  34}$ &  \\ [3pt]

  \hline
  \end{tabular} 
\end{minipage}

\begin{minipage}{17.5cm}

{\footnotesize 
$\dagger$ -- A literature distance is initially adopted to these stars, as log(g) cannot be determined from the spectra alone. $\ast$ -- Stars which have been placed on the ZAMS. References: (a) \citet{deZeeuw1999},
(b) \citet{vanLeeuwen2007},
(c) \cite{Dahm2005},
(d) \citet{Shevchenko1999},
(e) \citet[accepted]{Ababakr2015},
(f) \citet{Vieira2003},
(g) \citet{Fukagawa2006},
(h) \citet{Loinard2008}.} 

\end{minipage}

\end{table*}

Determining accurate stellar parameters is crucial for extracting an accretion rate, and for obtaining further information about the age, evolution, and ongoing processes in the environment around HAeBe stars. Many stars in this sample have had their stellar parameters determined previously, but this has often been done in smaller sub-sets using a variety of methods \citep[][see also the Appendix for additional references]{Mora2001,Hernandez2004,Manoj2006,Montesinos2009,Alecian2013}. For this reason a full treatment of determining stellar parameters is performed on the entire sample, in a homogeneous fashion, to provide better consistency between the stars. A comparison will also be made with the literature values to confirm the method employed; as most stars would be expected to have similar temperatures to the previous literature values.

The determination of parameters is performed in a three step process: 1) Spectral typing is performed using the X-Shooter spectra to provide accurate limits on the effective surface temperature, $T_{\rm eff}$, and where possible the surface gravity too, log(g); 2) KC-models and the photometry are used to assess the reddening, $A_V$, and distance/radius, $D/R_\star$, ratio towards the targets; 3) Finally, PMS evolutionary tracks are used to infer a mass, $M_\star $, and age (and other parameters if not determined yet). The stages of this process are now given in detail.\\

\subsection{Temperature and Surface Gravity Determination}
\label{sec:st_tng}

\begin{figure*}%
        \includegraphics[trim=1.0cm 1.0cm 0.5cm 0.5cm, width=0.45\textwidth]{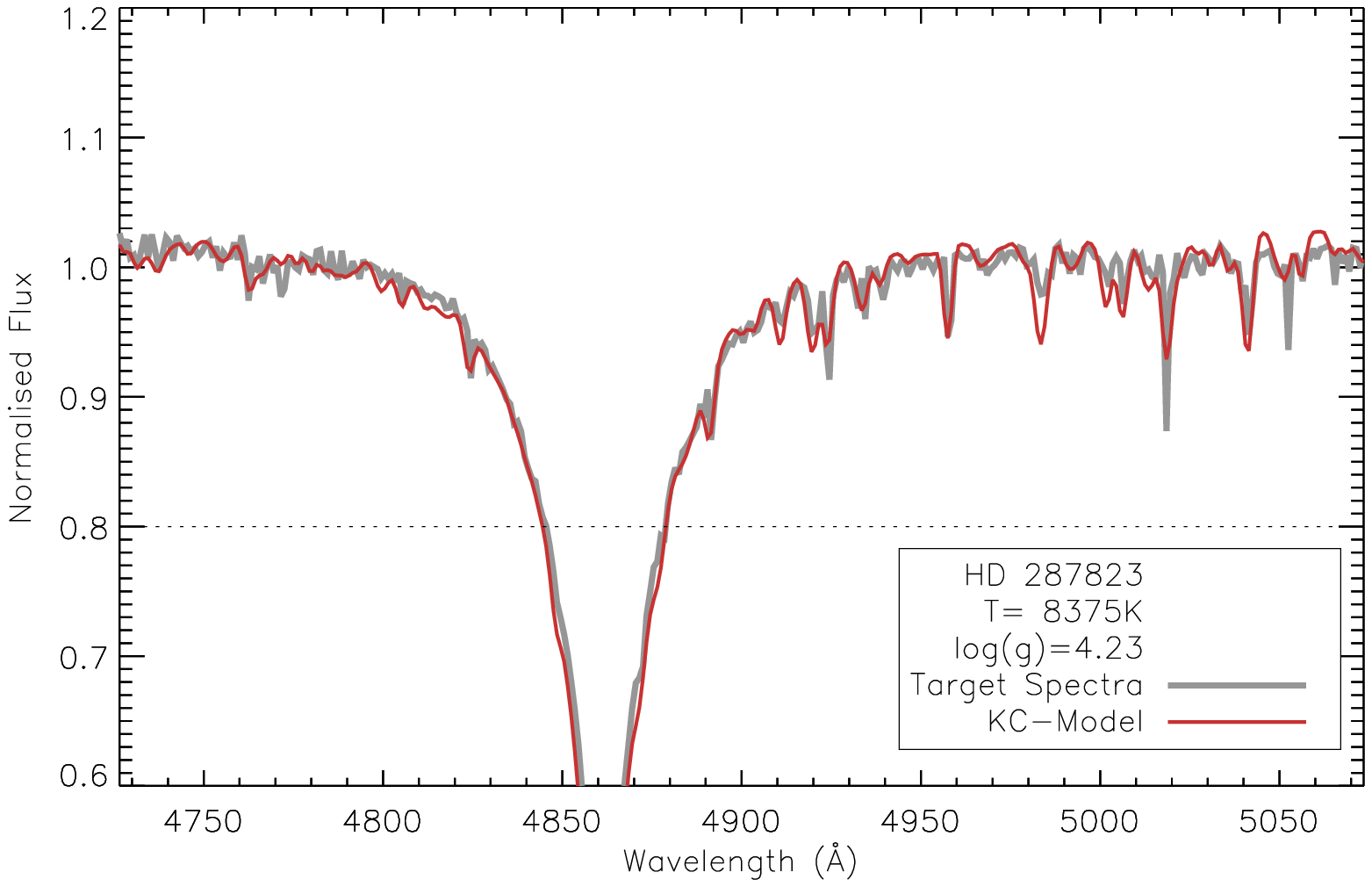}
        \includegraphics[trim=1.0cm 1.0cm 0.5cm 0.5cm,width=0.45\textwidth]{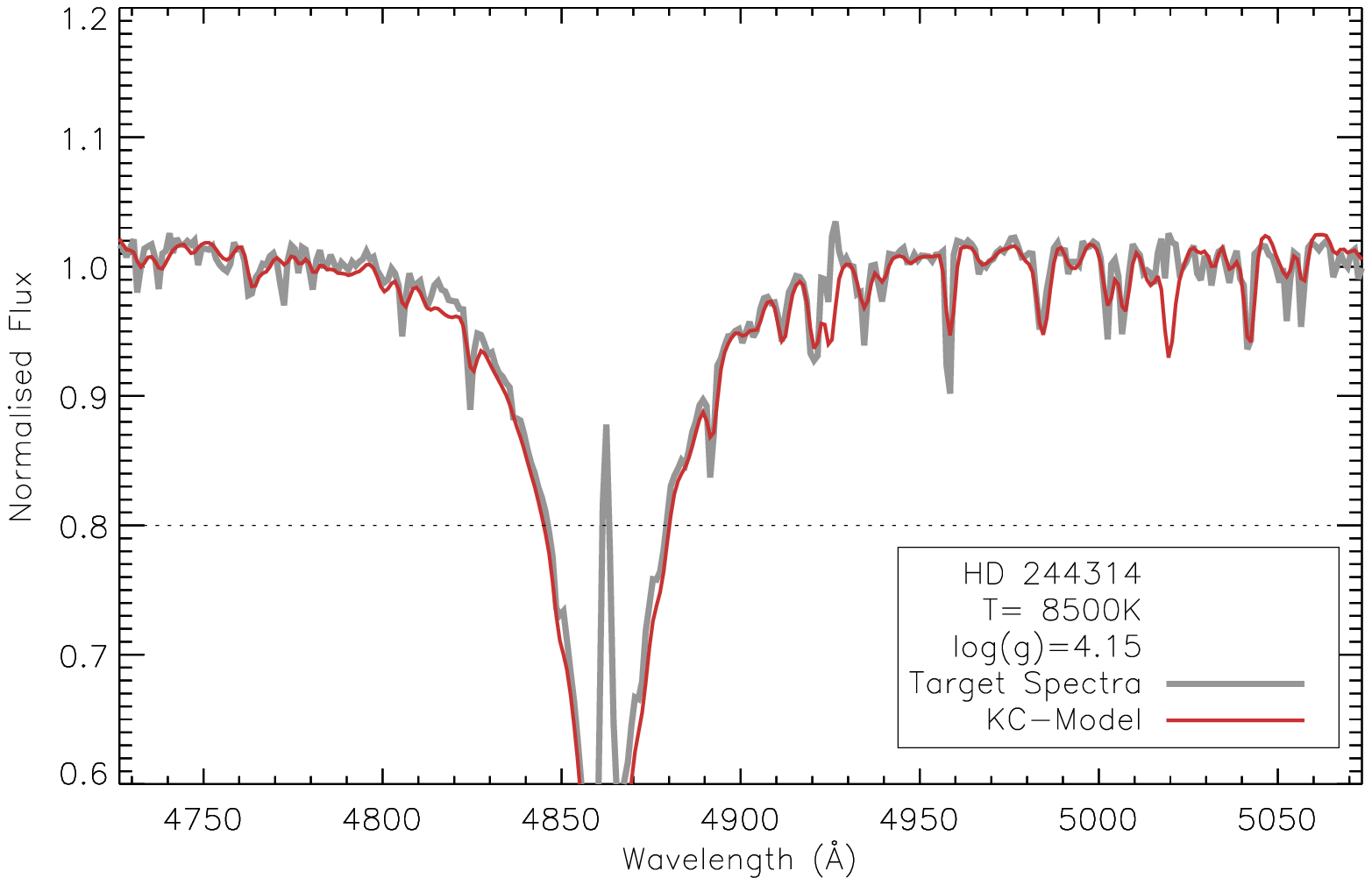}
        \includegraphics[trim=1.0cm 1.0cm 0.5cm 0.5cm,width=0.45\textwidth]{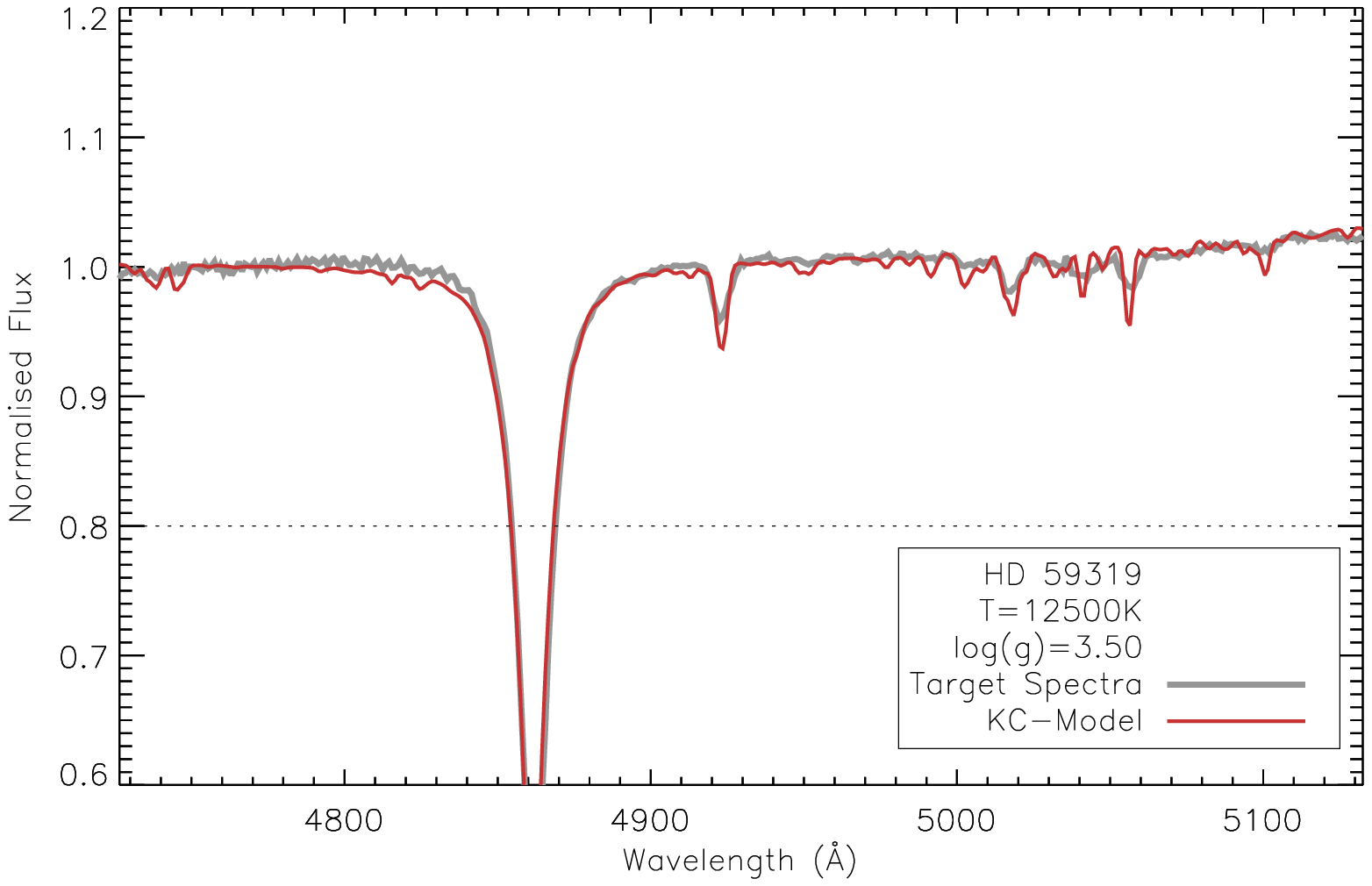}
        \includegraphics[trim=1.0cm 1.0cm 0.5cm 0.5cm,width=0.45\textwidth]{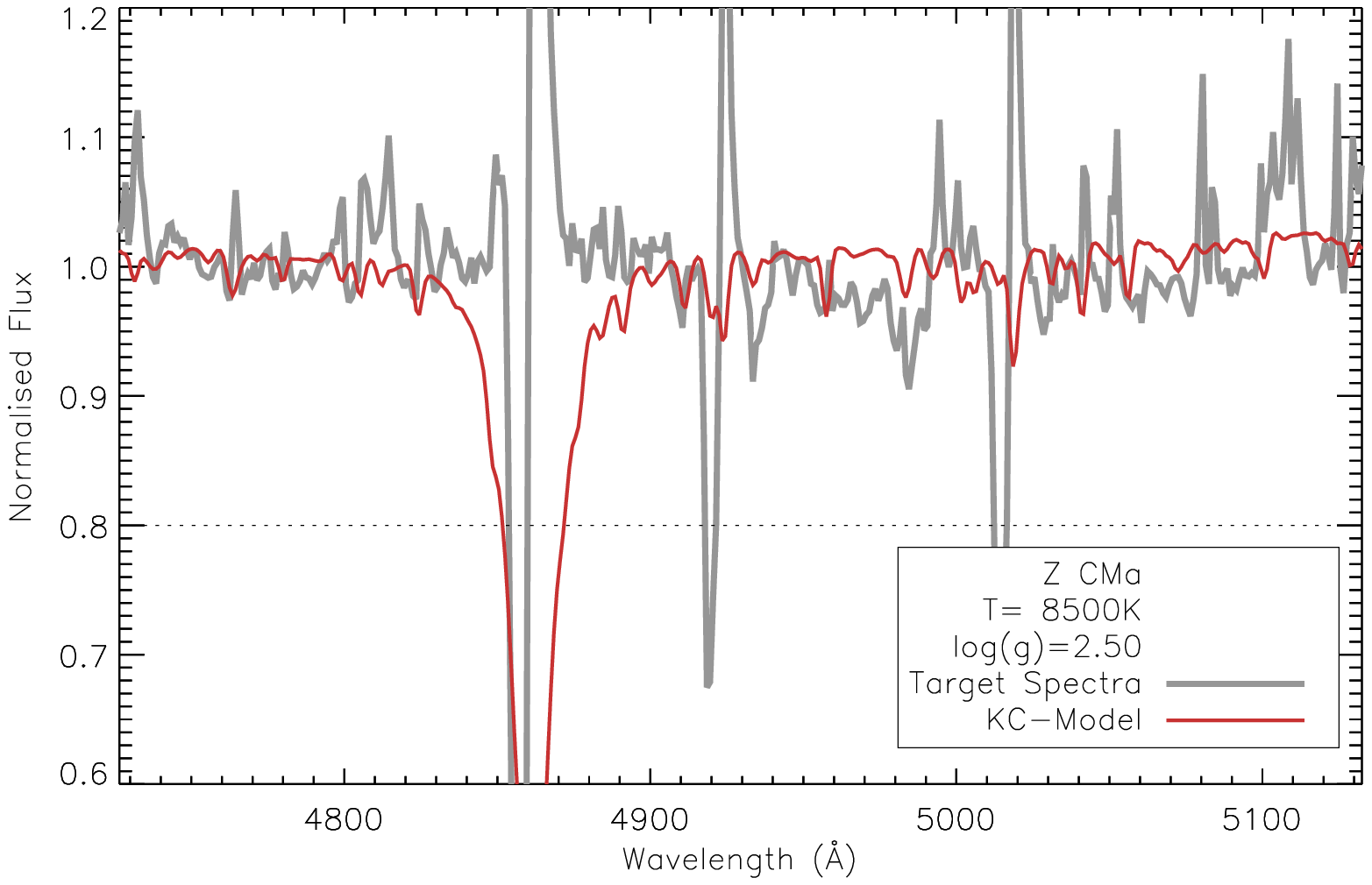}
 
    \caption{Examples of spectral typing for four targets are presented here. Each panel shows both the target spectra (grey), and a KC-model which denotes a good fit (red). The parameters for the KC-model are given for each fit. Also plotted is a dashed line, at 0.8 of the normalised intensity, which is used as a cut-off in the fitting. The two panels on the left show the cases for straight forward fit where there is no obvious emission. The figure in the top-right is a case where there is clear emission present; a good fit is still achieved. The bottom-right panel gives an example of one of the few objects which cannot be spectrally typed in this way due to extremely strong emission; this exceptional object, and others like it, are presented in detail in Appendix \ref{sec:app_except}.}
    \label{fig:spec_type}
\end{figure*}

The first stage takes advantage of the large wavelength coverage and good spectral resolution of X-Shooter to perform spectral typing; allowing us to narrow down the possible $T_{\rm eff}$ and log(g) of each target. This is done by following a similar method to \citet{Montesinos2009}, of spectral typing using the wings of the hydrogen Balmer series, and the continuum region 100--150$\rm \AA{}$ either side of the lines. These lines are favoured due to their sensitivity to changes in $T_{\rm eff}$ and log(g). Specifically, the H$\beta$, H$\gamma$, and H$\delta$ lines are used as they have the largest intrinsic absorption of the series, except for the H$\alpha$ line. H$\alpha$ is not used for spectral typing as it is often seen entirely in emission, with the emission being both the strongest and broadest of the Balmer series in HAeBes. This could in-turn affect the derived parameters. Therefore, fitting of models to the line wings is performed using the other lines in the series. To perform the fitting, each line is first normalised based on the continuum either side of the line. They are then compared against a grid of KC-model spectra, which have also been normalised in the same way using the same regions either side of the line. The resolution of the grid is set to be in steps of 250~K for $T_{\rm eff}$ and 0.1~dex in log(g). The metallicity is kept at $[\rm M/H]=0$ throughout, although it has been shown that the choice of metallicity can affect spectral typing in HAeBes \citep{Montesinos2009}. The fit of the synthetic spectra to the observed spectra is judged using the wings of each line, and continuum features, where the intensity is greater than 0.8; the line centre is excluded as it can often be found in emission. This approach avoids the problems of both emission, and rotational broadening in the line. Figure \ref{fig:spec_type} gives four examples of this fitting, highlighting the power for obtaining an accurate $T_{\rm eff}$ and log(g), where many errors are as small as the chosen step size. However, despite this reliable technique, issues arise for two cases. The first, is that there is a non-linear relationship between the Balmer line width and the surface gravity for objects which have $T_{\rm eff}<$~8000~K \citep{Guimaraes2006}. However, for temperatures up to 9000~K there is increased uncertainty due to the large presence of absorption features, which make normalising and comparing different surface gravity scenarios increasingly difficult. For these reasons we do not constrain log(g) using the spectra for stars with a suspected $T_{\rm eff}<$~9000~K. 

The widths of the Balmer lines are tightly correlated to $T_{\rm eff}$ and log(g), to the point where different combinations of the two can produce the same widths. However, this degeneracy can be broken when viewing the whole of the line profile and the absorption features within them (and also the photospheric absorption features outside of the wings).

The second issue concerns objects which display very strong emission lines; where the line strength is exceptionally strong across the Balmer series to the point where the width of the lines eclipse even the broad photospheric absorption wings. Extremely strong P-Cygni, or inverse P-Cygni, profiles can also affect the line shape in the wings. An example of extreme emission is shown in the bottom-right panel of Figure \ref{fig:spec_type}, where none of the intrinsic photospheric absorption lines can be seen due to the emission. P-Cygni absorption is also present in this example further complicating any possible analysis of the wings. Objects, like the example just given, where both $T_{\rm eff}$ and log(g) cannot be constrained by this method, will be treated separately on an individual basis and are detailed in Appendix \ref{sec:app_except}. The objects for which $T_{\rm eff}$ has been constrained can have all of their parameters determined in the next two steps.

For the telluric standards the same above steps are applied. This is because they are well-behaved stars for which a $T_{\rm eff}$ and log(g) determination is straight forward. These parameters are required for the data reduction discussed previously in Section \ref{sec:xs_data}. 

\begin{figure}
 \includegraphics[trim=0.5cm 0.5cm 0.0cm 0.5cm, width=0.5\textwidth]{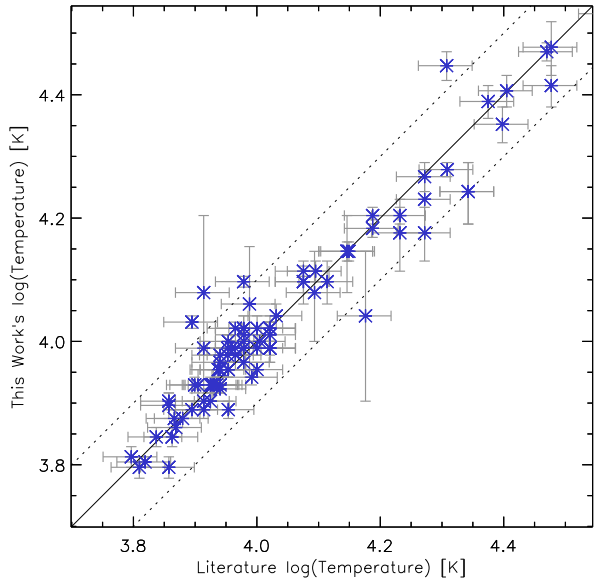}
 \caption{Shows the temperatures derived in this work in comparison to literature estimates. The solid black line is the expected line of correlation and the dashed lines are a 0.1 dex deviation from this. The standard deviation between the two is only 0.02, with a mean offset of 0.03 towards this work. The dashed lines therefore encompass 3$\sigma$, showing the two samples are well correlated. The literature temperatures used and their references are provided in Appendix \ref{sec:app_phot}.}
\label{fig:teff_lit_targets}
\end{figure}

Figure \ref{fig:teff_lit_targets} compares the temperatures derived in this work against previous estimates from the literature (see Table \ref{tab:phot} in Appendix \ref{sec:app_phot}).  The temperature is chosen for comparison as it is a key stellar parameter which can be determined more readily than log(g), and its appearance in the literature is more frequent than other parameters (allowing a greater number of comparisons to be made). The majority of literature works provide a spectral type rather than a precise temperature so we assign an error of 10\% for these. The figure shows over 95\% of the stars are in agreement, within the errors. Also, the temperature determinations in this work have been based on some of the best spectra available for these objects, which helps keep errors to a minimum. This serves as a justification for the homogeneous approach to determining temperatures and their use here, for both the target stars and the telluric standards alike.


\subsection{Photometry Fitting}
\label{sec:st_phot}

The second step of this process takes two directions: One case is where both $T_{\rm eff}$ and log(g) could be determined from the spectra, and the other case is for when only $T_{\rm eff}$ could be determined. In both cases fitting spectra of model atmospheres, based on the parameters determined in the previous step, to the observed optical photometry will be performed. The fitting will provide a level of reddening, $A_V$ to each star, and a scaling factor, $D/R_\star$, due to the fitting of surface flux models to observed photometry. An accurate temperature is paramount here in order to break any degeneracy of fitting models to the photometry.

To perform the fitting only the $BVRI$ points are used; the $U$-band can be influenced heavily by the Balmer Excess, and no photometry long-wards of the $I$-band is used due to the possible influence of the IR-excess (which itself would require dedicated modelling). The $B$-band can also be affected in the cases of extremely large flux excess. Fortunately, these cases are rare and the change in the $B$-band magnitude would not significantly affect the fitting (the fitting is far more sensitive to the input temperature). 

Another point to consider when looking at optical photometry is the effects of variability, as this has been observed in numerous HAeBes \citep{deWinter2001, Oudmaijer2001, Mendigutia2011a, Pogodin2012, Mendigutia2013}. However, variability information is not present for all of the targets,  but we estimate that the calculated parameters will not be affected significantly if the photometric variation is less than 0.2~mag. In all cases we use photometry when at maximum brightness, as this best reflects the scenario where we are mostly viewing the stellar photosphere. So an assumption is adopted here that the photometry we use is predominately photospheric and not highly variable. 

In order to fit the photometry a unique grid of KC-models is set up based on the limits derived in step 1 for each star; the grid follows the same step sizes used in the previous step too. Log(g) does not have a significant effect on the fitting to the photometry, as the spectral shape is overwhelmingly dominated by the temperature. This allows log(g)=4.0 to be adopted and used in this step for the stars where log(g) could not be determined from the spectra; this value will be revised in the next step.

The models are reddened until a best fit to the photometry is achieved; the best fit being when the reddened SED shape of the model is in-line with the photometry. The dereddening is performed using the reddening law of \citet{Cardelli1989}, with a standard R$_V$=3.1, in all cases. 
It is possible that the total-to-selective extinction may be higher, possibly R$_V=5.0$, for some of the stars in this sample based on previous analysis of HAeBes \citep{Hernandez2004,Manoj2006}. However, the choice of R$_V$ will only affect the targest with the most extinction and the changes this will have on the stellar parameters and accretion rates are minimal due to the observed colour excess remaining the same. The majority of the targets have a mean $E(B-V) \approx 0.4$.

\begin{figure}
 \includegraphics[trim=0.5cm 0.5cm 0.0cm 0.5cm, width=0.5\textwidth]{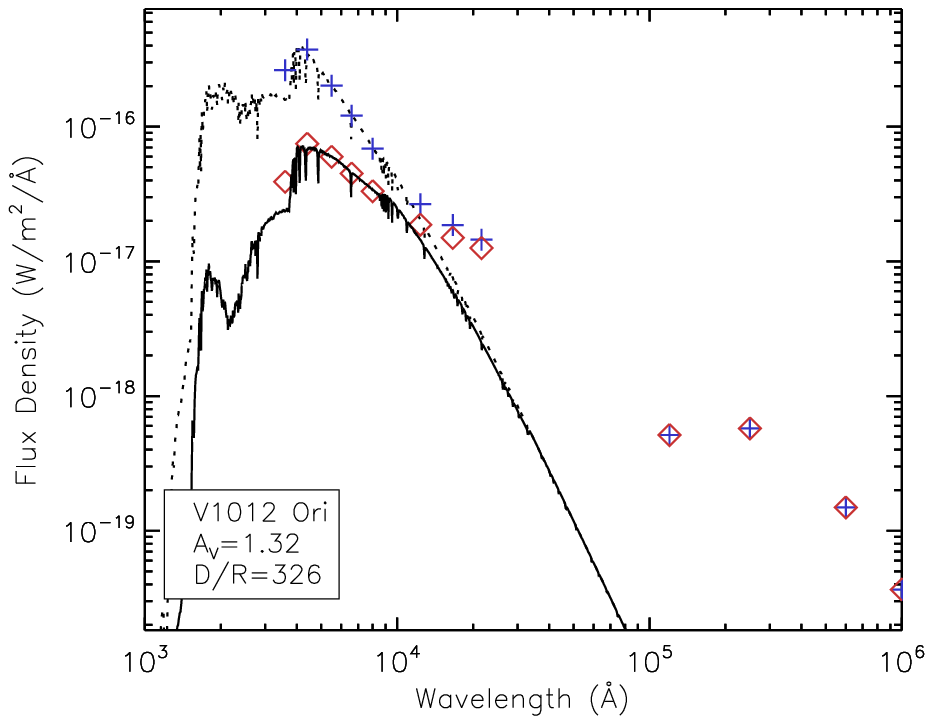}
 \caption{Here is an example of step 2 in the stellar parameter determinations (see Section \ref{sec:st_phot}), where a reddened KC-model (black) is fitted to the observed photometry (red diamonds). The opposite is also shown, of dereddened photometry (blue plus signs) fitted to a KC-model with no reddening applied (dashed line). The level of reddening, $A_V$, is displayed in the figure. The fit provides a ratio between the distance to the star and its radius, $D/R_\star $, as this is required to scale the model to fit the photometry. Also visible in this plot is how the $U$-band magnitude is higher than the KC-model used, a possible indication of Balmer Excess. A clear IR excess can also been seen, starting at around the J-band; a typical feature of PMS stars.}
\label{fig:photometry_sed}
\end{figure}

Returning to the fitting, the model is normalised to the $V$-band point by a scaling factor, which is  $(D/R_\star)^2$. This scaling factor arises from the fitting of models in units of surface flux to observed photometry. An advantage of knowing this scaling factor is that it allows either distance or radius to be determined provided the other is known. Figure \ref{fig:photometry_sed} shows an example of the above fitting for the case of V1012 Ori, along with a dereddened version of the photometry and the model spectra. This object is shown as it demonstrates a clear IR-excess, a noticeable $A_V$, and a $U$-band magnitude slightly higher than the KC-model spectra (possible Balmer Excess).

At this stage the techniques diverge between the stars for which a log(g) was determined, and for the ones in which it could not be. For the former no further action is taken in this step. For the latter a distance is adopted to the star based upon the location of the star on the sky and its possible associations with nearby star forming regions. The stars for which this is performed are noted in Table \ref{tab:params}; the literature distances adopted and references are both provided in the same table (and also in Table \ref{tab:phot} in Appendix \ref{sec:app_phot}). An error of 20\% is adopted for the distance, as this helps reflect the additional uncertainty on whether the star is truly part of the association, and the possible extent of the association. If the error is higher than 20\% then the higher error is adopted instead. By adopting a distance to these stars a radius can be determined from the scaling factor. Then, combining this radius with the temperature, the luminosity is calculated by a black-body relationship of $L_\star = 4 \pi R_\star ^2 \sigma T_{\rm eff}^4$ (this calculation is equivalent to the sum of the flux under the KC-model multiplied by  $4 \pi D^2$).

\subsection{Mass, Age, Radius, and log(g) Determination}
\label{sec:st_mrg}

In this third and final step, the remaining stellar parameters are now determined through the use of PMS tracks. The PARSEC tracks of \citet{Bressan2012} are used for the majority of this step as they cover a mass range of 0.1--12$\rm M_\odot$, which encompasses all of the theoretical HAeBe mass range, and a metallicity is chosen of Z=0.01 (this is close to solar metallicity \citep{Caffau2011}). Additionally, two tracks from \citet{Bernasconi1996} are used for objects greater than 12~$\rm M_\odot$.  Each track is of a fixed mass, with no accretion contribution, which evolves over time in $T_{\rm eff}$ and $L_\star $ as the star contracts. As $T_{\rm eff}$ and $L_\star $ change so do $R_\star$ and log(g) as a consequence. This allows each star to be plotted on either: a $L_\star $ vs. $T_{\rm eff}$ set of tracks, for the stars where $L_\star $ is known from the adopted distance; or on a log(g) vs. $T_{\rm eff}$ set of tracks, for the stars where both log(g) and $T_{\rm eff}$ were determined from the spectra. For the first scenario a mass and age are extracted from the PMS tracks. Then, log(g) is calculated using this mass and the radius from the previous step. For the second scenario, luminosity, mass, and age are all extracted from the tracks. These can then be used to obtain a radius from the temperature and luminosity; or the mass and log(g), both choices are equivalent. Finally, using the $D/R_\star$ factor a distance can be determined.

However, not all cases allow parameters to be extracted from the tracks. These few cases are where the stars are located below the zero-age main sequence, ZAMS. It should be noted that for a few of these cases, where the stars are only just below the ZAMS of the chosen tracks, then tracks relating to stars with a lower metallicity may be more appropriate. However, in general, it appears more likely that their placement is genuinely below the ZAMS and is due to the adopted literature distances used being incorrect; as their use provides small radii from the $D/R_\star$ ratio. The radius is deemed too small as it is less than the expected radius of a ZAMS star of the same temperature. Additionally, most of these stars have Diffuse Interstellar Bands, DIBs, in their spectra which suggest $A_V \sim 0.5-2.0$~mag \citep{Jenniskens1994}. Extinction due to DIBs follows a trend of $\sim 1.8$~mag/kpc \citep{Whittet2003}. This suggests that the distances should be greater than the adopted values and should be revised.
Previously, for these stars, the assumption had been made that the stars are associated with a star-forming region. It is now more probable from the spectral typing and position of the stars in relation to the PMS tracks, that some of the distances chosen are not valid i.e. the star may not be associated with the chosen region. Also, the spectrally determined $T_{\rm eff}$ is more likely to be correct as it comes form spectra which has been directly observed from the star itself, opposed to a distance inferred from a possible association. A solution to this problem is calculating new distances to these outlier targets; ones which provide more sensible radii and agree with the spectrally determined temperature. To do this the stars are placed on the ZAMS at a point appropriate for their derived temperature; essentially, this is a lower limit to the luminosity of the star. This provides values of $L_\star$, $R_\star$, $M_\star$, and an age. With the new ZAMS radii, revised distances are calculated from $D/R_\star$. All objects affected by these ZAMS changes are noted in Table \ref{tab:params}. At this point all basic stellar parameters, relevant to this work, have been determined.


\section[]{Balmer Excess Measurements}
\label{sec:be}

With knowledge of the stellar parameters obtained for all targets a measurement of the Balmer Excess, $\Delta D_B$, can now be made. $\Delta D_B$ is defined as the excess in flux above the intrinsic photospheric flux, seen across the Balmer Jump region (this region spans the wavelength range where the hydrogen Balmer series reaches its recombination limit $\sim 3640$--$3680$~${\rm \AA}$). The UV-excess is weaker, in terms of energy, in lower mass stars, but is more readily visible due to their cooler photospheres, on top of which the excess can be seen. This UV-excess has been measured in both brown dwarfs \citep{Herczeg2008,Herczeg2009,Rigliaco2012} and CTTs \citep{Calvet2004,Gullbring2000,Calvet2004,Ingleby2013}. From these past studies the current consensus to the origin of the excess is magnetospheric accretion. It has also been shown, in small samples, that an observable $\Delta D_B$ in HAeBes stars can be explained within the same context \citep{Muzerolle2004,Donehew2011,Mendigutia2011b,Pogodin2012}. We aim to further our understanding of accretion in HAeBes by testing accretion within the context of MA to a large sample of HAeBes; this includes numerous HBe stars for which little investigation has been done. The Balmer Excess is defined as:
\begin{equation}
 \Delta D_B=(U-B)_{0}-(U-B)_{\rm dered}
\label{eqn:db_original}
\end{equation}
where, $ (U-B)_{0}$ is the intrinsic colour of the target and ${(U-B)}_{\rm dered}$ is the dereddened observed colour index. Detailed below are the two best methods of measurement.


\begin{figure}
 \includegraphics[trim=0.5cm 0.5cm 0.5cm 0.5cm, width=0.5\textwidth]{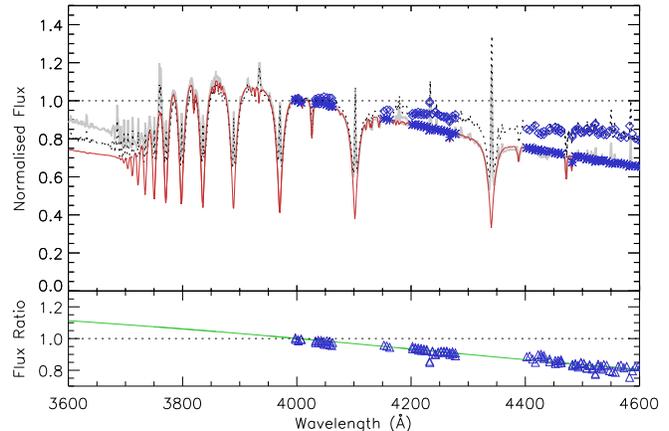}
 \caption{An example of the method 1 technique of measuring $\Delta D_B$ is shown here. Initially the observed spectrum (dashed-black) and the intrinsic spectrum (red, a KC-model matching the spectral type of the target) have been normalised to 4000~$\AA$. Continuum points are selected from these between 4000--4600~$\rm \AA$ (shown as blue points). A ratio of these are provided in the bottom panel and they are fitted by a reddening law, which is extrapolated to 3600~$\rm \AA$. This level of reddening correction is then applied to the original spectrum, with the result plotted in the top panel (grey). The SED of the model and the corrected spectra are now exactly the same between 4000-4600~$\rm \AA$, allowing measurement of the Balmer Excess to be performed using only the region around 3600~$\rm \AA$.}
 \label{fig:db_donehew}
\end{figure}


\subsection{Method 1 -- Spectral Matching: Single Point Measurement}
\label{sec:be_m1}

The first approach to measuring $\Delta D_B$ uses the spectral region of the the UVB arm from 3500--4600~$\rm \AA$, and adopts the same techniques employed by \citet{Donehew2011}. This method requires the spectrum of the target to be compared against the intrinsic spectrum of a star of the same spectral type. The KC-models mentioned earlier are used here as the intrinsic star spectra. Following the calibration in Section \ref{sec:xs}, the spectrum of each target shows the correct, reddened spectral shape. This allows both the target and model spectra to be normalised to 4000~$\rm \AA$, while preserving their spectral shape. Next, a correction for reddening present in the observed spectra is performed. To do this, the difference between the measured continuum of the target and the model between 4000--4600~$\rm \AA$~is fitted by a reddening law (the reddening law of \citet{Cardelli1989} is used here). This also provides a best-fit $A_V$. Extinction correction is applied to the whole spectrum, while maintaining the 4000~$\rm \AA$ pivot point for this correction. The result of this method is that the spectral shape of the target is adjusted such that the slope between the intrinsic model and the target spectrum now match. The success of this normalisation is independent of the amount of extinction towards the star \citep{Muzerolle2004,Donehew2011}. Fig \ref{fig:db_donehew} shows the application of this spectral slope matching technique along with an example output. 

To perform the measurement of $\Delta D_B$ attention must be drawn back to Equation \ref{eqn:db_original}, where the magnitudes are now converted into a flux:
\begin{equation}
 \Delta D_B=-2.5\:{\rm log}\left( \frac{F_U^{\rm phot}}{F_B^{\rm phot}} \right) +2.5{\rm log}\left( \frac{F_U^{\rm dered}}{F_B^{\rm dered}} \right)
\label{eqn:db_flux}
\end{equation}
where $F$ is the flux, with subscripts denoting the corresponding wavelength region, and the superscripts are: the intrinsic flux - denoted `phot', and the dereddened flux - denoted `dered'. For these measurements the fluxes are monochromatic. Now, consider the fact that the observed, dereddened flux includes an accretion contribution, such that $F_U^{\rm dered}=F_U^{\rm phot}+F_U^{\rm acc}$. This allows the above equation to be written as:
\begin{equation}
 \Delta D_B=2.5\:{\rm log}\left( \frac{F_U^{\rm phot}+F_U^{\rm acc}}{F_B^{\rm phot}+F_B^{\rm acc}}\times \frac{F_B^{\rm phot}}{F_U^{\rm phot}} \right)
\label{eqn:db_flux_acc}
\end{equation}
This equation can be reduced through the use of a normalisation factor $\alpha^{\rm norm}$, where $(F_B^{\rm phot} + F_B^{\rm acc})\times \alpha^{\rm norm}=F_B^{\rm phot}$. This normalisation across the $B$-band is performed automatically by matching the slope of the spectrum of the target to the intrinsic spectrum's slope (see the steps mentioned earlier). In essence $\alpha^{\rm norm}$ represents a reddening law. This gives us the final form of the $\Delta D_B$ equation:
\begin{equation}
 \Delta D_B=2.5\:{\rm log}\left( \frac{F_U^{\rm phot,norm}+F_U^{\rm acc,norm}}{F_U^{\rm phot}} \right)
\label{eqn:db_done}
\end{equation}
By these definitions, the $F_U^{\rm phot,norm}+F_U^{\rm acc,norm}$ is just the flux observed from the target spectra, and $F_U^{\rm phot}$ can be taken from a KC-model of the same spectral type. Since the spectrum obtained is of medium resolution we adopt a narrow, monochromatic, range over a typical broadband filter to represent the $U$-band magnitude. This also gives us better precision in measurements. 
The wavelength region of measurement is 3500--3680~$\rm \AA$. This is chosen as it is beyond the Balmer recombination limit. However, two of the echelle orders of X-Shooter overlap in this region, and the SNR in an echelle order decreases as wavelength decreases. Therefore, to minimise errors, the 3500--3600~$\rm \AA$ region from echelle order 21 and the 3600--3680~$\rm \AA$ region from echelle order 20 are measured and combined to give the most accurate result.


\subsection{Method 2 -- $B$-Band Normalised, Multi-point Measurements}
\label{sec:be_m2}

An alternate method of measuring $\Delta D_B$ is given by \citet{Mendigutia2013}, which also does not require the reddening towards a star to be known. This method covers a larger wavelength range, requiring measurements of both the $U$-band and $V$-band points. These two points are measured from the observed spectra and a KC-model of the same spectral type (the same model as in method 1), after normalisation to the $B$-band. Rather than correcting for $A_V$, as in the previous method, reddening independence is achieved by expanding Equation \ref{eqn:db_original} and substituting in an expression for each reddening component: $A_\lambda=A_V(k_\lambda/k_V)$, where $A_\lambda$ and $A_V$ are the extinction at any given wavelength and in the $V$-band, respectively. Similarly, $k_\lambda$ and $k_V$ are the opacities for any given wavelength and the $V$-band, respectively. Applying the expression for $A_V$ to Equation \ref{eqn:db_original} gives:
\begin{equation}
\Delta D_B = (U-B)^{\rm int}-(U-B)^{\rm obs}+A_V \left( \frac{k_U}{k_V}-\frac{k_B}{k_V} \right)
\label{eqn:mend1}
\end{equation}
The superscript `int' refers to the intrinsic magnitudes (from a KC-model in this case) while the superscript `obs' refers to the observed magnitudes (from the observed spectra).  The values of the opacities are determined by the reddening law adopted. The reddening law of \citet{Cardelli1989} is used here with an R$_V$=3.1, providing $k_U/k_V=1.57$ and $k_B/k_V=1.33$. To remove the $A_V$ term the relationship between $A_V$ and colour excess needs to be used: $A_V={\rm R}_V E(B-V)$. At this point it should be noted that the method is now reddening independent, since $A_V$ has been removed, but remains dependent on the reddening law adopted, as this affects the opacity ratios. This new form for the Balmer Excess is:
\begin{multline}
\Delta D_B = (U-B)^{\rm int}-(U-B)^{\rm obs}+ \\
{\rm R}_V \left( \frac{k_U}{k_V}-\frac{k_B}{k_V} \right)[ (B-V)^{\rm obs}-(B-V)^{\rm int} ]
\label{eqn:mend2}
\end{multline}
which can be expressed in terms of flux, instead of magnitudes, as follows:
\begin{multline}
\Delta D_B = 2.5\:{\rm log} \left( \frac{F_U^{\rm obs}\alpha^{\rm norm}F_B^{\rm phot}}{F_B^{\rm obs}\alpha^{\rm norm}F_U^{\rm phot}} \right)+ \\
2.5\:{\rm R}_V \left( \frac{k_U}{k_V}-\frac{k_B}{k_V} \right){\rm log}\left( \frac{F_V^{\rm obs}\alpha^{\rm norm}F_B^{\rm phot}}{F_B^{\rm obs}\alpha^{\rm norm}F_V^{\rm phot}} \right)
\label{eqn:mend3}
\end{multline}
where $\alpha^{\rm norm}$ has been added and is a normalising factor for the $B$-band, as seen in method 1, but the normalisation is instead performed such that the spectra will be unity at 4400~$\rm \AA$. Note that all these fluxes are considered monochromatic, with centres at the usual Johnson UBV wavelengths. This normalisation allows the equation to reduce to its final form:
\begin{multline}
\Delta D_B = 2.5\:{\rm log} \left( \frac{F_U^{\rm obs,norm}}{F_U^{\rm phot}} \right) + \\
2.5\:{\rm R}_V \left( \frac{k_U}{k_V}-\frac{k_B}{k_V} \right){\rm log}\left( \frac{F_V^{\rm obs,norm}}{F_V^{\rm phot}} \right)
\label{eqn:mend_final}
\end{multline}
in this form it can be seen that only four points need to be measured in order to obtain $\Delta D_B$ (two from the target spectra, two from the model).


\subsection{Comparisons and Checks}
\label{sec:be_cc}


\begin{figure}
 \includegraphics[trim=0.5cm 0.5cm 0.5cm 0.5cm, width=0.5\textwidth]{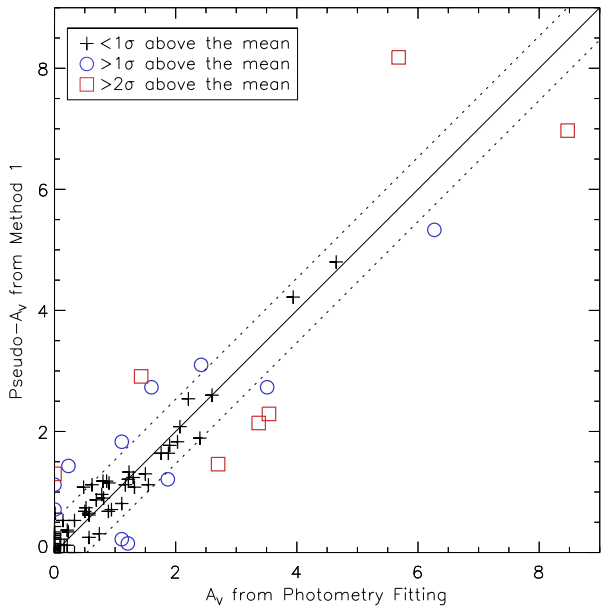}
 \caption{ This shows a comparison of a pseudo-$A_V$, extracted from the method 1 measurement of $\Delta D_B$ depending upon how much the spectra was adjusted, against the $A_V$ determined from the photometry fitting in Section \ref{sec:st_phot}. The solid black line is the line of correlation, while the dashed lines are 1$\sigma$ deviations, of 0.60~mag from this. Only 8\% of the targets are outside 2$\sigma$; these ones often have the largest $A_V$ values (these will be discussed in the text). Errors in the photometric $A_V$ are typically 0.05--0.15 (about the width of the points), and are given in Table \ref{tab:params}.}
 \label{fig:av_compare}
\end{figure}

\begin{figure}
 \includegraphics[trim=0.5cm 0.5cm 0.5cm 0.5cm, width=0.5\textwidth]{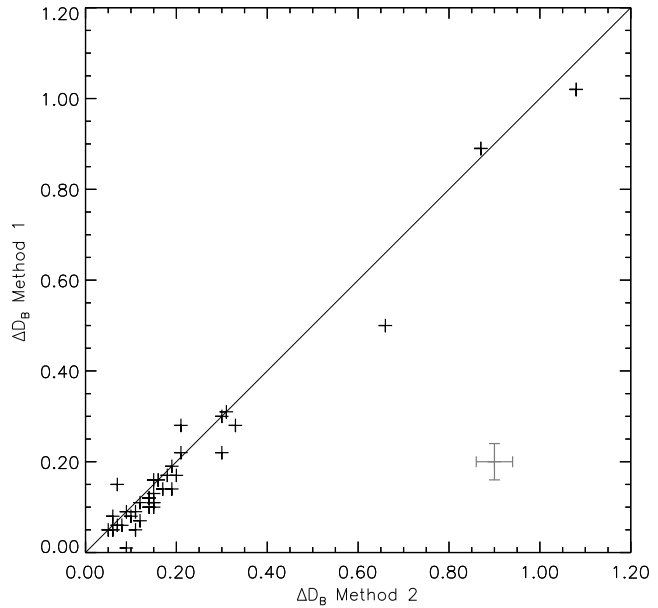}
 \caption{A comparison is made between the two different methods of measuring $\Delta D_B$ (detailed in Section \ref{sec:be_m1} \& \ref{sec:be_m2}). A line of expected 1:1 correlation is shown in black. For clarity individual error bars are not plotted due to the tightness of the points. Instead, a typical error bar of 0.04~mag is plotted in the bottom right corner. The actual errors range between 0.04--0.10~mag, and are provided in Table \ref{tab:db}. }
\label{fig:db_compare}
\end{figure}

\begin{figure}
\includegraphics[trim=1.0cm 0.5cm 0.25cm 0.25cm, width=\linewidth]{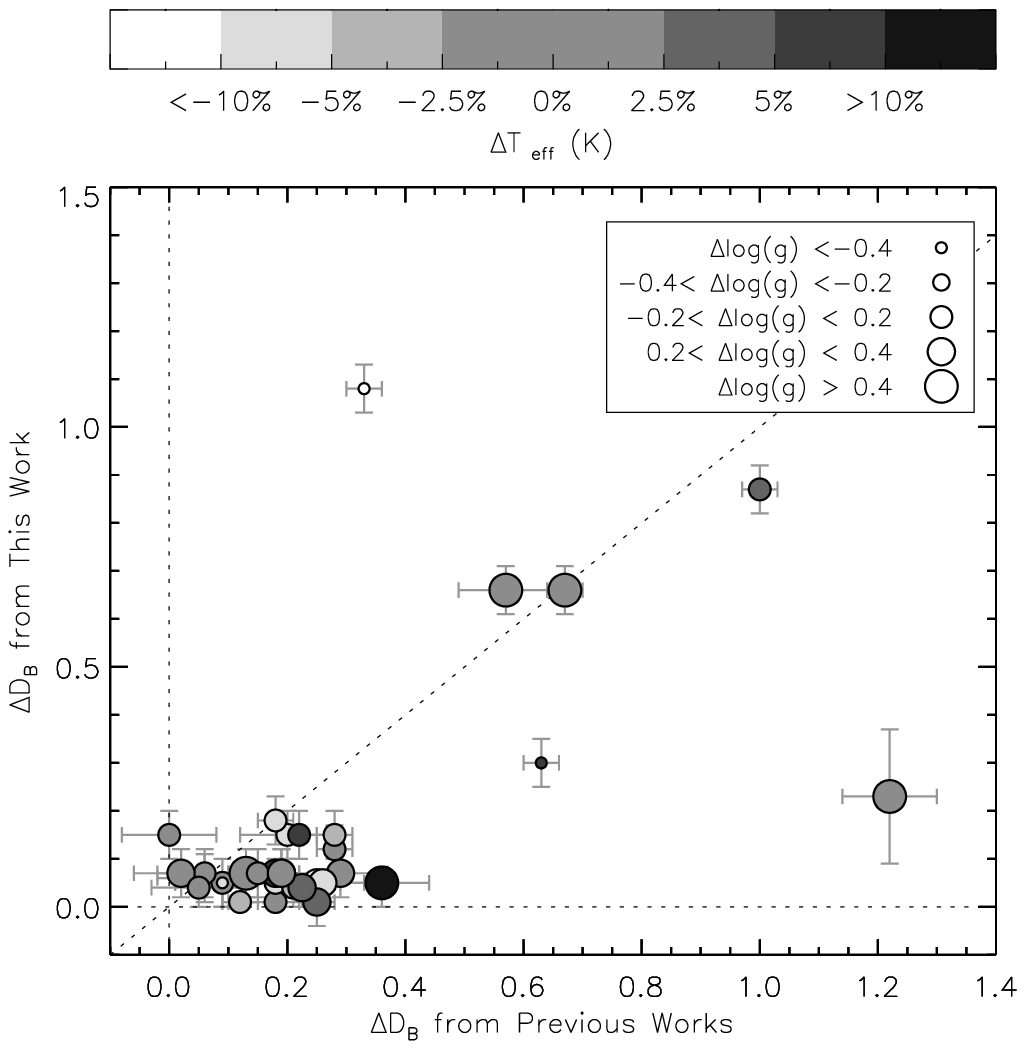}
\caption{ A comparison is drawn here between the final $\Delta D_B$ measured in this work versus the $\Delta D_B$ measured by other authors in the literature \citep{Donehew2011,Mendigutia2011b,Pogodin2012}. The difference in temperature between the two sources is calculated as a percentage of the total stellar temperature (cooler temperatures than the literature are white, while hotter ones are black). The size of each symbol reflects the difference in log(g) measured. Overall the largest deviations in $\Delta D_B$ are for the objects with the greatest differences in stellar parameters.}
\label{fig:db_lit_compare}
\end{figure}


The two methods used are similar but have some subtle differences. One, is that the central wavelength for the $B$-band normalisation is different between the two; it is centred at 4000~$\rm \AA$ for method 1, and is centred at 4400~$\rm \AA$ for method 2. The next difference is that method 1 performs a reddening correction using a section of the observed spectrum and relies on matching this to a stellar model. On the other-hand, method 2 avoids having to make a reddening correction by incorporating the adopted reddening law into the equation for $\Delta D_B$, and applies this over a much larger spectral region. Also, both approaches have been adapted from a definition which was based on broadband photometry. Therefore, some checks need to be made to see whether both approaches are comparable to each other.

The first check is between how the $A_V$ values determined in Section \ref{sec:st_title}, compare with the $A_V$ values extracted from method 1; as the fitting between 4000--4600~$\rm \AA$ can be used to infer an $A_V$ value. Figure \ref{fig:av_compare} displays this comparison. In the figure the standard deviation between the two is found to be 0.60~mag, and is represented by the dashed black lines. Within this 1$\sigma$ interval 79\% of the sample are included. This helps to highlight that the majority of the sample are tightly correlated while the outliers are more extreme and actually skew the standard deviation towards them. There are 7 stars showing differences greater than 2$\sigma$ from the mean. One of these is VY Mon which has the lowest SNR of the objects in the blue because it is very extinct. This makes the spectral shape adjustment more difficult and less accurate than other targets. The other outliers often have large $A_V$ and/or large $\Delta D_B$ values. This is not entirely unexpected as a significant excess can affect the SED shape of the spectra, which would complicate both photometry fitting and the spectral shape adjustments performed. In general, for HAeBes this is less likely as they are already very hot and the excesses need to be very strong to significantly affect the SED. One source of discrepancy lies in how the photometric method is coarse but covers the $BVRI$ points, while the spectral method covers a very narrow wavelength range of 4000~--~4600~$\rm \AA$, but with a greater accuracy in that region. The photometry used is also not simultaneous with the spectra; variability could therefore also play a role in the differences. Ultimately, this scatter is quite low with few outliers; this is more than acceptable considering the above factors and the standard reddening law adopted in both cases.

The next check is to see how $\Delta D_B$ varies between the two methods of measurement; Figure \ref{fig:db_compare} shows the comparison. There is a systematic offset of $\sim 0.02$ towards method 2 producing higher values, while the standard deviation of scatter between the two methods is $\sim 0.04$~mag. These differences are less than the systematic error on measurements. Since the original $\Delta D_B$ equation, Equation \ref{eqn:db_original}, can be seen to contain a dereddened term, the differences can be mostly attributed to how the reddening corrections are made in each case; though the normalisation of the spectra and points of measurement also influence the result. The method 1 covers a small wavelength range of 3600~--~4600~$\rm \AA$, of which only the 4000~--~4600~$\rm \AA$~region is used for the reddening correction. This means that this approach is not particularly sensitive to a given reddening law due to the small wavelength range it covers, and can be deemed reddening independent for low levels of extinction ($A_V<10$). On the other hand, method 2 depends more upon the adopted reddening law than method 1, because it covers a larger wavelength region of 3600~--~5500~$\rm \AA$.  Depending on the R$_V$ selected the resulting opacity ratios, seen in Equation \ref{eqn:mend2}, can change substantially, which in turn alters the measured $\Delta D_B$. Changing R$_V$ in method 1 does not noticeably affect $\Delta D_B$, for low $A_V$ values, as it is always the spectral profiles which are being matched. Through this matching the $A_V$ used will change to retain the SED shape and keeps $\Delta D_B$ the same. Returning to the figure, a few outliers can be seen between the two methods; the majority of these are objects with high extinction, or which were identified as having a discrepant $A_V$ between the photometric method and the spectral method in which they were determined. 

Overall, consistency is apparent between the methods employed here as the majority of measurements from each method lie within the errors of each other (see Table \ref{tab:db}).
 Based on the above analysis, we deem the methods equivalent. Therefore, in each case an average of the two will be taken for the final result; unless one method has a lower measurement error, then that method will be favoured over the other (this can occur depending on emission lines in both the measurement and normalisation regions). The $\Delta D_B$ value for each star, along with the errors and method(s) used to obtain it, are detailed in Table \ref{tab:db}. The errors given in $\Delta D_B$ appear large when compared with the value of $\Delta D_B$ itself. It should be noted that the detections are above 3$\sigma$ and the enhanced errors are mostly due to taking the logarithm of a ratio, see Equation \ref{eqn:db_original}, where an error of 1\% in the continuum detection can translate to more than a 30\% error in $\Delta D_B$ (depending upon how small the difference is between the intrinsic and observed spectra).

A comparison of the $\Delta D_B$ values determined in this work versus previous values published in the literature is shown in Figure \ref{fig:db_lit_compare}, as a consistency check. The majority of the measurements are clustered at values $<0.4$~mag, with literature values showing a slightly larger spread in $\Delta D_B$ than our sample. The main source of deviation between this work and the literature can be attributed to the $T_{\rm eff}$ and log(g) parameters used for each star; as these differ so will the intrinsic spectra form which $\Delta D_B$ is measured. The figure shows this clearly with a number of objects having deviant $\Delta D_B$ and stellar parameters, where the largest variations in $\Delta D_B$ are indeed the stars with the largest changes in $T_{\rm eff}$ and log(g), compared to the literature values. However, there is one star whose deviation in $\Delta D_B$ cannot be explained by the changes in $T_{\rm eff}$ and log(g) alone. Instead, the deviations may also be compounded by genuine variability of the star and/or accretion rate. Such variability can be seen within the literature, and in single stars themselves \citep{Pogodin2012, Mendigutia2013}. Additionally, intrinsic features in the spectra can contribute to differences too; the approach in this manuscript uses monochromatic points from spectra, whereas the majority of comparison stars primarily use broad-band photometry.
Overall, the majority of sources are in common, within the errors, and most discrepancies can be explained by the adoption of stellar parameters.

\section[]{Accretion Rates}
\label{sec:anal}

%
\begin{table*}
 \centering
 \begin{minipage}{130mm}
  \caption{Table of Accretion Rates. Column 1 gives the target name. Columns 2-4 give the measured Balmer Excess, filling factor and derived accretion rate. $\Delta D_B$ errors are rounded to the closest 0.01 and include all systematic errors too. Column 5 details by which method the values were obtained. Column 6 gives the accretion luminosity. Finally, Column 7 notes which stars can have their excess modelled successfully by MA.}
  \label{tab:db}
\begin{tabular}{l ccc ccc}
  \hline
     Name  & $\Delta D_B$ & $f$ & log($\dot{M}_{\rm acc}$) & Method(s) & log(${L}_{\rm acc}$) &  Achievable \\
  &  (mag) & (\%) & [M$_\odot$/yr] &  Used & [L$_\odot$] & by MA \\
  \hline

UX Ori &  $\leq$~ 0.04 &  $\leq$~   0.7 &  $\leq$~-7.26 & Method 1 \& 2  &  $\leq$~ 0.13 &  y  \\ [3pt]
PDS 174 &  $\leq$~ 0.02 &  $\leq$~   2.8 &  $\leq$~-6.76 & Method 2  &  $\leq$~ 0.92 &  y  \\ [3pt]
V1012 Ori &  0.19$^{+ 0.05}_{- 0.05}$ &    4.4$^{+   1.0}_{-   0.9}$ & -7.20$^{+ 0.21}_{- 0.28}$ & Method 1 \& 2  &  0.35$^{+ 0.26}_{- 0.32}$ &  y  \\ [3pt]
HD 34282 &  0.06$^{+ 0.05}_{- 0.05}$ &    1.7$^{+   0.9}_{-   0.8}$ & -7.69$^{+ 0.28}_{- 0.59}$ & Method 1  & -0.06$^{+ 0.32}_{- 0.61}$ &  y  \\ [3pt]
HD 287823 &  0.15$^{+ 0.05}_{- 0.05}$ &    3.0$^{+   0.7}_{-   0.7}$ & -7.13$^{+ 0.18}_{- 0.23}$ & Method 1 \& 2  &  0.37$^{+ 0.20}_{- 0.24}$ &  y  \\ [3pt]
HD 287841 &  $\leq$~ 0.05 &  $\leq$~   0.8 &  $\leq$~-7.82 & Method 1 \& 2  &  $\leq$~-0.32 &  y  \\ [3pt]
HD 290409 &  $\leq$~ 0.07 &  $\leq$~   2.1 &  $\leq$~-7.31 & Method 1 \& 2  &  $\leq$~ 0.25 &  y  \\ [3pt]
HD 35929 &  0.10$^{+ 0.05}_{- 0.05}$ &    1.0$^{+   0.4}_{-   0.3}$ & -6.37$^{+ 0.18}_{- 0.26}$ & Method 1 \& 2  &  0.87$^{+ 0.20}_{- 0.28}$ &  y  \\ [3pt]
HD 290500 &  0.21$^{+ 0.05}_{- 0.05}$ &    6.1$^{+   1.7}_{-   1.5}$ & -6.11$^{+ 0.17}_{- 0.17}$ & Method 1 \& 2  &  1.29$^{+ 0.35}_{- 0.21}$ &  y  \\ [3pt]
HD 244314 &  0.12$^{+ 0.05}_{- 0.05}$ &    2.4$^{+   0.7}_{-   0.7}$ & -7.12$^{+ 0.20}_{- 0.25}$ & Method 1 \& 2  &  0.35$^{+ 0.21}_{- 0.26}$ &  y  \\ [3pt]
HK Ori &  0.66$^{+ 0.05}_{- 0.05}$ &   27.7$^{+   6.0}_{-   3.8}$ & -6.17$^{+ 0.17}_{- 0.16}$ & Method 1 \& 2  &  1.33$^{+ 0.19}_{- 0.18}$ &  y  \\ [3pt]
HD 244604 &  0.05$^{+ 0.05}_{- 0.05}$ &    1.1$^{+   0.7}_{-   0.1}$ & -7.22$^{+ 0.26}_{- 0.32}$ & Method 1  &  0.22$^{+ 0.28}_{- 0.34}$ &  y  \\ [3pt]
UY Ori &  $\leq$~ 0.02 &  $\leq$~   0.6 &  $\leq$~-7.92 & Method 1 \& 2  &  $\leq$~-0.35 &  y  \\ [3pt]
HD 245185 &  $\leq$~ 0.07 &  $\leq$~   2.3 &  $\leq$~-7.29 & Method 1 \& 2  &  $\leq$~ 0.29 &  y  \\ [3pt]
T Ori &  $\leq$~ 0.05 &  $\leq$~   1.0 &  $\leq$~-6.54 & Method 1 \& 2  &  $\leq$~ 0.79 &  y  \\ [3pt]
V380 Ori &  0.87$^{+ 0.05}_{- 0.05}$ &   80.3$^{+  19.7}_{-  21.9}$ & -5.34$^{+ 0.10}_{- 0.15}$ & Method 1 \& 2  &  2.12$^{+ 0.31}_{- 0.16}$ &  y  \\ [3pt]
HD 37258 &  0.14$^{+ 0.05}_{- 0.05}$ &    4.5$^{+   1.5}_{-   1.3}$ & -6.98$^{+ 0.14}_{- 0.17}$ & Method 1 \& 2  &  0.58$^{+ 0.15}_{- 0.18}$ &  y  \\ [3pt]
HD 290770 &  0.15$^{+ 0.05}_{- 0.05}$ &    6.2$^{+   1.6}_{-   1.5}$ & -6.74$^{+ 0.12}_{- 0.14}$ & Method 1 \& 2  &  0.82$^{+ 0.16}_{- 0.17}$ &  y  \\ [3pt]
BF Ori &  0.15$^{+ 0.05}_{- 0.05}$ &    3.6$^{+   0.9}_{-   0.9}$ & -6.65$^{+ 0.17}_{- 0.25}$ & Method 2  &  0.77$^{+ 0.19}_{- 0.27}$ &  y  \\ [3pt]
HD 37357 &  0.30$^{+ 0.05}_{- 0.05}$ &   10.1$^{+   1.7}_{-   1.5}$ & -6.42$^{+ 0.09}_{- 0.06}$ & Method 1 \& 2  &  1.08$^{+ 0.09}_{- 0.06}$ &  y  \\ [3pt]
HD 290764 &  0.21$^{+ 0.05}_{- 0.05}$ &    3.5$^{+   0.8}_{-   0.7}$ & -6.56$^{+ 0.17}_{- 0.22}$ & Method 1 \& 2  &  0.80$^{+ 0.21}_{- 0.24}$ &  y  \\ [3pt]
HD 37411 &  0.15$^{+ 0.05}_{- 0.05}$ &    4.9$^{+   1.3}_{-   1.2}$ & -7.13$^{+ 0.24}_{- 0.34}$ & Method 1 \& 2  &  0.47$^{+ 0.29}_{- 0.38}$ &  y  \\ [3pt]
V599 Ori &  $\leq$~ 0.01 &  $\leq$~   0.1 &  $\leq$~-7.67 & Method 2  &  $\leq$~-0.33 &  y  \\ [3pt]
V350 Ori &  0.15$^{+ 0.05}_{- 0.05}$ &    3.7$^{+   1.0}_{-   0.9}$ & -6.95$^{+ 0.18}_{- 0.23}$ & Method 1 \& 2  &  0.55$^{+ 0.19}_{- 0.24}$ &  y  \\ [3pt]
HD 250550 &  0.30$^{+ 0.05}_{- 0.05}$ &   17.1$^{+   4.3}_{-   3.1}$ & -5.63$^{+ 0.14}_{- 0.11}$ & Method 1 \& 2  &  1.82$^{+ 0.32}_{- 0.17}$ &  y  \\ [3pt]
V791 Mon &  0.19$^{+ 0.05}_{- 0.05}$ &   27.5$^{+  10.2}_{-   8.0}$ & -6.16$^{+ 0.28}_{- 0.35}$ & Method 1 \& 2  &  1.55$^{+ 0.32}_{- 0.39}$ &  y  \\ [3pt]
PDS 124 &  0.11$^{+ 0.05}_{- 0.05}$ &    4.1$^{+   1.3}_{-   1.2}$ & -7.11$^{+ 0.13}_{- 0.19}$ & Method 2  &  0.50$^{+ 0.14}_{- 0.20}$ &  y  \\ [3pt]
LkHa 339 &  0.13$^{+ 0.05}_{- 0.05}$ &    5.3$^{+   1.5}_{-   2.1}$ & -6.81$^{+ 0.12}_{- 0.22}$ & Method 2  &  0.75$^{+ 0.13}_{- 0.22}$ &  y  \\ [3pt]
VY Mon &  0.23$^{+ 0.14}_{- 0.14}$ &   17.5$^{+  22.7}_{-  13.0}$ & -5.50$^{+ 0.42}_{- 0.64}$ & Method 1 \& 2  &  1.96$^{+ 0.60}_{- 0.66}$ &  y  \\ [3pt]
R Mon &   0.86$^{+  0.05}_{-  0.05}$ &  -  &  -  & Method 1 \& 2  &  -  &  n  \\ [3pt]
V590 Mon &  -  &  -  &  -  & -  &  -  &  -  \\ [3pt]
PDS 24 &  $\leq$~ 0.05 &  $\leq$~   1.9 &  $\leq$~-7.25 & Method 2  &  $\leq$~ 0.31 &  y  \\ [3pt]
PDS 130 &  0.16$^{+ 0.05}_{- 0.05}$ &    6.6$^{+   1.6}_{-   1.5}$ & -6.23$^{+ 0.12}_{- 0.13}$ & Method 1 \& 2  &  1.22$^{+ 0.17}_{- 0.15}$ &  y  \\ [3pt]
PDS 229N &  0.09$^{+ 0.05}_{- 0.05}$ &    6.4$^{+   2.4}_{-   3.6}$ & -6.67$^{+ 0.14}_{- 0.36}$ & Method 2  &  0.96$^{+ 0.14}_{- 0.36}$ &  y  \\ [3pt]
GU CMa &  0.14$^{+ 0.05}_{- 0.05}$ &   60.3$^{+  33.4}_{-  18.5}$ & -5.00$^{+ 0.23}_{- 0.13}$ & Method 1 \& 2  &  2.72$^{+ 0.41}_{- 0.36}$ &  y  \\ [3pt]
HT CMa &  0.11$^{+ 0.05}_{- 0.05}$ &    4.3$^{+   1.5}_{-   1.4}$ & -6.61$^{+ 0.16}_{- 0.19}$ & Method 1 \& 2  &  0.89$^{+ 0.20}_{- 0.19}$ &  y  \\ [3pt]
Z CMa &  1.08$^{+ 0.05}_{- 0.05}$ &   48.0$^{+  17.0}_{-   9.4}$ & -3.01$^{+ 0.20}_{- 0.19}$ & Method 1 \& 2  &  4.05$^{+ 0.22}_{- 0.22}$ &  y  \\ [3pt]
HU CMa &  0.14$^{+ 0.05}_{- 0.05}$ &   12.2$^{+   3.4}_{-   4.9}$ & -6.35$^{+ 0.11}_{- 0.22}$ & Method 1 \& 2  &  1.27$^{+ 0.11}_{- 0.22}$ &  y  \\ [3pt]
HD 53367 &   0.10$^{+  0.05}_{-  0.05}$ &  -  &  -  & Method 1 \& 2  &  -  &  n  \\ [3pt]
PDS 241 &  0.05$^{+ 0.05}_{- 0.05}$ &   21.6$^{+  20.3}_{-   1.3}$ & -5.56$^{+ 0.29}_{- 0.06}$ & Method 1 \& 2  &  2.25$^{+ 0.37}_{- 0.23}$ &  y  \\ [3pt]
NX Pup &  0.08$^{+ 0.05}_{- 0.05}$ &    0.9$^{+   0.4}_{-   0.3}$ & -6.96$^{+ 0.21}_{- 0.33}$ & Method 2  &  0.34$^{+ 0.23}_{- 0.34}$ &  y  \\ [3pt]
PDS 27 &  0.17$^{+ 0.13}_{- 0.16}$ &   40.0$^{+  55.0}_{-  39.0}$ & -3.96$^{+ 0.76}_{- 1.32}$ & Method 1 \& 2  &  3.49$^{+ 0.78}_{-0.82}$ &  y  \\ [3pt]
PDS 133 &   1.26$^{+  0.05}_{-  0.05}$ &  -  &  -  & Method 1 \& 2  &  -  &  n  \\ [3pt]
HD 59319 &  0.05$^{+ 0.05}_{- 0.05}$ &    3.4$^{+   2.3}_{-   0.4}$ & -5.76$^{+ 0.22}_{- 0.11}$ & Method 1 \& 2  &  1.65$^{+ 0.26}_{- 0.15}$ &  y  \\ [3pt]

  \hline
  \end{tabular}
\end{minipage}
\end{table*}

\begin{table*}
 \centering
 \begin{minipage}{130mm}
  \contcaption{}
  \begin{tabular}{l ccc ccc}
  \hline
     Name  & $\Delta D_B$ & $f$ & log($\dot{M}_{\rm acc}$) & Method(s) & log(${L}_{\rm acc}$) & Achievable \\
  &  (mag) & (\%) & [M$_\odot$/yr] &  Used & [L$_\odot$] & by MA \\
  \hline

PDS 134 &  $\leq$~ 0.03 &  $\leq$~   3.0 &  $\leq$~-5.60 & Method 2  &  $\leq$~ 1.82 &  y  \\ [3pt]
HD 68695 &  0.05$^{+ 0.05}_{- 0.05}$ &    1.3$^{+   0.8}_{-   0.1}$ & -7.78$^{+ 0.30}_{- 0.38}$ & Method 2  & -0.17$^{+ 0.34}_{- 0.41}$ &  y  \\ [3pt]
HD 72106 &  0.31$^{+ 0.05}_{- 0.05}$ &    7.7$^{+   1.4}_{-   1.3}$ & -6.21$^{+ 0.15}_{- 0.18}$ & Method 1 \& 2  &  1.20$^{+ 0.18}_{- 0.20}$ &  y  \\ [3pt]
TYC 8581-2002-1 &  0.15$^{+ 0.05}_{- 0.05}$ &    4.6$^{+   1.2}_{-   1.0}$ & -6.58$^{+ 0.10}_{- 0.13}$ & Method 1 \& 2  &  0.88$^{+ 0.11}_{- 0.13}$ &  y  \\ [3pt]
PDS 33 &  $\leq$~ 0.04 &  $\leq$~   1.2 &  $\leq$~-7.84 & Method 1 \& 2  &  $\leq$~-0.21 &  y  \\ [3pt]
HD 76534 &  $\leq$~ 0.01 &  $\leq$~   1.7 &  $\leq$~-6.95 & Method 1 \& 2  &  $\leq$~ 0.77 &  y  \\ [3pt]
PDS 281 &  -  &  -  &  -  & -  &  -  &  -  \\ [3pt]
PDS 286 &  0.07$^{+ 0.05}_{- 0.05}$ &   64.6$^{+ 35.4}_{-  39.6}$ & -5.41$^{+ 0.20}_{- 0.69}$ & Method 1 \& 2  &  2.55$^{+ 0.62}_{- 0.71}$ &  y  \\ [3pt]
PDS 297 &  $\leq$~ 0.01 &  $\leq$~   0.4 &  $\leq$~-7.60 & Method 2  &  $\leq$~-0.12 &  y  \\ [3pt]
HD 85567 &   0.55$^{+  0.05}_{-  0.05}$ &  -  &  -  & Method 1 \& 2  &  -  &  n  \\ [3pt]
HD 87403 &  0.05$^{+ 0.05}_{- 0.05}$ &    1.5$^{+   0.9}_{-   0.2}$ & -5.82$^{+ 0.20}_{- 0.07}$ & Method 2  &  1.48$^{+ 0.21}_{- 0.09}$ &  y  \\ [3pt]
PDS 37 &  0.16$^{+ 0.22}_{- 0.15}$ &   40.0$^{+  54.0}_{-  39.0}$ & -3.56$^{+ 0.60}_{- 1.62}$ & Method 1  &  3.85$^{+ 0.65}_{- 1.62}$ &  y  \\ [3pt]
HD 305298 &   0.06$^{+  0.05}_{-  0.05}$ &  -  &  -  & Method 2  &  -  &  n  \\ [3pt]
HD 94509 &  -  &  -  &  -  &  -  &  -  &  -  \\ [3pt]
HD 95881 &  $\leq$~ 0.05 &  $\leq$~   1.5 &  $\leq$~-5.65 & Method 1 \& 2  &  $\leq$~ 1.63 &  y  \\ [3pt]
HD 96042 &  0.12$^{+ 0.05}_{- 0.05}$ &   94.4$^{+  5.6}_{-  38.5}$ & -4.57$^{+ 0.03}_{- 0.28}$ & Method 1 \& 2  &  3.18$^{+ 0.32}_{- 0.30}$ &  y  \\ [3pt]
HD 97048 &  $\leq$~0.01 &    $\leq$~0.4 &-$\leq$~-8.16 & Method 2  & $\leq$~ -0.55 &  y  \\ [3pt]
HD 98922 &  $\leq$~ 0.01 &  $\leq$~   0.4 &  $\leq$~-6.97 & Method 2  &  $\leq$~ 0.41 &  y  \\ [3pt]
HD 100453 &  $\leq$~ 0.01 &  $\leq$~   0.1 &  $\leq$~-8.31 & Method 1 \& 2  &  $\leq$~-0.92 &  y  \\ [3pt]
HD 100546 &  0.18$^{+ 0.05}_{- 0.05}$ &    6.1$^{+   1.6}_{-   1.5}$ & -7.04$^{+ 0.13}_{- 0.15}$ & Method 1 \& 2  &  0.56$^{+ 0.14}_{- 0.15}$ &  y  \\ [3pt]
HD 101412 &  $\leq$~ 0.04 &  $\leq$~   1.2 &  $\leq$~-7.61 & Method 1 \& 2  &  $\leq$~-0.04 &  y  \\ [3pt]
PDS 344 &  $\leq$~ 0.03 &  $\leq$~   3.5 &  $\leq$~-7.02 & Method 1 \& 2  &  $\leq$~ 0.68 &  y  \\ [3pt]
HD 104237 &  0.17$^{+ 0.05}_{- 0.05}$ &    2.8$^{+   0.7}_{-   0.6}$ & -6.68$^{+ 0.15}_{- 0.20}$ & Method 1 \& 2  &  0.70$^{+ 0.18}_{- 0.22}$ &  y  \\ [3pt]
V1028 Cen &  0.10$^{+ 0.05}_{- 0.05}$ &   10.6$^{+   3.7}_{-   3.3}$ & -5.76$^{+ 0.16}_{- 0.22}$ & Method 1  &  1.76$^{+ 0.26}_{- 0.24}$ &  y  \\ [3pt]
PDS 361S &  0.12$^{+ 0.05}_{- 0.05}$ &   26.2$^{+   9.9}_{-   7.7}$ & -5.26$^{+ 0.17}_{- 0.20}$ & Method 1 \& 2  &  2.35$^{+ 0.27}_{- 0.23}$ &  y  \\ [3pt]
HD 114981 &  $\leq$~ 0.06 &  $\leq$~   8.1 &  $\leq$~-5.48 & Method 1 \& 2  &  $\leq$~ 2.03 &  y  \\ [3pt]
PDS 364 &  0.28$^{+ 0.05}_{- 0.05}$ &   26.8$^{+   8.4}_{-   6.4}$ & -6.05$^{+ 0.13}_{- 0.12}$ & Method 1 \& 2  &  1.58$^{+ 0.13}_{- 0.12}$ &  y  \\ [3pt]
PDS 69 &  0.31$^{+ 0.05}_{- 0.05}$ &   62.8$^{+  29.8}_{-  20.8}$ & -5.32$^{+ 0.21}_{- 0.21}$ & Method 1 \& 2  &  2.28$^{+ 0.30}_{- 0.27}$ &  y  \\ [3pt]
DG Cir &   0.79$^{+  0.05}_{-  0.05}$ &  -  &  -  & Method 1 \& 2  &  -  &  n  \\ [3pt]
HD 132947 &  0.06$^{+ 0.05}_{- 0.05}$ &    2.1$^{+   1.0}_{-   1.0}$ & -6.71$^{+ 0.17}_{- 0.42}$ & Method 1 \& 2  &  0.73$^{+ 0.18}_{- 0.42}$ &  y  \\ [3pt]
HD 135344B &  0.07$^{+ 0.05}_{- 0.05}$ &    0.7$^{+   0.3}_{-   0.3}$ & -7.37$^{+ 0.24}_{- 0.41}$ & Method 1 \& 2  & -0.04$^{+ 0.26}_{- 0.42}$ &  y  \\ [3pt]
HD 139614 &  0.09$^{+ 0.05}_{- 0.05}$ &    1.5$^{+   0.6}_{-   0.5}$ & -7.63$^{+ 0.20}_{- 0.30}$ & Method 1  & -0.10$^{+ 0.21}_{- 0.31}$ &  y  \\ [3pt]
PDS 144S &  $\leq$~ 0.01 &  $\leq$~   0.1 &  $\leq$~-8.35 & Method 1  &  $\leq$~-0.90 &  y  \\ [3pt]
HD 141569 &  0.05$^{+ 0.05}_{- 0.05}$ &    1.5$^{+   0.9}_{-   0.1}$ & -7.65$^{+ 0.33}_{- 0.47}$ & Method 1  & -0.05$^{+ 0.37}_{- 0.50}$ &  y  \\ [3pt]
HD 141926 &   0.20$^{+  0.05}_{-  0.05}$ &  -  &  -  & Method 1 \& 2  &  -  &  n  \\ [3pt]
HD 142666 &  $\leq$~ 0.01 &  $\leq$~   0.1 &  $\leq$~-8.38 & Method 1 \& 2  &  $\leq$~-0.93 &  y  \\ [3pt]
HD 142527 &  0.06$^{+ 0.05}_{- 0.05}$ &    0.6$^{+   0.3}_{-   0.3}$ & -7.45$^{+ 0.19}_{- 0.48}$ & Method 1  & -0.09$^{+ 0.19}_{- 0.48}$ &  y  \\ [3pt]
HD 144432 &  0.07$^{+ 0.05}_{- 0.05}$ &    1.0$^{+   0.5}_{-   0.4}$ & -7.38$^{+ 0.22}_{- 0.40}$ & Method 1 \& 2  &  0.02$^{+ 0.24}_{- 0.41}$ &  y  \\ [3pt]
HD 144668 &  0.20$^{+ 0.05}_{- 0.05}$ &    3.9$^{+   0.9}_{-   0.8}$ & -6.25$^{+ 0.16}_{- 0.19}$ & Method 1 \& 2  &  1.10$^{+ 0.19}_{- 0.22}$ &  y  \\ [3pt]
HD 145718 &  $\leq$~ 0.01 &  $\leq$~   0.2 &  $\leq$~-8.51 & Method 1  &  $\leq$~-1.01 &  y  \\ [3pt]
PDS 415N &  $\leq$~ 0.04 &  $\leq$~   0.5 &  $\leq$~-8.45 & Method 1 \& 2  &  $\leq$~-0.91 &  y  \\ [3pt]
HD 150193 &  0.07$^{+ 0.05}_{- 0.05}$ &    1.6$^{+   0.8}_{-   0.7}$ & -7.45$^{+ 0.25}_{- 0.43}$ & Method 2  &  0.10$^{+ 0.26}_{- 0.44}$ &  y  \\ [3pt]
AK Sco &  $\leq$~ 0.04 &  $\leq$~   0.4 &  $\leq$~-7.90 & Method 1  &  $\leq$~-0.52 &  y  \\ [3pt]
PDS 431 &  0.11$^{+ 0.05}_{- 0.05}$ &    4.3$^{+   1.5}_{-   1.4}$ & -6.06$^{+ 0.16}_{- 0.22}$ & Method 1 \& 2  &  1.34$^{+ 0.21}_{- 0.24}$ &  y  \\ [3pt]
KK Oph &  $\leq$~ 0.05 &  $\leq$~   1.0 &  $\leq$~-7.84 & Method 2  &  $\leq$~-0.29 &  y  \\ [3pt]
HD 163296 &  0.07$^{+ 0.05}_{- 0.05}$ &    1.8$^{+   0.8}_{-   0.8}$ & -7.49$^{+ 0.14}_{- 0.30}$ & Method 2  &  0.08$^{+ 0.14}_{- 0.30}$ &  y  \\ [3pt]
MWC 297 &  0.11$^{+ 0.08}_{- 0.08}$ &   56.3$^{+  43.7}_{-  26.5}$ & -5.16$^{+ 0.25}_{- 0.43}$ & Method 1 \& 2  &  2.62$^{+ 0.40}_{- 0.48}$ &  y  \\ [3pt]

  \hline
  \end{tabular}
\end{minipage}
\end{table*}



\begin{figure}
\includegraphics[trim=1.0cm 0.5cm 0.25cm 0.25cm, width=\linewidth]{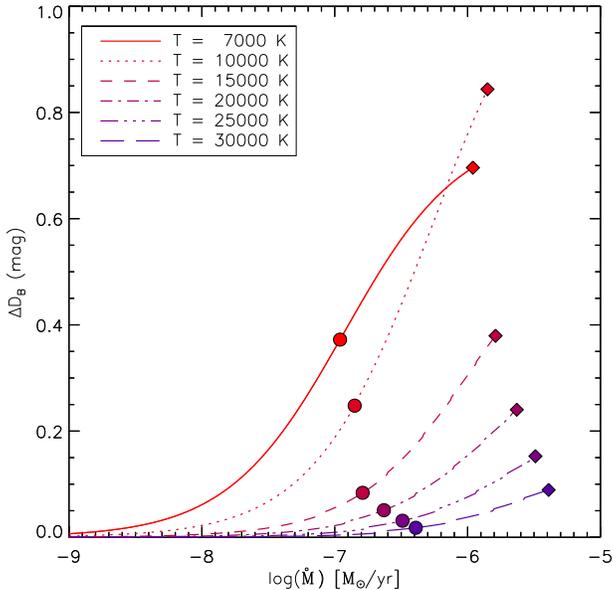}
\caption{ The relationship between $\Delta D_B$ and $\dot{M}_{\rm acc}$ is shown here, for a series of stars with different temperatures (labelled in the legend). The stellar parameters used for each case are ones typical for a ZAMS star of the temperature in question. It is apparent that the same $\Delta D_B$ value would result in a higher accretion rate in hotter stars. The filled circles display the point at which $f=$0.1, where the accretion column covers 10\% of the surface. Similarly, the filled diamonds are where $f=$1.0 (full coverage). In all cases $\mathscr{F}$=10$^{12}$~erg~cm$^{-2}$~$\rm \AA ^{-1}$. }
\label{fig:db_curves}
\end{figure}

\begin{figure*}
\includegraphics[trim=1.0cm 0.5cm 0.25cm 0.25cm, width=\linewidth]{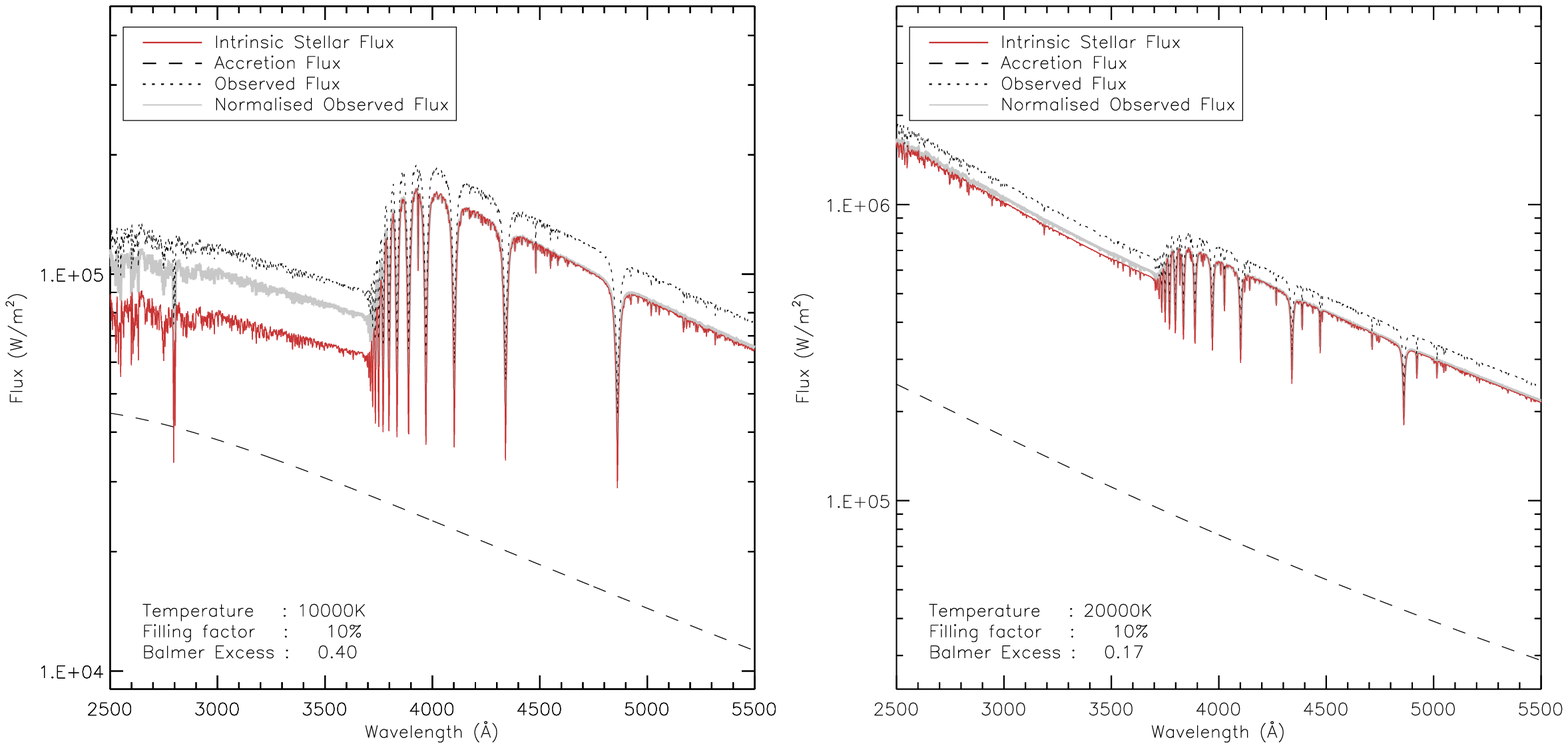}
\caption{This figure shows KC-model atmospheres (red) for a 10000~K star, on the left, and a 20000~K star, on the right. In both cases $\mathscr{F}$=10$^{12}$~erg~cm$^{-2}$~$\rm \AA ^{-1}$ and  log($\dot{M}_{\rm acc}$)=-6.5. These allow the accretion flux (black, dashed) to be calculated. Adding this flux to the intrinsic photosphere provides the observed spectra (black, dotted). This spectra is then normalised to 4000~$\rm \AA$~(grey), so that its continuum matches the intrinsic between 4000--4600~$\rm \AA$, allowing $\Delta D_B$ to be measured via the method 1 approach. The resulting $\Delta D_B$ values are given in the plot, demonstrating how they vary depending on the temperature of the star.}
\label{fig:db_with_f}
\end{figure*}


Accretion rates are an important parameter of pre-main sequence stars. They provide an insight into how the stars are evolving, along with the impact this will have on disc-star interactions, and may even have repercussions on planet formation. 

\subsection{Magnetospheric Modelling}
\label{sec:anal_calc}

In this work the measured $\Delta D_B$ is used to calculate $\dot{M}_{\rm acc}$ using accretion shock-modelling within the context of MA. This theory is adopted in order to test its applicability to a wide sample of HAeBes. The main assumption here is that the excess flux visible over the Balmer Jump region is produced by shocked emission from an in-falling accretion column. A detailed description of the magnetospherically driven accretion column and shock-modelling is given by \citet[hereafter CG98]{Calvet1998}, while a description of its application to HAeBe stars is given in \citet{Muzerolle2004} and \citet{Mendigutia2011b}. Here we summarise the key points of those papers and detail how they work in regard to this sample.

Firstly, the magnetic field lines of the star interact with the disc and truncate it at the truncation radius, $R_i$. It is generally accepted that the truncation radius is close to, or inside, the co-rotation radius, $R_{\rm cor}$, \citep[CG98]{Koenigl1991, Shu1994}. For this work $R_i$ is chosen to be 2.5~$R_\star $, as this has been shown to be an appropriate value which is often less than $R_{\rm cor}$ \citep{Muzerolle2004, Mendigutia2011b}. $R_{\rm cor}$ can be smaller than the adopted 2.5~$R_\star $, as is the case is for fast rotators, but this will not affect the derived accretion rate significantly i.e. for a very small truncation radius of $R_{i}$=1.5~R$_\odot$ the resulting accretion rate would be less than a factor of two different from one where $R_{i}=$2.5~R$_\odot$. 

At the truncation radius material is funnelled by the field lines and falls at speeds close to free-fall towards the stellar surface, where it shocks the photosphere upon impact. The velocity of the infalling material, $v_s$, is given as:
\begin{equation}
v_s= \left( \frac{2{\rm G}M_\star }{R_\star }\right)^{1/2} \left( 1 - \frac{R_\star }{R_i} \right)^{1/2}
\label{eqn:cg_vs}
\end{equation}
The velocity can be related to the accretion rate via the density. This is because $\dot{M}_{acc}$ is flowing at the same rate as the velocity through an accretion column, which also covers a given area of the star. Therefore, the density can be expressed as:
\begin{equation}
\rho=\frac{ \dot{M}_{\rm acc} }{ A v_s }
\label{eqn:cg_rho}
\end{equation}
where $A$ is the area of the star covered by the accretion column, defined as $A=f4\pi R_\star ^2$, and $f$ is a filling factor such that $f=0.1$ would be 10\% surface coverage. The filling factor is required as we consider the accretion to be funnelled though a column, rather than being evenly distributed over the entire stellar surface. Putting this in terms of energy, the total inward flux of energy of the accretion column is: 
\begin{equation}
\mathscr{F}=(1/2)\rho v_s^3
\label{eqn:bigf}
\end{equation}
This amount of energy is carried into the column and must be re-emitted back out of the star (see \citetalias{Calvet1998} for details on this energy balance). This means the total luminosity from the accretion column, as given in \citetalias{Calvet1998}, can be written as:
\begin{equation}
L_{\rm col}=(\mathscr{F}+F_{*})A = \zeta \left( \frac{G\dot{M}_{\rm acc}M_\star }{R_\star } \right) +F_\star A = \zeta L_{\rm acc} + F_\star A
\label{eqn:cg_lum}
\end{equation}
where $F_\star $ is the intrinsic flux of the stellar photosphere, $L_{\rm acc}$ is the accretion luminosity, and $\zeta=1-(R_\star /R_i)$. The accretion luminosity is defined as $L_{\rm acc}=G\dot{M}_{\rm acc}M_\star /R_\star $.

As shown in \citet{Mendigutia2011b}, the column luminosity is $L_{\rm col}=F_{\rm col}A$, where $F_{\rm col}$ is the flux produced by the accretion column. This total amount of flux can be expressed as a blackbody function, where $F_{\rm col}=\sigma T^4_{\rm col}$. Similarly the same can be done for the photosphere, $F_{*}=\sigma T^4_{*}$. This results in $\sigma T_{\rm col}^4=\mathscr{F}+\sigma T_\star ^4$. 

At this point the unknowns are $f$, $\mathscr{F}$, $T_{\rm col}$, and $\dot{M}_{\rm acc}$. $T_{\rm col}$ has just been shown to be governed by the amount of energy flowing onto the photosphere, $\mathscr{F}$, and by the temperature of the photosphere itself, $T_\star $. For each star $T_{\rm col}$ is determined using the temperatures we derived, and by fixing  $\mathscr{F}=10^{12}$~erg~cm$^{-2}$~$\rm \AA ^{-1}$, as this has been shown to provide appropriate filling factors of $\leq 0.15$ in the majority of cases in HAeBes studied so far \citep{Muzerolle2004,Mendigutia2011b}. This leaves only $f$ and $\dot{M}_{\rm acc}$ remaining. $\dot{M}_{\rm acc}$ can be determined from $\Delta D_B$ by making use of the equations above; for which there is a unique $\Delta D_B$ vs. $\dot{M}_{\rm acc}$ combination for each star due to its stellar parameters. To obtain this curve, $\dot{M}_{\rm acc}$ values are tested between $10^{-3}$--$10^{-10}{\rm M}_\odot{\rm yr}^{-1}$. With $\mathscr{F}$ fixed, and all the other stellar parameters known, the filling factor corresponding to each $\dot{M}_{\rm acc}$ value is found through the following equation (which is a rearrangement of the $2^{\rm nd}$~and~$3^{\rm rd}$ terms in Equation \ref{eqn:cg_lum}):
\begin{equation}
f = \zeta \left( \frac{ G \dot{M}_{\rm acc}M_\star }{ R_\star  } \right) \frac{1}{4 \pi R_\star ^2 \mathscr{F} }
\label{eqn:f}
\end{equation}
The $T_{\rm col}$ determined previously is used to make a black-body, which represents the accretion hotspot, and multiplying this by $f$ gives the excess flux. The excess flux is then combined with a KC-model, determined using the relevant stellar parameters. From this, $\Delta D_B$ can be measured. This is repeated for all $\dot{M}_{\rm acc}$ and $f$ combinations. The result provides a unique $\Delta D_B$ vs. $\dot{M}_{\rm acc}$ curve, which the accretion rate can be read from.

Figure \ref{fig:db_curves} gives the $\Delta D_B$ vs. $\dot{M}_{\rm acc}$ curves for a series of different temperature stars (for simplicity in the figure their other parameters are taken from the ZAMS). The figure demonstrates how the same $\Delta D_B$, measured in two different temperature stars, can refer to wildly differing accretion rates. Also, the T=10000~K curve has the highest $\Delta D_B$ value as the size of the Balmer Jump peaks at around this temperature. 

Figure \ref{fig:db_with_f} demonstrates the same concept of $\Delta D_B$ vs. $\dot{M}_{\rm acc}$ changing as a function of temperature, as shown by the curves in Figure \ref{fig:db_curves}. Although, this figure also highlights how the excess flux impacts the appearance of the spectra too. There are two cases in the figure, one for a star of 10000~K, and the other for a star of 20000~K. It can be seen for $\dot{M}_{\rm acc}=10^{-6.5}$ that the resulting $\Delta D_B$ changes from 0.41 for the 10000 K star, to only 0.17 for the 20000~K star. This is why the calculation of separate $\Delta D_B$ vs. $\dot{M}_{\rm acc}$ curves, for each star, are crucial. It also demonstrates that the SED shape is not significantly affected by the excess longwards of 4000~$\rm \AA$, which means that the approach of methods 1 and 2 remain valid (as does the photometry fitting). Table \ref{tab:db} contains the $\dot{M}_{\rm acc}$ values for each star using an individual curve for each star.


\begin{figure}
\includegraphics[trim=1.0cm 0.5cm 0.25cm 0.25cm, width=\linewidth]{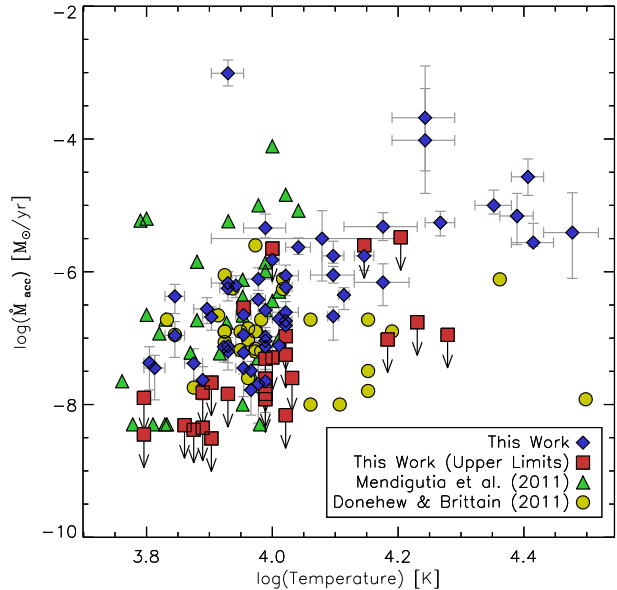}
\caption{ $\dot{M}_{\rm acc}$ versus $T_{\rm eff}$ is shown for each object, along with literature values for a comparison. In general it can be seen that $\dot{M}_{\rm acc}$ increases with temperature, with a scatter of 2-3 orders of magnitude in $\dot{M}_{\rm acc}$ throughout. }
\label{fig:macc_vs_teff_lit}
\end{figure}

\begin{figure}
\includegraphics[trim=1.0cm 0.5cm 0.25cm 0.25cm, width=\linewidth]{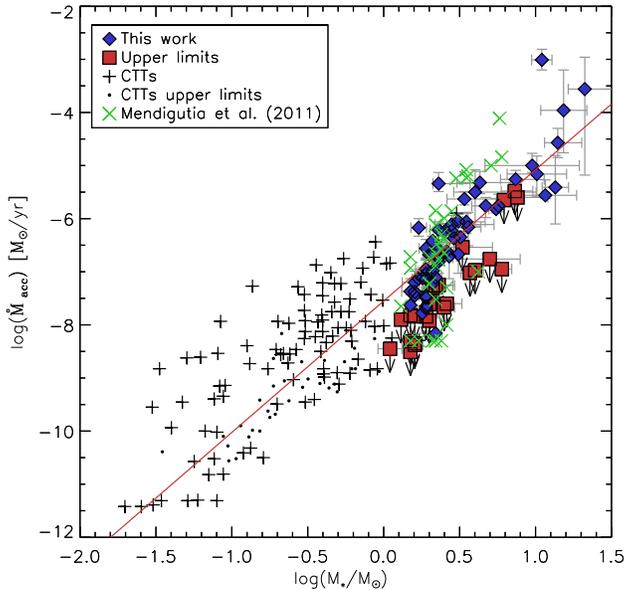}
\caption{$\dot{M}_{\rm acc}$ versus $M_\star$ is shown for each star, where possible, along with additional HAeBe sources from \citet{Mendigutia2011b}, and CTTs from \citet{Natta2006}. A red solid line fit to all of the points, excluding upper limits, is shown of $\dot{M}_{\rm acc} \propto M_\star^{2.47 \pm 0.07}$. }
\label{fig:macc_vs_mass_ctts}
\end{figure}

\begin{figure}
\includegraphics[trim=1.0cm 0.5cm 0.25cm 0.25cm, width=\linewidth]{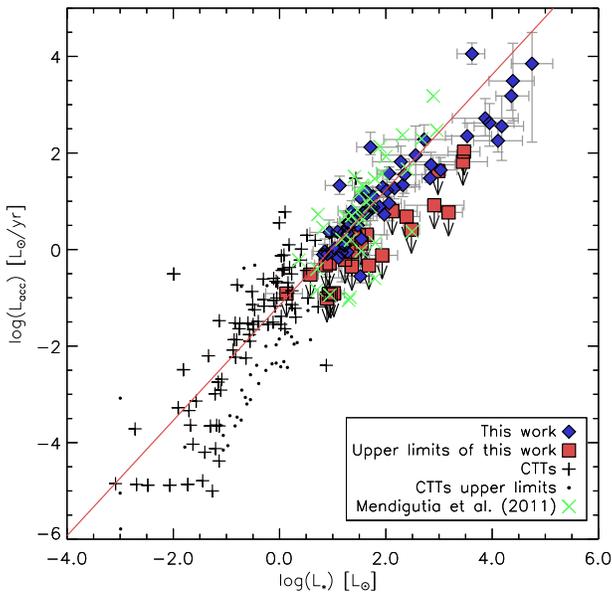}
\caption{$L_{\rm acc}$ versus $L_\star$  is shown for each star in this work, along with additional HAeBe sources from \citet{Mendigutia2011b}, and CTTs from \citet{Natta2006}. A best fit is obtained of  $L_{\rm acc} \propto L_\star^{1.19\pm0.03}$, which is plotted as a solid red line and excludes the upper limits.}
\label{fig:lacc_vs_logl_ctts}
\end{figure}


\subsection{Literature Comparisons}
\label{sec:anal_calc}

Comparisons of $\dot{M}_{\rm acc}$ derived in this work are made against previous detections in HAeBes and CTTs. In the first comparison, Figure \ref{fig:macc_vs_teff_lit} places this sample against other stars from the literature in which $\dot{M}_{\rm acc}$ has also been determined directly using  $\Delta D_B$. For the HAes, $\sim$10000~K and lower, the range in this work appears similar to previous works, with $\dot{M}_{\rm acc}$ spanning from anywhere between $10^{-8}$--$10^{-5}$~${\rm M}_\odot{\rm yr}^{-1}$, with the exception of one star at $\sim 10^{-3}$~${\rm M}_\odot{\rm yr}^{-1}$ (Z CMa, which is likely a very young HBe star based on it's mass of 11~M$_\odot$ ). The HBes closest to the HAes show a similar range in magnitude of $10^{-7}$--$10^{-4}$~${\rm M}_\odot{\rm yr}^{-1}$. The scatter then decreases once the temperature has increased beyond 20000~K, where $\dot{M}_{\rm acc}$ spans $10^{-6}$--$10^{-4}$~${\rm M}_\odot{\rm yr}^{-1}$. This decrease can be partially attributed to a detection effect, as the temperature of the star increases the observable $\Delta D_B$ will decrease. Therefore, if the temperature of the star is very high then low accretion rates will be undetectable via the Balmer Excess method. This is supported by the $\Delta D_B$ vs. log($\dot{M}_{\rm acc}$) curves in Figure \ref{fig:db_curves}. Returning to Figure \ref{fig:macc_vs_teff_lit} comparisons are also drawn against previously published accretion rates. The \citet{Mendigutia2011b} sample has a slightly larger scatter showing some $\dot{M}_{\rm acc}$ detections below our findings, this can again be attributed to detection limits in this work. But it can also be seen that there are many stars in \citet{Mendigutia2011b} which have  accretion rates about an order of magnitude higher than our findings. The exact reason for the discrepancies is unknown, but it is likely to be a combination of the two different types of dataset, spectra and photometry, and the different methods of measurement used i.e. the photometric method requires dereddening to be performed prior to measurement of $\Delta D_B$. Variability may also play a role.

 Comparing our results with the work of \citet{Donehew2011} we find a systematically higher accretion rate for objects hotter than 10000~K, the HBes, of around 1--2 orders of magnitude. This can be attributed to their adoption of a single $\Delta D_B$ vs. $\dot{M}_{\rm acc}$ relationship for all of their objects. Whereas in this work, the relationship between the two has been calculated on an individual basis for each star, based on its stellar parameters (see Figure \ref{fig:db_curves}). Therefore, they are not directly comparable.

A comparison is made of $\dot{M}_{\rm acc}$ vs. $M_\star$ in Figure \ref{fig:macc_vs_mass_ctts}, which includes literature values too. More specifically, this comparison looks at how the results of this sample compare to the  HAeBes from \citet{Mendigutia2011b}, along with a look at lower luminosity CTTs from \citet{Natta2006}. A trend is seen of increasing accretion rate with increasing stellar mass; the fit shown in the figure gives $\dot{M}_{\rm acc} \propto M_\star^{2.47 \pm 0.07}$. The position of the HAeBes obtained in this work show agreement with the values obtained from \citet{Mendigutia2011b}. However, it can be seen at around the HAe mass range, of $\sim 1$--2.5~M$\odot$, that there is a dip in the trend. Whether this is due to the physical mass of the stars, or is an observational effect from different sample is unclear. This dip will be discussed further in Section \ref{sec:dis_accretion} in regard to the HAeBes of this sample. An investigation into the meaning of this dip, and how the relationships behave in CTTs and HAeBes, is presented in a dedicated paper by this group to the topic \citep{Mendigutia2015}. Overall, the figure shows a trend that covers a large mass range spanning low mass CTTs to high mass HBes, with only some slight deviation in the HAe mass range.

Figure \ref{fig:lacc_vs_logl_ctts} instead shows a relationship of the luminosities instead of the mass; specifically of how $L_{\rm acc}$ changes as a function of $L_\star$. Again, comparisons are made against HAeBes and CTTs from the literature. A positive correlation between the two is also seen here of $L_{\rm acc} \propto L_\star^{1.19 \pm 0.03}$. This trend in the data shows a scatter of around 2 dex in $L_{\rm acc}$ throughout the luminosity range covered; this scatter is comparable to the scatter in $\dot{M}_{\rm acc}$ shown in Figure \ref{fig:macc_vs_teff_lit}.

In total, accretion rates, and therefore accretion luminosities, have been calculated for 81 stars in the sample. Their values are seen to agree with previous literature estimates of accretion in HAeBes. The accretion rates obtained are observed to increase with both temperature and luminosity; this trend is seen in the literature for CTTs and HAeBes alike.

\section[]{Discussion}
\label{sec:dis}


\begin{figure*}
\includegraphics[trim=1.0cm 0.5cm 0.25cm 0.25cm, width=\linewidth]{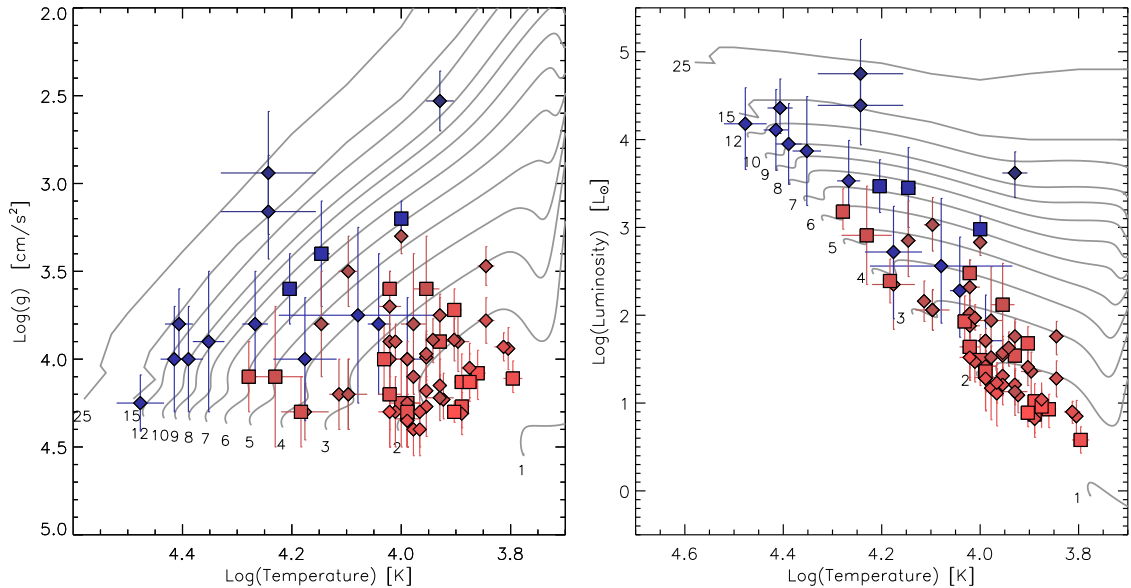}
\caption{ The left panel shows all the stars for which log(g) and $T_{\rm eff}$ could be determined from the spectra. These are translated into an HR-diagram in the right hand panel. In both panels the colour of the points reflect the strength of the accretion rate determined in each star; dark-blue symbols are the strongest accretors while light-red ones are the weakest accretors. The squares denote objects where $\dot{M}_{\rm acc}$ is an upper limit. The PMS evolutionary mass  tracks of \citet{Bressan2012} and \citet{Bernasconi1996} are also plotted as solid grey lines, and are labelled according to mass. Stars which were moved onto the ZAMS are not included in this plot.}
\label{fig:hr_diagram}
\end{figure*}


\subsection{Overall Results}

$\Delta D_B$ is clearly detected in 62 of the stars, while a further 26 stars have upper limits placed on them. The remaining 3 stars are measured as having a negative or zero $\Delta D_B$. The possible reasons for this for each star are now discussed: V590 Mon is observed to have $\Delta D_B=0$ within the errors, which is acceptable as it may not be accreting; PDS 281 has been listed previously as a possible evolved star \citep{Vieira2003}, as such the parameters derived in this work may be incorrect based on our assumptions, and if it is evolved it is unlikely to be accreting; HD 94509 has very narrow and deep absorption lines in its spectrum which suggest it is a super-giant star with a low log(g), as supported by past observations \citep{Stephenson1971}, while such low values are not covered by our adopted model atmospheres. Futher investigation into the accretion properties of these 3 stars through emission lines will be presented in a future paper by the authors, though their questionable nature as PMS objects should be noted.

There are 7 objects for which the measured $\Delta D_B$ value cannot be reproduced though magnetospheric accretion shock-modelling, using the method we adopt. This is because the appropriate $\Delta D_B$ vs. $\dot{M}_{\rm acc}$ curve calculated for each of the stars, based on its stellar parameters, cannot reach the observed $\Delta D_B$ before a 100\% filling factor is achieved (see Figure \ref{fig:db_curves} for the points at which a 100\% filling factor is seen for different temperatures). Within this subset, 3 stars have a very large $\Delta D_B$ of $> 0.85$ (PDS 133, R Mon, and DG Cir), 3 have temperatures exceeding 20000~K (HD 141926, HD 53367, and HD 305298), while the final star lies in between these two scenarios having a strong $\Delta D_B$ value and is mid-B spectral type (HD 85567). These stars are all HBes. 

Additionally, there are 12 stars whose measured $\Delta D_B$ are modelled by filling factors of greater than 25\% of the stellar surface. This is allowed, but it is an unusual occurrence under MA \citep{Valenti1993,Long2011}. A filling factor greater than 1 is an unphysical value, as it implies that the accretion column covers more than the total surface area of the star. This suggests that the MA scenario adopted here needs to be revised, or discarded, for the stars with unphysical filling factors. Caution should be exercised when considering the $\dot{M}_{\rm acc}$ values of stars with high filling factors.
This amounts to 9\% of $\Delta D_B$ detections being non-reproducible though the adopted MA shock-modelling, with a further 15\% having unusually high filling factors. All of this gives a possible indication that MA may not be applicable in all HAeBes; particularly for stars with a large $\Delta D_B$, or which have high temperatures i.e. the HBes. The remaining 76\% can be fitted successfully within the context of MA. 

\subsection{HR-diagram}

Using the spectra, log(g) could be determined for the majority of the sample in addition to $T_{\rm eff}$; for these stars their stellar parameters were determined using PMS tracks.
Their placement on these tracks confirms the young nature of these stars and is shown in Figure \ref{fig:hr_diagram}, in the left-hand panel, while the right-hand panel shows the corresponding HR-diagram. The stars which required revised distances to be calculated for them (see Section \ref{sec:st_mrg}) are not included in Figure \ref{fig:hr_diagram} as their placement is artificial compared to the other stars. 
A large proportion of the sample are clustered between 2--3~$M_\odot$, which is likely caused by a combination of two effects. The first being that lower mass sources are more numerous, as described by the initial mass function, IMF, \citep{Salpeter1955}. The range of masses determined in this work agrees fairly well with a typical Salpeter-IMF distribution, particularly when considering the selection criteria (the criteria skews our sample towards high mass objects, as these are the ones of greater interest in this work). Table \ref{tab:imf} shows the comparison of the mass distribution in this work versus the distribution given in \citet{Zinnecker2007} for a typical IMF function.

\begin{table}
\centering
\begin{minipage}{60mm}
\caption{Compares the number of HAeBes found in different mass bins with the theoretical IMF distribution}
\label{tab:imf}

\begin{tabular}{lcc}
\hline
Mass Bin & Theoretical & This Work \\
\hline
1--2~$M_\odot$ & 99 & 31$^\dagger$ \\
2--4~$M_\odot$ & 39 & 36 \\
4--8~$M_\odot$ & 15 & 11 \\
8--16~$M_\odot$ & 6 & 11 \\
$>16$~$M_\odot$ & 4 & 2 \\
\hline
\end{tabular}
  
\begin{minipage}{60mm}
  {\footnotesize $^\dagger$ This sample is focused on HAeBes and does not represent the 1--2~$M_\odot$ bin well, as HAeBes are generally more massive.}
\end{minipage} 
  
\end{minipage}
\end{table}

The second aspect, which may be contributing to the clustering, is a visibility effect due to low mass stars being more evolved and less extinct than younger high mass stars (as predicted by a comparison between the Kelvin-Helmholtz time-scale and the free-fall time-scale). This second point is supported by the $A_V$ values measured in this sample where, in general, the lower mass objects tend to have lower $A_V$ values. However, it should be noted that a high $A_V$ does not necessarily mean that the HAeBe has a high mass, as many low mass stars of young ages also have high extinction values (e.g. V599 Ori has an $A_V=4.65$, but only has $T_{\rm eff}=8000$~K and $M_\star =2.5$~M$_\odot$). 
Another point to note is that clustering of the stars in the figure could be attributed to the stars actually belonging to the same cluster. The main star forming regions in which some of the HAeBes in this work appear to be associated with are the Orion-OB1, Mon-OB1, CMa-R1 and Sco-OB2 regions \citep{deZeeuw1999,Shevchenko1999,Dahm2005,vanLeeuwen2007}. Since the regions are located at fixed distances, clustering of luminosities will occur if the stars are of similar spectral type. It is worth noting that the number of stars in each mass bin of a cluster is governed by the initial mass of the cloud in which they form. By looking at just a few star forming regions we naturally get clusters of similar mass stars in each one; resulting in clustered regions in an HR-diagram. However, only a small number of distances are adopted from the literature as an input parameter in this work, and they are drawn from various catalogues and regions on the sky. The spread on the HR-diagram can simply be attributed to relatively low number statistics.

\subsection{Age}
\label{sec:dis_age}


\begin{figure}
\includegraphics[trim=1.0cm 0.5cm 0.25cm 0.25cm, width=\linewidth]{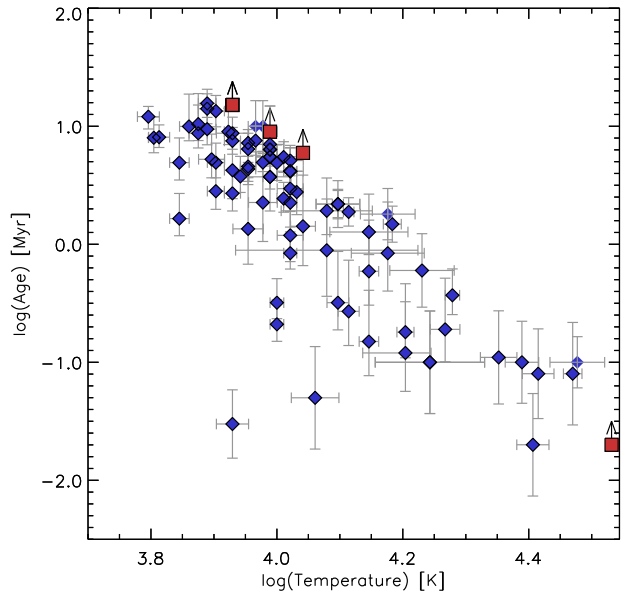}
\caption{The PMS tracks in Figure \ref{fig:hr_diagram} are used to obtain an age for each star. The ages determined are plotted here against the temperature of the star. The plot shows how the older stars are always the cooler stars i.e. the ones with a lower mass, which evolve towards the main sequence slower than their high mass counter parts, as expected. However, all of the hottest objects, $T_\star $~$>$~20000~K, are seen to be the youngest ones, age$ < 0.5$~Myr. Some cool and young stars are also present, which are likely to be in the early stages of their PMS evolution.}
\label{fig:age_vs_teff}
\end{figure}


\begin{figure}
\includegraphics[trim=1.0cm 0.5cm 0.25cm 0.25cm, width=\linewidth]{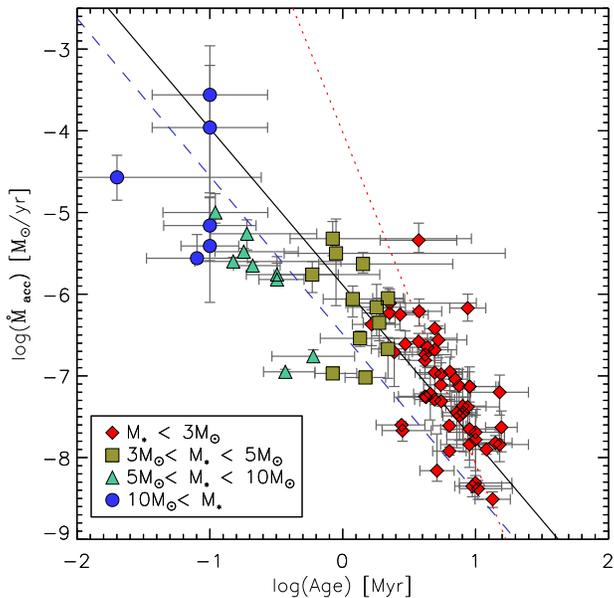}
\caption{ Plotted here are the derived accretion rates, from Table \ref{tab:db}, against the age of the star, in a log-log plot. Various mass bins are noted by symbol and colour. Various fits are made to the different mass bins. All of the HAes are shown in red ($<3$~M$_\odot$), and can be fitted by a relationshiip of $ \dot{M}_{\rm acc} \propto t^{-4.06 \pm 0.53}$, shown as a dotted red line; the remaining HBes ($>3$~M$_\odot$) are all fitted with the dashed blue line where $ \dot{M}_{\rm acc} \propto t^{-1.93 \pm 0.24}$. Finally, a fit to all of the HAeBes is shown in black of $ \dot{M}_{\rm acc} \propto t^{-1.92 \pm 0.09}$. Z CMa has been excluded from this fit (see the text for discussion).}
\label{fig:age_vs_macc}
\end{figure}


Next, the ages of the stars are investigated. Generally for PMS stars, the higher the mass, the younger the object is. Figure \ref{fig:age_vs_teff} shows age against temperature ($T_{\rm eff}$ is roughly proportional to $M_\star$ for MS stars, and stars close to the ZAMS). The plot shows an inverse relationship between age and temperature where increasing temperature results in younger ages. This is as expected as the hot objects, which evolve faster, will move away from the PMS stage of their lives quicker than the lower mass stars, allowing the higher mass stars only to be seen at an early age. This point is worth stressing when it comes to looking at HAeBes statistically, as the HBes will always be much younger than the majority of HAes, but they can also be much closer to the main sequence, as this is relative to their mass.

Figure \ref{fig:age_vs_macc} shows how $\dot{M}_{\rm acc}$ changes with the age of a star. As the age increases the accretion rate diminishes, much like what has been seen for the temperature. A fit to the data provides a relationship of $\dot{M}_{\rm acc} \propto t^{-\eta}$, were $t$ is the age in Myrs, and $\eta=1.92 \pm 0.09$. The figure also shows the stars split into separate mass bins too. A fit to just the HBes, where $M_\star > 3$~M$_\odot$, obtains $\eta=1.93 \pm 0.24$, which is very similar to the result for all of the HAeBes, only slightly offset. However, for the HAes alone in this work a much steeper relationship is obtained of $\eta=4.06 \pm 0.53$. The HAeBes as a whole, and the HBe case, agree with the HAeBes investigated in the work of \citet{Mendigutia2012}, where the authors obtain $\eta=1.8$. For CTTs a relationship has been observed where $\eta = 1.5\mbox{--}2.8$ \citep{Hartmann1998}. This range also encompasses the case for the HAeBes as a whole, and the HBes. However, more recent studies suggest that the relationship for CTTs is actually lower than this, where $\eta = 1.2$ \citep{SiciliaAguilar2010,CarattioGaratti2012}, which suggests that there is a difference in the $\dot{M}_{\rm acc} \propto t^{-\eta}$  relationship between the CTTs and the HAeBes. Some caution should be noted for the ages of the HBes as their ages are less accurate than the HAes, since the HBes are younger. In particular, there are some stars which are suspected to have ages $<1$~Myr, the uncertainty in the ages of these stars are taken into account in the fitting.

Overall, these relationships indicate that $\dot{M}_{\rm acc}$ could be an evolutionary property of HAeBes, which decreases as the star evolves; possibly accreting all of its material or dispersing its disc with time. Modelling of disc dispersion through photoevaporation suggests that the disc lifetimes are indeed shorter for more massive stars \citep{Gorti2009}. This offers an explanation for the steep exponent observed in the HAes, in which we could be observing the transition stage of disc dispersion as they approach the main sequence, resulting in a decreased accretion rate. The HBes, on the other hand, are younger and may not be dispersing their disc yet, which allows them to retain a more shallow relationship between $\dot{M}_{\rm acc}$ and age.

\subsection{Accretion Rate vs. Stellar Parameters}
\label{sec:dis_accretion}


\begin{figure}
\includegraphics[trim=1.0cm 0.5cm 0.25cm 0.25cm, width=\linewidth]{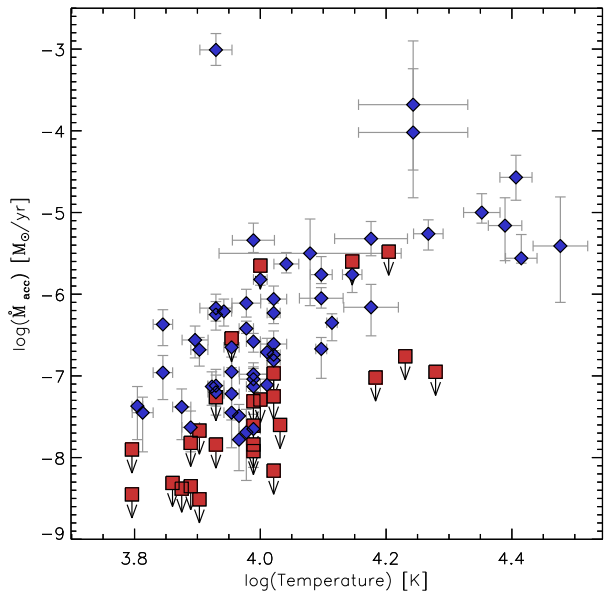}
\caption{ $\dot{M}_{\rm acc}$ is shown versus $T_{\rm eff}$ in a log-log plot here, where $\dot{M}_{\rm acc}$ appears to be increasing steadily with increasing temperature. The red squares denote upper limits. The outlier at ${\rm log(}\dot{M}_{\rm acc}) \sim -3.0$ and log($T_{\rm eff}) \sim 3.95$ is Z CMa. }
\label{fig:macc_vs_teff_no_lit}
\end{figure}


Moving on from the age, the next natural questions are: how is $\dot{M}_{\rm acc}$ related to the stellar parameters of the star; and are they influenced by it or vice-versa? 
In Figure \ref{fig:macc_vs_teff_no_lit} a comparison is made between $\dot{M}_{\rm acc}$ and $T_{\rm eff}$. The scatter in  $\dot{M}_{\rm acc}$ remains constant at about 2 orders of a magnitude throughout. There is one object, Z CMa, which can be seen as an outlier from the general scatter. This star is cool, 8500~K,  very massive, $M_\star=11$~M$_\odot$, and has a very large Balmer Excess, $\Delta D_B=1.05$. Its placement on the HR-diagram and PMS tracks puts it at a very early stage of evolution, in which it appears to be able to accrete at great rates. This star appears to be an exception to the majority of other stars and is excluded in all fitting because of this. The overall trend is that $\dot{M}_{\rm acc}$ increases steadily with temperature; the temperature of a star is generally proportional to its mass leading to the next relationship.


\begin{figure}
\includegraphics[trim=1.0cm 0.5cm 0.25cm 0.25cm, width=\linewidth]{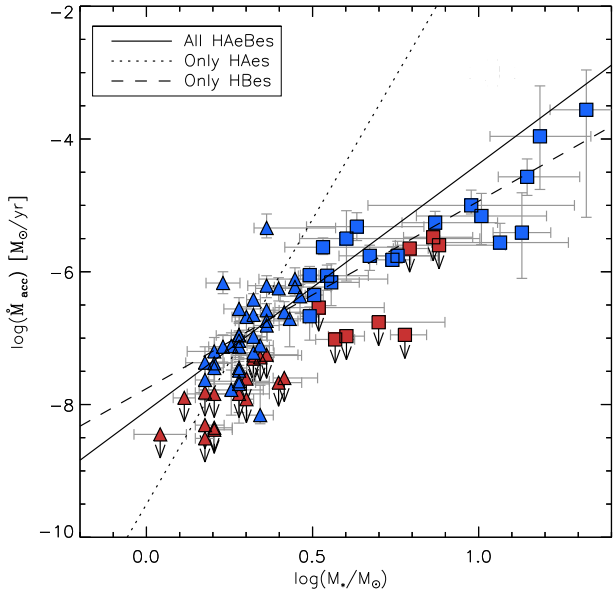}
\caption{ $\dot{M}_{\rm acc}$ versus $M_\star$ is plotted here, for all stars in which $\dot{M}_{\rm acc}$ could be determined. The stars are split into HAes (as triangles, where $M_\star < 3$~M$_\odot$) and HBes (as squares where $M_\star > 3$~M$_\odot$). Upper limits are denoted as the points in red with downward arrows from them. Separate fits are made to the HAes, HBes, and the group as a whole, of the form $\dot{M}_{\rm acc} \propto M_\star ^a$, where $a$ is found to be $8.59\pm 1.46$, $2.82\pm 0.51$, and $3.72\pm 0.27$, respectively. A discussion of the fits is provided in the text.}
\label{fig:macc_vs_mass}
\end{figure}

\begin{figure}
\includegraphics[trim=1.0cm 0.5cm 0.25cm 0.25cm, width=\linewidth]{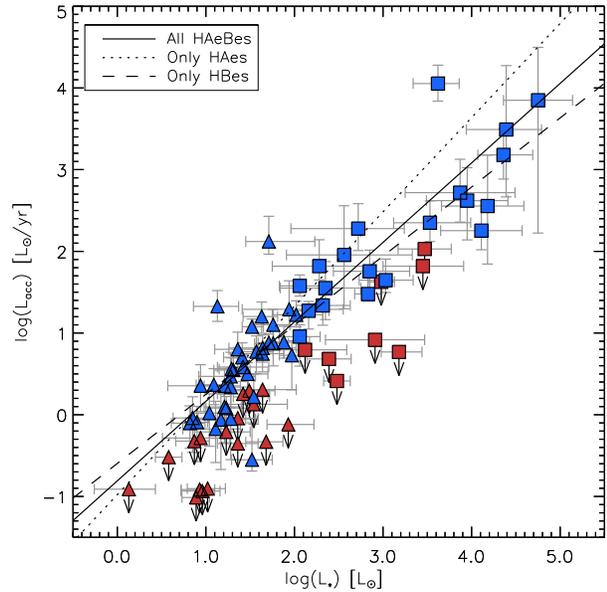}
\caption{$L_{\rm acc}$ versus $L_\star$ is plotted here, for all stars in which $\dot{M}_{\rm acc}$ could be determined. The stars are split into HAes (as triangles, where $M_\star < 3$~M$_\odot$) and HBes (as squares where $M_\star > 3$~M$_\odot$). Upper limits are denoted as the points in red with downward arrows from them. Separate fits are made to the HAes, HBes, and the group as a whole, of the form $L_{\rm acc} \propto L_\star ^a$, where $a$ is found to be $1.15 \pm 0.20$, $0.85 \pm 0.12$, and $0.97 \pm 0.06$, respectively. A discussion of the fits is provided in the text.}
\label{fig:lacc_vs_logl}
\end{figure}


Figure \ref{fig:macc_vs_mass} compares the  log($\dot{M}_{\rm acc}$) vs. log($M_\star$) relationship, along with a series of fits to the data. In the figure the stars are  split into two groups, the HAes and the HBes, which comprise 60\% and 40\% respectively of the total sample (the split between the two regimes is made at 3~$\rm M_\odot$). A best fit to the HAes is made of $\dot{M}_{\rm acc} \propto M_\star ^{8.59 \pm 1.40}$, while for the HBes a shallower relationship of $\dot{M}_{\rm acc} \propto M_\star ^{2.82 \pm 0.41}$ is seen. An overall fit to the HAeBes is obtained of $\dot{M}_{\rm acc} \propto M_\star ^{3.72 \pm 0.27}$, which lies between the HAe and HBe regime and favours the HBe case, which covers a greater mass distribution. 
When considering the HAeBes as a whole, the relationship found between $\dot{M}_{\rm acc}$ and  $M_\star$ is a factor of $\sim$2.0 larger in exponent than in low mass PMS stars, where $\dot{M}_{\rm acc} \propto M_\star ^{2.0\pm0.2}$ \citep{Muzerolle2005,Natta2006}. The trend we observe of a steeper relationship between accretion rate and stellar mass, over CTTs, agrees with the findings of \citet{Mendigutia2011b}, who also obtained a relationship with a much steeper exponent; they obtained $\dot{M}_{\rm acc} \propto M_\star ^{4.6-5.2}$. Although, the exponent for the HAeBes as a whole obtained in this work is slightly shallower than \citet{Mendigutia2011b}, this could be attributed to our sample containing more HBes. 
Overall, it is apparent that HAeBes have higher accretion rates and a steeper relationship to $M_\star $ than CTTs. This could be due to HAeBes being younger stars, which are in earlier stages of accretion. The different relationships observed between HAes and HBes, could be due to the HAes crossing into a transitional disc phase, in which accretion rates may be lower (see the discussion above in Section \ref{sec:dis_age}). This is also possible due to the HAes having a longer stage of evolution compared to HBes.\\

\begin{table*}
\centering
\begin{minipage}{100mm}
\caption{Summarises the exponents describing accretion relationships for various PMS mass groups. The relationships are of the form $A=B^C$, with $C$ being the exponent.  }
\label{tab:exponents}

\begin{tabular}{lccc}
\hline
Group & $\dot{M}_{\rm acc}$ vs. $M_{\rm acc}$  & $L_{\rm acc}$ vs. $L_{\rm acc}$ & $\dot{M}_{\rm acc}$ vs. age \\
\hline
CTTs & 2.00 $\pm$ 0.20 & $\sim$1.50 &  -1.20,  -1.50 -- -2.80 \\
CTTs+HAeBes & 2.47 $\pm$ 0.07 & 1.19 $\pm$ 0.03 &  N/A \\ 
HAes & 8.59 $\pm$ 1.46  & 1.15 $\pm$ 0.20 & -4.06 $\pm$ 0.53 \\ 
HBes & 2.82 $\pm$ 0.51  & 0.85 $\pm$ 0.12 & -1.93 $\pm$ 0.24 \\
HAeBes & 3.72 $\pm$ 0.27  & 0.97 $\pm$ 0.06 & -1.92 $\pm$ 0.09\\
\hline
\end{tabular}
  
\begin{minipage}{100mm}
  {\footnotesize References for CTTs values are from: \citet{Muzerolle2005, Natta2006} for Column 2 -- \citet{Natta2006, Tilling2008} for Column 3 -- and \citet{SiciliaAguilar2010, CarattioGaratti2012}  for the first value in Column 4, with \citet{Hartmann1998} for the second value.}
\end{minipage} 
  
\end{minipage}
\end{table*}

Alternatively, luminosities can be compared against each other instead of masses; Figure \ref{fig:lacc_vs_logl} shows the luminosity plot of log($L_{\rm acc}$/L$_\odot$) vs. log($L_\star$/L$_\odot$). A best fit to all the HAeBes is found of $L_{\rm acc} \propto L_\star^{0.97 \pm 0.06}$. This fit is in agreement with the work of \citet{Mendigutia2011b}, where these authors also found a shallower relationship for HAeBes of $L_{\rm acc} \propto L_\star^{1.2}$. As in the previous comparisons presented, when looking at the masses, the HAes and HBes are divided into two groups in the figure. A best fit to the HAes is obtained of $L_{\rm acc} \propto L_\star^{1.15 \pm 0.20}$, which is slightly shallower than the trends seen in CTTs of $L_{\rm acc} \propto L_\star^{1.5}$ \citep{Natta2006,Tilling2008}. The HBes, on the other hand, demonstrate an even shallower relationship of $L_{\rm acc} \propto L_\star^{0.85 \pm 0.12}$, this is turn shifts the weighting when looking at the HAeBes as a whole. 
The data is suggestive of $L_{\rm acc}$ being tightly correlated with $L_\star$, but the exact relationship changes in exponent between the HAe and HBe regime.

Table \ref{tab:exponents} is provided as a condensation of the various accretion relationships extracted from the graphs. It demonstrates clearly the changes between the different mass groups when looking at accretion as either a function of mass, luminosity, or age.


\section[]{Conclusions and Final Remarks}
\label{sec:conc}

To conclude, we have presented the largest spectroscopic survey  of HAeBes to date, obtaining the following results:

\begin{itemize}

\item Basic stellar parameters are determined for the whole sample, in a homogeneous fashion, by initially constraining $T_{\rm eff}$ and log(g) using some of the best spectra available for these stars. As such, the parameters are more consistent between objects, while only a handful of objects require specialist treatment. The findings are in agreement with previous works in the literature.

\item A UV-excess, $\Delta D_B$, is clearly detected in 62 stars of the sample, with upper limits allowed on a further 26 stars. $\dot{M}_{\rm acc}$ is determined for 81 of these stars through modelling within the context of magnetospheric accretion. However, 7 of the $\Delta D_B$ detections cannot be reproduced in this context. These 7 stars are all HBe stars, often with very large $\Delta D_B$ values of $> 0.85$, or high temperatures exceeding 20000~K. This suggests a possible breakdown in the MA regime for HBes, particularly for early-type HBes.

\item A clear trend is observed of $\dot{M}_{\rm acc}$ increasing as a function of stellar mass. The relationship obtained is a power law of the form: $\dot{M}_{\rm acc} \propto M_\star ^{3.72 \pm 0.27}$. This is a steeper law than previously observed in CTTs, which is only $\dot{M}_{\rm acc} \propto M_\star^{2.0}$. We interpret this increased exponent, in the relationship between $\dot{M}_{\rm acc}$ and $M_\star$, for HAeBes as a possible combination of them being younger and therefore more active in formation than older CTTs. Deviations are seen between the HAes, where $\dot{M}_{\rm acc} \propto M_\star ^{8.59 \pm 1.40}$, and the HBes, where $\dot{M}_{\rm acc} \propto M_\star ^{2.82 \pm 0.0.41}$. An explanation could be that the HAes are crossing into a transitional disc phase, in which accretion rates may be lower.

\item There is also a trend between the accretion luminosity and the stellar luminosity, which is found to be $L_{\rm acc} \propto L_\star^{10.97 \pm 0.06}$ for the sample. This is lower than found in CTTs where $L_{\rm acc} \propto L_\star^{1.5}$ \citep{Natta2006,Tilling2008}. However, for a subset of the HAes the relationship is much closer to the CTTs case, where we observe $L_{\rm acc} \propto L_\star^{1.15 \pm 0.20}$. In contrast a shallower relation of $L_{\rm acc} \propto L_\star^{0.85 \pm 0.12}$ is seen in the HBes. This demonstrates that the stellar luminosity of a star appears to be a good indicator of the accretion luminosity for a huge range of stellar luminosities, upto the HAe mass range, but there may be deviations in the HBe mass range.

\item A trend is also seen in the HAeBes between the age of the star and $\dot{M}_{\rm acc}$, where the accretion rate decreases with increasing age, characterised by the form $ \dot{M}_{\rm acc} \propto t^{-\eta}$, with $\eta=1.92 \pm 0.09$. This implies that the accretion rate decreases as stars approach the main sequence. However, this result is affected by two factors. The first is the most massive stars, with the higher accretion rates ,are only observable at young ages due to their rapid evolution. Secondly, the less massive stars have a longer PMS lifetime, which could allow their accretion rate to diminish within this time. These factors could explain the change in the relationship for the HAes case where $\eta=4.06 \pm 0.53$. Overall, this suggests that the younger objects are indeed accreting at a faster rate, and that the accretion rate diminishes more quickly  for older HAeBe stars, which could be a consequence of disc dissipation.

\end{itemize}

This study has led to three main findings. Firstly, the HAeBes display relationships in accretion which are similar but different to CTTs. $\dot{M}_{\rm acc}$ is observed to have a steeper relationship with $M_\star$ than seen in CTTs, while $L_{\rm acc}$ shows a shallower relationship with $L_\star$ than the CTTs case.
Secondly, there are also notable differences within the HAeBe group; when separating the HAes and HBes. Most notably, the HAes display a much steeper relationship in $\dot{M}_{\rm acc}$  when related to both age and $M_\star$. In both cases the steepness of the relationship is approximately double that seen in HBes. Although, the HAes also display a  $L_{\rm acc}$ relationship to $L_\star$ which is comparable to CTTs.
The third, and final, finding is that multiple early-type HBes, and stars with an observable $\Delta D_B$ of $> 0.85$, cannot be modelled successfully though magnetospheric accretion. This suggests that there is a possible change in accretion mechanisms in these stars which requires further investigation. To further these findings, the next steps are to look at emission lines which are known tracers of accretion in CTTs and test their applicability in HAeBes. This will be presented in Paper II of the series.


\section*{Acknowledgments}

The authors wish to thank the referee Benjam{\'{i}}n Montesinos for thorough and constructive comments, which have helped improve the clarity of the manuscript.
JRF gratefully acknowledges a studentship from the Science and Technology Facilities Council of the UK. JDI gratefully acknowledges funding from the European Union FP7-2011 under grant agreement no. 284405. This work has made use of NASA's Astrophysics Data System; it has also made use of the SIMBAD database, operated at CDS, Strasbourg, France.


\bibliographystyle{mn2e}

 \newcommand{\noop}[1]{}


\appendix

\section[]{Photometry}
\label{sec:app_phot}

This appendix serves as a reference source for the sample. Provided here is Table \ref{tab:phot}, which contains all of the photometry used in this work, along with references to the sources. Additionally, previously assigned literature values of distance and temperature are included for the whole sample (where possible, some do not have previous distance estimates).\\


\begin{table*}
 \centering
 \begin{minipage}{130mm}
 \label{tab:phot} 
  \caption{Photometry from the literature.}
\begin{tabular}{l cccccc cc cc}
  \hline
   Name   & U & B & V & R & I & Phot & Teff & Teff & D & D \\
        & (mag) & (mag) & (mag) & (mag) & (mag) & Ref & (K) & Ref & (pc) & ref \\
\hline
 
UX Ori & 10.94 & 10.71 & 10.34 & 10.12 &  9.88 & a &  8410 & i &  340 & ap \\ 
PDS 174 & 13.54 & 13.65 & 12.84 & 12.18 & 11.42 & b & 18700 & b &   340 & ap \\ 
V1012 Ori & 12.62 & 12.46 & 12.04 & 11.61 & 11.25 & c &  8600 & c &   340 & ap \\ 
HD 34282 & 10.15 & 10.05 &  9.89 &  9.81 &  9.71 & a &  8720 & i &   340 & ap \\ 
HD 287823 &  9.98 &  9.90 &  9.68 &  9.59 &  9.48 & b &  8720 & j &  340 & ap \\ 
HD 287841 & 10.63 & 10.50 & 10.21 & 10.06 &  9.89 & a &  8990 & i &   340 & ap \\ 
HD 290409 & 10.20 & 10.11 & 10.02 &  9.96 &  9.89 & b & 10500 & b &   340 & ap \\ 
HD 35929 &  8.71 &  8.53 &  8.12 &  7.87 &  7.61 & a &  6870 & k &  360 & aq \\ 
HD 290500 & 11.41 & 11.35 & 11.04 &  - &  - & d &  8970 & b &   470 & ap \\ 
HD 244314 & 10.42 & 10.30 & 10.10 &  9.96 &  9.80 & b &  8720 & l &  440 & ap \\ 
HK Ori & 11.72 & 11.79 & 11.41 & 11.05 & 10.66 & a &  8460 & m &  440 & ap \\ 
HD 244604 &  9.68 &  9.57 &  9.38 &  9.27 &  9.12 & a &  8720 & l &   440 & ap \\ 
UY Ori & 13.38 & 13.16 & 12.79 & 12.56 & 12.19 & b & 10500 & b &   510 & ap \\ 
HD 245185 & 10.02 & 10.00 &  9.91 &  9.87 &  9.82 & a &  9520 & l &   440 & ap \\ 
T Ori & 11.38 & 10.98 & 10.43 & 10.10 &  9.63 & a &  8660 & i &   510 & ap \\ 
V380 Ori & 10.80 & 11.04 & 10.53 & 10.11 &  9.50 & a &  9230 & n &  510 & ap \\ 
HD 37258 &  9.84 &  9.80 &  9.67 &  9.59 &  9.49 & a &  8970 & o &  510 & ap \\ 
HD 290770 &  9.18 &  9.30 &  9.27 &  9.23 &  9.18 & b & 10500 & b &   470 & ap \\ 
BF Ori & 10.34 & 10.05 &  9.82 &  9.68 &  9.48 & a &  8990 & i &   510 & ap \\ 
HD 37357 &  9.00 &  8.95 &  8.84 &  8.79 &  8.72 & a &  9230 & l &   510 & ap \\ 
HD 290764 & 10.29 & 10.20 &  9.88 &  9.68 &  9.44 & b &  7200 & b &   470 & ap \\ 
HD 37411 & 10.07 &  9.95 &  9.82 &  9.72 &  9.58 & a &  9100 & l &   510 & ap \\ 
V599 Ori & 17.07 & 15.41 & 13.76 & 12.69 & 11.55 & a &  7200 & b &   510 & ap \\ 
V350 Ori & 11.39 & 11.15 & 10.82 & 10.62 & 10.34 & a &  8990 & i &   510 & ap \\ 
HD 250550 &  9.34 &  9.61 &  9.54 &  9.32 &  9.54 & e & 10750 & l &   280 & ar \\ 
V791 Mon & 10.31 & 10.68 & 10.38 & 10.12 &  9.87 & b & 18700 & p & 1100 & as \\ 
PDS 124 & 13.15 & 12.97 & 12.44 & 12.15 & 11.81 & b &  9520 & b &   830 & w \\ 
LkHa 339 & 14.59 & 14.24 & 13.47 & 12.80 & 11.93 & a &  9230 & q &  830 & w \\ 
VY Mon & 15.28 & 14.56 & 12.97 & 11.82 & 10.60 & a &  8200 & i &   800 & at \\ 
R Mon & 12.17 & 12.53 & 11.93 & 11.41 & 10.87 & a & 12400 & i &   800 & at \\ 
V590 Mon & 12.52 & 12.75 & 12.60 & 12.42 & 12.12 & a & 13000 & r &  800 & at \\ 
PDS 24 & 13.94 & 13.62 & 13.26 & 12.98 & 12.69 & b & 10500 & b &   590 & ap \\ 
PDS 130 & 14.42 & 14.06 & 13.40 & 12.96 & 12.44 & b & 10500 & b &   830 & w \\ 
PDS 229N & 13.82 & 13.70 & 13.13 & 12.74 & 12.24 & b &  9520 & b &   830 & w \\ 
GU CMa &  5.88 &  6.56 &  6.54 &  6.47 &  6.37 & a & 25000 & s & 1050 & au \\ 
HT CMa & 12.55 & 12.29 & 11.87 & 11.38 & 11.87 & e &  9520 & q &  1050 & au \\ 
Z CMa & 11.20 & 10.50 &  9.25 &  8.40 &  7.65 & f & 30000 & f &  1050 & au \\ 
HU CMa & 11.72 & 11.84 & 11.55 & 11.32 & 11.16 & a & 11900 & q &  1050 & au \\ 
HD 53367 &  6.80 &  7.37 &  6.95 &  6.67 &  6.30 & a & 29500 & s &  1050 & au \\ 
PDS 241 & 12.33 & 12.71 & 12.06 & 11.45 & 11.11 & b & 30000 & b &  7000 & av \\ 
NX Pup &  9.93 &  9.96 &  9.63 &  9.38 &  9.07 & a &  7290 & t &  410 & ap \\ 
PDS 27 & 14.61 & 14.32 & 13.00 & 12.00 & 10.98 & b & 17500 & u & 2900 &u \\ 
PDS 133 & 13.57 & 13.61 & 13.13 & 12.79 & 12.50 & b & 14000 & b &  2500 & b \\ 
HD 59319 &  7.86 &  8.23 &  8.31 &  8.34 &  8.42 & a & 11900 & v & - & -  \\ 
PDS 134 & 12.50 & 12.61 & 12.20 & 11.92 & 11.65 & b & 14000 & b &  - & -  \\ 
HD 68695 & 10.00 &  9.92 &  9.82 &  9.76 &  9.66 & b &  9520 & w &  410 & ap \\ 
HD 72106 &  8.39 &  8.50 &  8.50 &  8.49 &  8.49 & b &  9810 & x &  370 & ap \\ 
TYC 8581-2002-1 & 12.18 & 11.94 & 11.48 & 11.19 & 10.95 & b & $\dagger$ 8200 & b &   145 & ap \\ 
PDS 33 & 12.85 & 12.63 & 12.34 & 12.16 & 11.97 & b &  9520 & b &   370 & ap \\ 
HD 76534 &  7.68 &  8.18 &  8.07 &  7.97 &  7.84 & a & 20350 & y &  370 & ap \\ 
PDS 281 &  9.43 &  9.46 &  8.87 &  8.50 &  8.08 & b & 17050 & b &   370 & ap \\ 
PDS 286 & 14.39 & 13.91 & 12.15 & 10.91 &  9.76 & b & 30000 & b &   370 & ap \\ 
PDS 297 & 12.50 & 12.34 & 12.03 & 11.83 & 11.59 & b &  7850 & b &   145 & ap \\ 
HD 85567 &  8.11 &  8.65 &  8.51 &  8.33 &  8.08 & a & 12450 & z &  650 & aq \\ 
HD 87403 &  9.28 &  9.31 &  9.26 &  9.22 &  9.16 & b & 10100 & aa &  145 & ap \\ 
PDS 37 & 15.56 & 15.06 & 13.54 & 12.38 & 11.21 & b & 17500 & u & 3700 &u \\ 
HD 305298 & 10.36 & 11.07 & 10.86 & 10.66 & 10.47 & b & 36900 & ab & - & -  \\ 
HD 94509 &  9.01 &  9.15 &  9.12 &  9.10 &  9.10 & a &  9730 & ac & - & -  \\ 
HD 95881 &  8.53 &  8.36 &  8.19 &  - &  - & g &  8990 & ad &  118 & ap \\ 
HD 96042 &  7.89 &  8.60 &  8.47 &  8.36 &  8.23 & b & 25400 & ad &  - & -  \\ 
HD 97048 &  8.96 &  8.80 &  8.44 &  8.20 &  7.95 & a & 10010 & ae &  160 & aq \\ 

  \hline
  \end{tabular}
\end{minipage}
\end{table*}


\begin{table*}
 \centering
 \begin{minipage}{130mm}
  \contcaption{}
 \begin{tabular}{l cccccc cc cc}
  \hline
   Name   & U & B & V & R & I & Phot & Teff & Teff & D & D \\
        & (mag) & (mag) & (mag) & (mag) & (mag) & Ref & (K) & Ref & (pc) & ref \\
  \hline

HD 98922 &  6.74 &  6.82 &  6.77 &  6.69 &  6.61 & a & 10500 & w &   850 & aq \\ 
HD 100453 &  8.10 &  8.07 &  7.78 &  7.60 &  7.42 & b &  7390 & aa &   122 & aq \\ 
HD 100546 &  6.60 &  6.70 &  6.69 &  6.67 &  6.66 & a & 10500 & af &   97 & aq \\ 
HD 101412 &  9.57 &  9.42 &  9.24 &  9.13 &  9.00 & b & 10010 & aa &   118 & ap \\ 
PDS 344 & 13.09 & 13.40 & 13.15 & 12.95 & 12.77 & b & 15400 & b &  - & -  \\ 
HD 104237 &  6.64 &  6.73 &  6.52 &  6.38 &  6.23 & a &  8410 & z &   115 & aq \\ 
V1028 Cen & 10.39 & 10.70 & 10.61 & 10.48 & 10.33 & a & 14100 & z &   130 & aq \\ 
PDS 361S & 13.10 & 13.35 & 12.85 & 12.49 & 12.09 & b & 18700 & b &  - & -  \\ 
HD 114981 &  6.55 &  7.13 &  7.23 &  7.27 &  7.33 & b & $\dagger$15400 & b &   550 & aq \\ 
PDS 364 & 13.85 & 13.93 & 13.46 & 13.05 & 12.63 & b & 11900 & ag &  118 & ap \\ 
PDS 69 &  9.92 & 10.12 &  9.80 &  9.50 &  9.12 & b & 17050 & ah &  630 & ah \\ 
DG Cir & 15.96 & 15.87 & 14.75 & 13.96 & 13.06 & a & 15000 & ai &  700 & aw \\ 
HD 132947 &  8.87 &  8.96 &  8.91 &  8.89 &  8.89 & a & 10500 & af &  - & -  \\ 
HD 135344B &  9.14 &  9.14 &  8.63 &  8.16 &  7.83 & h &  6590 & aj &  140 & ap \\ 
HD 139614 &  8.67 &  8.64 &  8.40 &  8.26 &  8.11 & b &  7850 & aj &   140 & ap \\ 
PDS 144S & 13.59 & 13.28 & 12.79 & 12.49 & 12.16 & b &  8200 & b &   1000 & b \\ 
HD 141569 &  7.23 &  7.20 &  7.10 &  7.03 &  6.95 & a &  9520 & aj &   116 & aq \\ 
HD 141926 &  8.72 &  9.20 &  8.64 &  8.21 &  7.77 & b & 20300 & b &  - & -  \\ 
HD 142666 &  9.42 &  9.17 &  8.67 &  8.35 &  8.01 & b &  7580 & aj &   145 & ap \\ 
HD 142527 &  9.20 &  9.15 &  8.27 &  - &  - & g &  6260 & ak &  140 & ak \\ 
HD 144432 &  8.64 &  8.53 &  8.17 &  7.94 &  7.72 & a &  7350 & i &   160 & aq \\ 
HD 144668 &  7.28 &  7.11 &  6.78 &  6.57 &  6.38 & a &  7930 & al &  160 & aq \\ 
HD 145718 & 10.00 &  9.62 &  9.10 &  8.79 &  8.45 & b &  8200 & ag &   145 & ax \\ 
PDS 415N & 13.43 & 12.96 & 12.04 & 11.47 & 10.85 & b &  7200 & b &   120 & ay \\ 
HD 150193 &  9.69 &  9.33 &  8.80 &  8.41 &  7.97 & a & 10010 & af &   120 & ay \\ 
AK Sco &  9.56 &  9.53 &  8.90 &  8.54 &  8.18 & a &  6450 & am &  130 & aq \\ 
PDS 431 & 14.20 & 13.99 & 13.42 & 13.02 & 12.59 & b &  9520 & b &   145 & ap \\ 
KK Oph & 13.17 & 12.97 & 12.36 & 11.83 & 11.03 & a &  8030 & an &  145 & ap \\ 
HD 163296 &  7.00 &  6.96 &  6.85 &  6.80 &  6.71 & a &  8720 & l &   119 & aq \\ 
MWC 297 & 14.94 & 14.27 & 12.03 & 10.18 &  8.80 & a & 23700 & ao &  250 & ao \\ 

  \hline
  \end{tabular}
\end{minipage}


\begin{minipage}{130mm}

{\footnotesize 
$\dagger$ These two stars are listed as objects QT3 (TYC 8581-2002-1) and QT4 (HD 114981) in the first Table of \citet{Vieira2003}. However, their places appear swapped in the second table by these authors. This swap is supported by additional photometry of HD 114981 and by the authors observed temperatures and the temperatures derived in this work. Based on this we have swapped the photometry from \citet{Vieira2003} around for these two stars.
References: 
(a) \citet{deWinter2001}, 
(b) \citet{Vieira2003}, 
(c) \citet{Miroshnichenko1999}, 
(d) \citet{Guetter1979}, 
(e) \citet{Herbst1999}, 
(f) \citet{vandenAncker2004}, 
(g) \citet{Malfait1998}, 
(h) \citet{Coulson1995}, 
(i) \citet{Mora2001}, 
(j) \citet{Hernandez2005}, 
(k) \citet{Miroshnichenko2004}, 
(l) \citet{Gray1998}, 
(m) \citet{Baines2004}, 
(n) \citet{Finkenzeller1984}, 
(o) \citet{Gray1993}, 
(p) \citet{Cidale2001}, 
(q) \citet{Hernandez2004}, 
(r) \citet{Perez2008}, 
(s) \citet{TjinADjie2001}, 
(t) \citet{Finkenzeller1985}, 
(u) \citet[accepted]{Ababakr2015}, 
(v) \citet{Houk1988}, 
(w) \citet{Herbst1976}, 
(x) \citet{Houk1982}, 
(y) \citet{Valenti2000}, 
(z) \citet{vandenAncker1998}, 
(aa) \citet{Guimaraes2006}, 
(ab) \citet{Graham1970}, 
(ac) \citet{Stephenson1971}, 
(ad) \citet{Houk1975}, 
(ae) \citet{Whittet1987}, 
(af) \citet{Levenhagen2006}, 
(ag) \citet{Carmona2010}, 
(ah) \citet{Reipurth1993}, 
(ai) \citet{Gahm1980}, 
(aj) \citet{Dunkin1997}, 
(ak) \citet{Fukagawa2006}, 
(al) \citet{TjinADjie1989}, 
(am) \citet{Andersen1989}, 
(an) \citet{Herbig2005}, 
(ao) \citet{Drew1997}, 
(ap) \citet{deZeeuw1999}, 
(aq) \citet{vanLeeuwen2007}, 
(ar) \citet{Canto1984}, 
(as) \citet{Hilton1995}, 
(at) \citet{Dahm2005}, 
(au) \citet{Shevchenko1999}, 
(av) \citet{Avedisova2000}, 
(aw) \citet{Franco1990},
(ax) \citet{Preibisch2008},
(ay) \citet{Loinard2008}.  }

\end{minipage}

\end{table*}

\section[]{Exceptional Stars}
\label{sec:app_except}

Seven of the stars in the sample cannot be assigned a temperature from the spectra alone (see Section \ref{sec:st_tng}). For these objects a different approach must be undertaken on an individual basis in order to assign a limiting temperature. This is done by drawing upon as many literature sources on these objects as possible. Fortunately, there are very few objects in the sample which require this specialist treatment. The stars, and steps taken towards them are detailed below:\\

\textit{VY Mon} -- This star is included here because it has the worst SNR of the sample. This makes accurate spectral typing difficult, but a cautious estimate of around ~12000~K can be made for the temperature. This agrees with literature estimates of 8200-12000~K \citep{Mora2001,Manoj2006}. A generous error of 4000~K is adopted.

\textit{R Mon} -- In the spectra of R Mon all lines are seen in emission or as P-Cygni profiles, making any temperature estimate impossible from spectra alone. The temperature has been previously listed as around 12000~K in past works \citep{Mora2001,Manoj2006}. We adopt this literature temperature.

\textit{Z CMa} -- Has lots of P-Cygni and emission lines in its spectra, but lacks absorption features for spectral typing. Again we must turn to the literature. In the literature this star is seen to have the largest spread in listed temperatures; ranging from 30000~K \citep{vandenAncker2004,Manoj2006}, down to 11500~K \citep{Donehew2011} and 8500~K \citep{Hinkley2013}. We choose to adopt, and test, the most recent temperature from \citet{Hinkley2013}. This is because their work spatially resolves the Herbig star in this system from its FU Or-like companion. In addition to this they provide SED fitting to the observed photometry to determine the temperature.

\textit{PDS 27 and PDS 37} -- These two objects display very strong emission and P-Cygni profiles. They are also the focus of a recent paper by \citet[accepted]{Ababakr2015} who determine distances and stellar parameters of the objects. We adopt their stellar parameters and distances in our work as they follow a similar methodology. The temperatures they found of $\sim$21000~K, for both objects, are in agreement with the values found by \citet{Vieira2003}.

\textit{PDS 133} -- Another star devoid of any photospheric absorption in its spectra, and has extremely strong emission lines (the equivalent width of H$\alpha$ is $\sim$-100$\rm \AA$). Therefore, using the spectra to assign a temperature is impossible. For this reason we adopt a temperature around 14000~K, based on the literature \citep{Vieira2003}.

\textit{DG Cir} -- Another star with reasonably strong emission; the Balmer series are seen as P-Cygni profiles. A broad spectral type of class B has previously been assigned to this star by \citet{Sanduleak1973} and \citet{Vieira2003}. The authors \citet{Gahm1980} do not give a spectral type but note its similarities to V380 Ori. A small indication of absorption lines can be seen around 5200~$\rm \AA$, but they appear close to many emission lines making an exact temperature determination difficult. We therefore agree with a B spectral type, and based on the absorption would narrow this to a late-B type star of $\sim$11000~K, with a generous error of 3000~K.

\label{lastpage}


\end{document}